%% file: main.tex
\pdfoutput=1
\documentclass[sigconf, nonacm]{acmart}

\usepackage{xspace}
\usepackage[vlined, ruled, linesnumbered]{algorithm2e}
\usepackage{multirow}
\usepackage[shortlabels]{enumitem}
\usepackage{subcaption}
\usepackage[skins, fitting]{tcolorbox}
\usepackage{balance}

\newenvironment{tightitemize}{\begin{list}{$\bullet$}{\setlength{\rightmargin}{0em}\setlength{\leftmargin}{1em}\setlength{\topsep}{0.5mm}\setlength{\itemsep}{-0.6mm}\setlength{\itemindent}{1em}}}{\end{list}}

\newif\iftr
\trtrue 

\iftr
\newcommand{\ptr}[1]{#1}
\newcommand{\ppaper}[1]{}
\else
\newcommand{\ptr}[1]{}
\newcommand{\ppaper}[1]{#1}
\fi

\newcommand{\mon}{monotonicity\xspace}
\newcommand{\Mon}{Monotonicity\xspace}
\newcommand{\MonS}{\code{$M$}\xspace}
\newcommand{\AvaS}{\code{$V$}\xspace}
\newcommand{\con}{consistency\xspace}
\newcommand{\Con}{Consistency\xspace}

\newcommand{\ava}{visibility\xspace}
\newcommand{\Ava}{Visibility\xspace}

\newcommand{\unavailable}{invisible\xspace}

\newcommand{\conC}{\con-committed\xspace}
\newcommand{\ConC}{\Con-committed\xspace}
\newcommand{\cc}{\code{$C_c$}\xspace}

\newcommand{\conM}{\con-minimal\xspace}
\newcommand{\ConM}{\Con-minimal\xspace}
\newcommand{\cm}{\code{$C_m$}\xspace}

\newcommand{\conF}{\con-fresh\xspace}
\newcommand{\ConF}{\Con-fresh\xspace}
\newcommand{\cf}{\code{$C_f$}\xspace}

\newcommand{\viewG}{view graph\xspace}

\newcommand{\ViewGG}{View Graph\xspace}

\newcommand{\writeTxn}{write transaction\xspace}
\newcommand{\WriteTxn}{Write transaction\xspace}
\newcommand{\readTxn}{read transaction\xspace}
\newcommand{\ReadTxn}{Read transaction\xspace}
\newcommand{\txn}{transaction\xspace}

\newcommand{\WTxnMng}{WriteTxn~manager\xspace}

\newcommand{\RTxnMng}{ReadTxn~manager\xspace}
\newcommand{\pRTxn}{ReadTxn\xspace}

\newcommand{\fullTdb}{Transactional Panorama\xspace}
\newcommand{\tdb}{transactional panorama\xspace}
\newcommand{\Tdb}{Transactional panorama\xspace}
\newcommand{\TDB}{Transactional Panorama\xspace}

\newcommand{\macThe}{MVC~Theorems\xspace}
\newcommand{\macPro}{MVC~properties\xspace}
\newcommand{\MacPro}{MVC~Properties\xspace}

\newcommand{\lastread}{LastRead\xspace}
\newcommand{\metaInfo}{MetaInfo\xspace}
\newcommand{\ts}{t_s\xspace}
\newcommand{\tc}{t_c\xspace}
\newcommand{\civ}{c_{\code{uc}}\xspace}
\newcommand{\tr}{t_r\xspace}
\newcommand{\itemlist}{item list\xspace}
\newcommand{\pitem}{item\xspace}

\newcommand{\stitle}[1]{\vspace{0.2em}\noindent\textbf{#1.}}

\newcommand{\code}[1]{\textsf{\small\mdseries #1}}
\newcommand{\serial}{serializability\xspace}

\newcommand{\uaMetric}{invisibility\xspace}
\newcommand{\UaMetric}{Invisibility\xspace}
\newcommand{\uaFunc}{I\xspace}
\newcommand{\snMetric}{staleness\xspace}
\newcommand{\SnMetric}{Staleness\xspace}
\newcommand{\snFunc}{S\xspace}
\newcommand{\vt}{n\xspace}
\newcommand{\Vt}{N\xspace}
\newcommand{\vs}{v\xspace}
\newcommand{\Vs}{V\space}

\newcommand{\bt}{b\xspace}

\newcommand{\WS}{N\xspace}
\newcommand{\RR}{H\xspace}
\newcommand{\RG}{O\xspace}

\newcommand{\TxnT}{Time\xspace}
\newcommand{\RTime}{D\xspace}
\newcommand{\ExecTime}{Q\xspace}
\newcommand{\Metric}{P\xspace}
\newcommand{\vdef}{view definition\xspace}
\newcommand{\qres}{view result\xspace}
\newcommand{\Qres}{View result\xspace}
\newcommand{\pqres}{result\xspace}
\newcommand{\res}{result\xspace}

\newcommand{\vstate}{view state\xspace}
\newcommand{\pvstate}{state\xspace}

\newcommand{\vd}{view definition\xspace}

\newcommand{\TxW}{W\xspace}
\newcommand{\txw}{w\xspace}
\newcommand{\TxR}{R\xspace}
\newcommand{\txr}{r\xspace}

\newcommand{\GCFB}{Globally-Consistent Fully-Blocking\xspace}
\newcommand{\pGCFB}{GCFB\xspace}
\newcommand{\GCPB}{Globally-Consistent Partially-Blocking\xspace}
\newcommand{\pGCPB}{GCPB\xspace}
\newcommand{\GCNB}{Globally-Consistent Non-Blocking\xspace}
\newcommand{\pGCNB}{GCNB\xspace}
\newcommand{\LCNB}{Locally-Consistent Non-Blocking\xspace}
\newcommand{\pLCNB}{LCNB\xspace}
\newcommand{\LCMB}{Locally-Consistent Minimum-Blocking\xspace}
\newcommand{\pLCMB}{LCMB\xspace}
\newcommand{\ICNB}{Inconsistent Non-Blocking\xspace}
\newcommand{\pICNB}{ICNB\xspace}

\newcommand{\kGCNB}{$k$-GCNB\xspace}
\newcommand{\kLCNB}{$k$-LCNB\xspace}
\newcommand{\kLCMB}{$k$-LCMB\xspace}
\newcommand{\relaxed}{relaxed\xspace}

\newcommand{\regMove}{Regular~Move\xspace}
\newcommand{\waitMove}{Wait~and~Move\xspace}
\newcommand{\ranMove}{Random~Move\xspace}
\newcommand{\exploreRange}{explore range\xspace}
\newcommand{\Explorerange}{Explore range\xspace}

\newcommand{\Viewportsize}{Viewport~size\xspace}
\newcommand{\viewportSize}{viewport~size\xspace}

\newcommand{\Refreshinterval}{Refresh~interval\xspace}
\newcommand{\prefreshinterval}{refresh~interval\xspace}

\newcommand{\readb}{read behavior\xspace}
\newcommand{\Readb}{Read behavior\xspace}

\newcommand{\noOpt}{NoOpt\xspace}

\newcommand{\metricOpt}{MetricOpt\xspace}
\newcommand{\execOpt}{Antifreeze\xspace}
\newcommand{\bothOpt}{transactional panorama\xspace}
\newcommand{\pBothOpt}{TP\xspace}
\newcommand{\model}{model\xspace}
\newcommand{\framework}{framework\xspace}
\newcommand{\undercompute}{under-computation\xspace}
\newcommand{\uc}{\code{UC}\xspace}
\newcommand{\ucs}{\code{UCs}\xspace}
\newcommand{\ua}{\code{UC}\xspace}
\newcommand{\uas}{\code{UCs}\xspace}
\newcommand{\version}{version\xspace}
\newcommand{\versions}{versions\xspace}
\newcommand{\pG}{graph\xspace}
\newcommand{\oldAndNew}{committed and latest\xspace}
\newcommand{\oldOrNew}{committed or latest\xspace}
\newcommand{\oldG}{committed graph\xspace}
\newcommand{\newG}{latest graph\xspace}

\newcommand{\pSN}[1]{S{#1}}
\newcommand{\pIV}[1]{I{#1}}
\newcommand{\bBaseApproach}{Base Lens\xspace}

\newcommand{\baseApproach}{base lens\xspace}

\newcommand{\approach}{lens\xspace}
\newcommand{\Approach}{Lens\xspace}
\newcommand{\approaches}{lenses\xspace}
\newcommand{\Approaches}{Lenses\xspace}

\newenvironment{denselist}{
    \begin{list}{\tiny{$\bullet$}}%
    {\setlength{\itemsep}{0ex} \setlength{\topsep}{0ex}
    \setlength{\parsep}{0pt} \setlength{\itemindent}{0pt}
    \setlength{\leftmargin}{1em}
    \setlength{\partopsep}{0pt}}}%
    {\end{list}}

\newenvironment{itemize*}{
    \vspace*{-0.05in}
    \begin{itemize}
        \setlength{\itemsep}{0pt}
        \setlength{\parsep}{1pt}
        \setlength{\topsep}{1pt}
        \setlength{\partopsep}{0pt}
        \setlength{\leftmargin}{1em}
        \setlength{\labelwidth}{1em}
        \setlength{\labelsep}{0.5em}
    }{\end{itemize}
    \vspace*{-0.05in}
}
\newenvironment{enumerate*}{
    \vspace*{-0.05in}
    \begin{enumerate}
        \setlength{\itemsep}{0pt}
        \setlength{\parsep}{3pt}
        \setlength{\topsep}{3pt}
        \setlength{\partopsep}{0pt}
        \setlength{\leftmargin}{2em}
        \setlength{\labelwidth}{1.5em}
        \setlength{\labelsep}{0.5em}
    }{\end{enumerate}
    \vspace*{-0.05in}
}

\definecolor{darkgreen}{rgb}{0.0, 0.2, 0.13}
\definecolor{auburn}{rgb}{0.43, 0.21, 0.1}
\definecolor{antiquefuchsia}{rgb}{0.57, 0.36, 0.51}
\definecolor{armygreen}{rgb}{0.29, 0.33, 0.13}
\definecolor{ao}{rgb}{0.0, 0.5, 0.0}
\definecolor{purple}{rgb}{0.75, 0.0, 1.0}
\definecolor{yellow}{rgb}{0.99, 0.76, 0.0}

\iftr
\newenvironment{reviewone}{\par}{\par}
\newenvironment{reviewtwo}{\par}{\par}
\newenvironment{reviewthree}{\par}{\par}
\newenvironment{reviewmulti}{\par}{\par}
\newcommand{\rone}[1]{#1}
\newcommand{\rtwo}[1]{#1}
\newcommand{\rthree}[1]{#1}
\else
\newenvironment{reviewone}{\par\color{blue}}{\par}
\newenvironment{reviewtwo}{\par\color{blue}}{\par}
\newcommand{\rone}[1]{\textcolor{blue}{#1}}
\newcommand{\rtwo}[1]{\textcolor{blue}{#1}}
\newcommand{\rthree}[1]{\textcolor{blue}{#1}}
\fi

\newcommand{\varname}[1]{\textit{#1}\xspace}
\DeclareMathOperator*{\argmax}{arg\,max}

\newcommand{\AceAuthors}{Dixin Tang, Indranil Gupta, Aditya G. Parameswaran}

\usepackage[english]{babel}
\newtheorem{theorem}{Theorem}

\newtheorem{definition}{Definition}

\newcommand\vldbdoi{XX.XX/XXX.XX}
\newcommand\vldbpages{XXX-XXX}
\newcommand\vldbvolume{14}
\newcommand\vldbissue{1}
\newcommand\vldbyear{2020}

\newcommand\vldbtitle{\shorttitle} 
\newcommand\vldbavailabilityurl{https://github.com/totemtang/ACE}

\renewcommand{\baselinestretch}{0.96}

\begin{document}

\title{\fullTdb: A Conceptual Framework for \\ User Perception in Analytical Visual Interfaces\vspace{-4mm}}

\author{Dixin Tang$^1$, Alan Fekete$^2$, Indranil Gupta$^3$, Aditya G. Parameswaran$^1$}
\affiliation{UC Berkeley$^1$ $|$ The University of Sydney$^2$ $|$ University of Illinois Urbana-Champaign$^3$}
\email{{totemtang, adityagp}@berkeley.edu, alan.fekete@sydney.edu.au, indy@illinois.edu}



\begin{abstract}
Many tools empower analysts and data scientists
to consume analysis results in a visual interface.
When the underlying data changes, these results
need to be updated,
but this update can take a long time---all
while the user continues to explore the results.
Tools can either (i) hide away
results that haven't been updated, hindering exploration;
(ii) make the updated results immediately available to the user
(on the same screen as old results), leading to confusion and incorrect insights;
or (iii) present old---and therefore stale---results to the user during the update. 
To help users reason about these options and others, and make appropriate trade-offs,
we introduce \TDB,
a formal framework that adopts \txn{s}
to jointly model the system refreshing 
the analysis results and 
the user interacting with them.
We introduce three key properties that
are important for user perception in this context:  \ava (allowing
users to continuously explore results),
\con (ensuring that results presented are from the
same version of the data), and \mon (making sure
that results don't ``go back in time''). 
Within \tdb, we characterize all 
feasible property combinations, 
design new mechanisms (that we call {\em \approaches}) for  
presenting analysis results to the user 
while preserving a given property combination, formally prove their relative
orderings for various performance criteria, and discuss their use cases. 
We propose novel algorithms to preserve each property combination 
and efficiently present fresh analysis results. 
We implement our framework
into a popular, open-source BI 
tool, 
illustrate the relative performance implications of different \approach{es}, and 
demonstrate the benefits of the novel \approach{es} and our optimizations.  
\end{abstract}

\maketitle

\pagestyle{plain}

\input{intro}

\input{model}

\input{system}

\input{prototype}
\input{experiment}
\input{related_conclusion}

\bibliographystyle{abbrv}
\bibliography{ref}

\end{document}

%% file: intro.tex
\section{Introduction}
\label{sec:intro}

Many data-centric tools
empower a user to visually
organize, present, and consume
multiple data analysis results  
within a single interface, such as a dashboard.
Each such analysis result is represented on this interface as
a scalar value, table, or visualization, 
and is computed using the 
source data or other analysis results, in turn, as {\em views}.
This pattern appears in a variety of contexts:

\vspace{1pt}
\noindent
{\em Visual analytics} or Business Intelligence (BI)  tools,
like Tableau~\cite{tableau} or PowerBI~\cite{powerBI},
empower a user to embed visualizations on a dashboard, each via a SQL query 
on an underlying database;

\vspace{1pt}
\noindent
{\em Spreadsheet tools}, such as Microsoft Excel~\cite{msexcel} and
Google Sheets~\cite{sheets}, allow a user to add derived computation
in the form of spreadsheet formulae, visualizations,
and pivot tables;

\vspace{1pt}
\noindent
{\em Data application builder 
tools}, such as Streamlit~\cite{streamlit}, Plotly~\cite{plotly}, 
and Redash~\cite{redash}, enable a user to efficiently develop
interactive dashboards,
employing computation done in Python UDFs and pandas dataframe functions, 
and SQL; and

\vspace{1pt}
\noindent
{\em Monitoring and observability tools}, 
such as Datadog~\cite{datadog}, Kibana~\cite{kibana},
 and Grafana~\cite{grafana},
 empower a user to make sense of their telemetry 
 data and logs via a combination of automatically
 defined and customizable dashboard widgets.

\vspace{1pt}
\noindent 
In all of these contexts, there is {\em a network of views
defined on underlying data, each of
which is then visualized on an interface}. 
These views and the corresponding visualizations often
need to be refreshed when the source data is modified.
For example, a dashboard in a BI tool
is refreshed with respect to 
regular changes to the underlying database tables
(e.g., new batches of data). 
However, this refresh is rarely instantaneous, especially on large datasets.
This represents a challenge, since the user is continuously
exploring the visualizations during the refresh. 
On the one hand, refreshing visualizations arbitrarily can be jarring to the user, since different
visualizations on the screen may be in different stages of being refreshed. 
On the other hand,
not refreshing them in a timely manner can lead to stale results.
The question we explore is:
{\bf \em How do we allow users to continuously explore results in a visual interface,
while ensuring that the results are not confusing
or stale?}

\begin{table}[t]
\fontsize{6.5}{8}\selectfont
\begin{tabular}{l@{\hspace{1mm}}l@{\hspace{-3mm}}c@{\hspace{-3mm}}c@{\hspace{-3mm}}c}
\Approach name                                                                  & Example Tools                                                            & \multicolumn{1}{l}{Monotonicity} & \multicolumn{1}{l}{Visibility} & \multicolumn{1}{l}{Consistency} \\ \hline
\begin{tabular}[c]{@{}l@{}}Globally-Consistent\\ Fully-Blocking (\pGCFB)\end{tabular}                    & \begin{tabular}[c]{@{}l@{}}MS Excel~\cite{msexcel}\\ Libre Calc~\cite{libreCalc}\\ Tableau~\cite{tableau}\end{tabular}  & Yes                              & No                             & Yes                             \\ \hline
\begin{tabular}[c]{@{}l@{}}Globally-Consistent\\ Partially-Blocking (\pGCPB)\end{tabular}         & \begin{tabular}[c]{@{}l@{}}Power BI~\cite{powerBI}\\ Superset~\cite{superset}\\ Dataspread~\cite{Antifreeze}\end{tabular} & Yes                              & No                             & Yes                             \\ \hline
\begin{tabular}[c]{@{}l@{}}Inconsistently\\ Non-Blocking (\pICNB)\end{tabular}       & Google Sheets~\cite{sheets}                                                            & Yes                              & Yes                            & No                              \\ \hline
\end{tabular}
\caption{\small Properties maintained by existing tools}
\label{tbl:existing_work}
\vspace{-8mm}
\end{table}

\begin{figure}[!t]
    \centering
    \includegraphics[width=85mm]{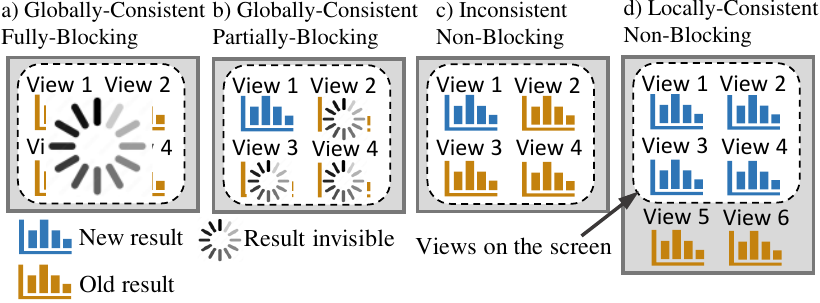}
    \vspace{-8mm}
    \caption{Visual examples of different \approaches for refreshing views in a dashboard}
    \label{fig:existing_tools}
    \vspace{-7mm}
\end{figure}

Unfortunately, existing tools make fixed, and somewhat arbitrary
decisions on how to address this question.
For example, Excel~\cite{msexcel}, Calc~\cite{libreCalc}, and Tableau~\cite{tableau}
block the user from exploring  the interface
until all of the views are refreshed (Figure~\ref{fig:existing_tools}a).
Other tools, like PowerBI~\cite{powerBI}, Superset~\cite{superset}, 
and Dataspread~\cite{Antifreeze},
improve on this approach by hiding (or greying) away any views 
that have not yet been refreshed,
while still letting the user explore the other up-to-date views (Figure~\ref{fig:existing_tools}b).
Yet other tools, like Google Sheets~\cite{sheets},
opt for not hiding any views, and instead just progressively
make them available as they are refreshed---this approach
has the downside of different results on the screen being
in different stages of being refreshed, leading
to incorrect insights (Figure~\ref{fig:existing_tools}c).

\stitle{\TDB and Underlying Properties} In this paper, 
we introduce a formal framework, 
named 
{\em \tdb}\footnote{We call this framework as such because it
involves adapting transactions to a problem of 
fidelity across various viewpoints (screens) over space and time, i.e., a panorama.}, to 
enable users and system designers to reason about 
the aforementioned 
question in a more principled manner. 
We adopt \txn{s} to jointly model the system 
concurrently updating visualizations,
with the user consuming these visualizations, over time
and space (i.e., across screens).
To the best of our knowledge, \emph{\tdb is the first framework that leverages 
\txn{s} to reason about correct user perception in visual interfaces}.
In this setting, we define three key desirable properties
{\em monotonicity}, {\em visibility}, and {\em consistency}, 
which we call the {\em \macPro}. 
\Mon guarantees that if a user reads the result 
for a view, any subsequent read will always 
return the same or more recent result (i.e., monotonic read~\cite{DSBook})
so that results never go ``back in time''. 
\Ava guarantees that the user can always explore
the result of any visualization
on the screen---instead of them being greyed out. \Con guarantees that the results 
displayed on the screen should be consistent 
with the same snapshot of source data~\cite{GolabJ11Consistency, ZhugeGW97Multiview}---enabling correct derivation of 
relationships between results on the same screen. 

\begin{figure}[!t]
    \centering
    \includegraphics[height=35mm]{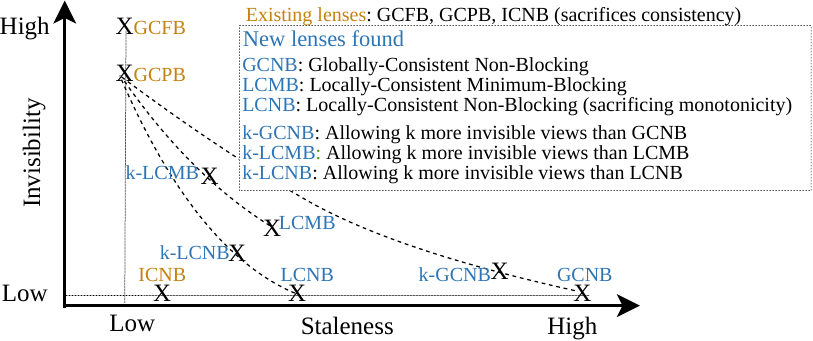}
    \vspace{-4mm}
    \caption{Trade-off between \uaMetric and \snMetric across different \approach{es}}
    \label{fig:trade-off_full}
    \vspace{-8mm}
\end{figure}

\stitle{Concrete Property Combinations via \Approaches} 
There are various mechanisms we can use to 
present results to the user in a visual interface,
resulting
in concrete selections for the aforementioned properties,
that we call {\em \approaches}\footnote{These are called lenses
since they capture various instantiations of our \tdb framework.}.
Consider our examples of existing systems (Figure~\ref{fig:existing_tools}a--c);
we list the corresponding three \approaches in Table~\ref{tbl:existing_work}---\pGCFB, \pGCPB, and \pICNB (the acronyms will be explained later). While \pGCFB and \pGCPB opt for \mon and \con, instead of \ava, \pICNB opts for \mon and \ava, but not \con.
In this work, we study the feasibility of
different property combinations and \approaches, 
and characterize their performance trade-offs. 
In particular, we 
explore the trade-off between
{\em \uaMetric}, i.e., the duration when the user
is unable to interact with visualizations, 
and {\em \snMetric}, i.e., the duration when visualizations displayed to the user 
have not been refreshed, as shown in Figure~\ref{fig:trade-off_full}.
For example, \pGCFB blocks the user from 
exploring the interface until all of the new \res{s} are computed, 
so it has high \uaMetric. 
But \pGCFB 
also has zero \snMetric since 
it does not present stale results. 
On the other hand \pGCPB reduces \uaMetric (vs. \pGCFB) by presenting the newly computed 
\res{s} to the user whenever available 
while also not showing stale results. 
\pICNB, which sacrifices \con, 
has higher \snMetric because the user can read stale results, 
but none of the views shown are \unavailable, 
i.e., visualizations that are greyed out. 


\stitle{Novel Property Combinations: Exploring the Trade-off}
As we also show in Figure~\ref{fig:trade-off_full} (in green),
we discover a number of novel \approaches, resulting
in new property combinations
and associated performance implications. 
We introduce three new \approaches: 
\GCNB (\pGCNB), \LCNB (\pLCNB), and \LCMB (\pLCMB),
none of which are dominated by the three existing \approaches.
For example, \pLCNB always allows the user to inspect the results 
of any visualizations (i.e., preserving \ava), 
and refreshes the visualizations on the screen 
when all of their new results are computed (i.e., preserving \con). 
Figure~\ref{fig:existing_tools}d shows an example of \pLCNB, 
where the user can quickly read the new results 
on the screen (i.e., $View_{1-4}$)  without waiting 
for computing the new results that 
are not on the current screen (i.e., $View_{5-6}$). 
\pLCNB can be used when a user wants to always see 
and interact with consistent results on the screen. 
However, as we prove later, \pLCNB needs to sacrifice \mon 
when the user explores different visualizations 
(e.g., by scrolling). 
In fact, we demonstrate one can 
achieve both \con and \ava simultaneously 
only by either sacrificing \mon or suffering from 
high \snMetric.
For the aforementioned new \approaches, 
we further introduce $k$-\relaxed variants
(i.e., \kGCNB, \kLCNB, and \kLCMB), where 
$k$ represents the number of additional \unavailable views 
allowed for each \approach. 

\stitle{Usage Scenarios} 
This suite of \approaches 
allow a user or a system designer to determine 
their desired properties 
and gracefully 
explore the trade-off between \snMetric and \uaMetric. 
Current tools, 
while enabling users to customize
their dashboards extensively (in terms of 
the placement of visualizations and selection of visualization 
queries and encodings), make
fixed choices in this regard.
A user has no say in how results
are refreshed and presented,
and a system designer opts for whatever is easiest.
The \tdb framework is intended to address
this gap.
From an end-user standpoint, they may
be able to make appropriate performance 
trade-offs via various customization knobs.
A system designer may similarly 
be able to make appropriate 
selections during tool design, with 
end-users and use-cases in mind.

\stitle{Translating to Practice: Challenges}
Translating our \tdb framework to practice
in real data analysis and BI systems requires addressing 
several challenges:

\vspace{2pt}
\noindent 
{\em (1)} In a visual interface, the user does not explicitly submit 
\txn{s} as in traditional systems, 
but reads the views by looking at the screen. 
In addition, the user can read different subsets of views 
by scrolling to different screens. 
Therefore, a challenge is to 
adapt \txn{s} to model user behavior, 
an aspect not considered in classical transaction processing literature. 

\vspace{2pt}
\noindent 
{\em (2)} In a visual interface, 
the user may want to quickly read new results for some views 
before the system computes all of the new results.  
If we model an update along with refreshing the related views 
as a \txn (to preserve \con), the user essentially wants to read the 
results of an \emph{uncommitted \txn}. 
The \macPro for reading uncommitted results are not considered by 
traditional systems and needs to be defined in our model. 

\vspace{2pt}
\noindent 
{\em (3)} Finally, instantiating \tdb requires 
designing new algorithms for efficiently maintaining different property 
combinations for different \approach{es} while reducing \uaMetric and \snMetric. 
Traditional concurrency control protocols, such as 2PL or OCC~\cite{BernsteinHG87CCBook}, 
do not apply here because they do not consider maintaining \con 
on uncommitted results and the other user-facing properties, 
\mon and \ava.


\stitle{Summary of Contributions} 
We address these challenges as part of \tdb and make the following contributions:
\begin{denselist}
\item We present the \tdb \model in Section~\ref{sec:model}. 
We model reads and writes 
on the views as operations within \txn{s} and 
introduce a special read-only \txn to model a user's behavior of reading the views on the screen. We define the \macPro 
and use a series of theorems to exhaustingly explore  
the possible property combinations. 
We formally order different \approach{es} based on 
\uaMetric and \snMetric, 
and provide guidance for selecting the right
\approach for specific use cases. 
\item We design efficient algorithms for maintaining  properties 
for 
\approach{es} in Section~\ref{sec:system}, 
and propose optimizations to reduce \uaMetric and \snMetric 
while refreshing  analysis results in Section~\ref{sec:scheduler}. 
\item We implement
\tdb within Apache Superset, a popular open-source BI tool~\cite{superset}, in Section~\ref{sec:prototype}.
We perform extensive experiments to 
characterize the relative benefits of different \approaches 
on various workloads, demonstrate the benefits
of the new \approaches, 
and show the performance benefit of our optimizations on reducing 
\uaMetric and \snMetric compared to the baselines---by up to {\bf 70\%} and {\bf 75\%}, respectively---in Section~\ref{sec:experiment}. 
\end{denselist}




%% file: model.tex
\section{\fullTdb \model}
\label{sec:model}

We present our \tdb \model in this section. 
We introduce various key concepts in Section~\ref{sec:bg}. 
Then, we discuss modeling reading/writing views 
using \txn{s} in Section~\ref{sec:abstraction}. 
Next, we formalize our \model in Section~\ref{sec:formalize}, 
define the \macPro in Section~\ref{sec:macPro}, 
and evaluate the feasibility of various property 
combinations in Section~\ref{sec:feasibility}. 
We introduce new property combinations and 
the corresponding \approaches in Section~\ref{sec:new_property}. 
Afterwards, we define performance metrics in Section~\ref{sec:metrics}
and order the \approach{es} by 
the performance metrics in Section~\ref{sec:order}.
\ptr{Finally, we discuss selecting appropriate 
\approach{es} for specific use cases in Section~\ref{sec:use_cases} 
and extensions of the \model in Section~\ref{sec:extension}.} 
\ppaper{\rone{Due to space limitations,
we had to omit the discussion for 
selecting appropriate \approach{es} for specific use cases 
and a number of extensions to our 
framework; these can be found in our
technical report~\cite{tpTR}.}}

\begin{figure}[!t]
    \centering
    \includegraphics[width=75mm]{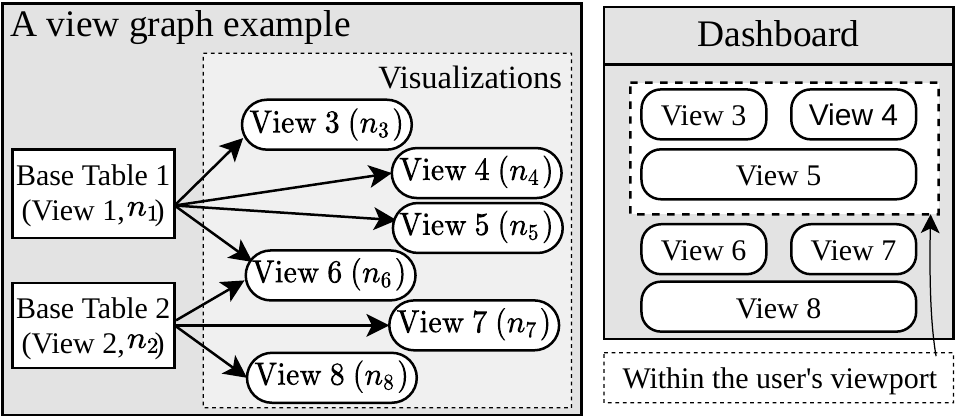}
    \vspace{-4mm}
    \caption{An example of a dashboard and its view graph}
    \label{fig:graph_example}
    \vspace{-5mm}
\end{figure}

\subsection{Preliminaries}
\label{sec:bg}

\stitle{View and \viewG} 
In our context, we define a \emph{view}
to represent arbitrary computation, 
expressed in any manner, including SQL, 
pandas dataframe expressions, spreadsheet formulae, or UDFs, 
taking other views and/or source data as input. 
A \viewG is a directed acyclic graph (DAG) 
that captures the dependencies across views and source data, 
both represented as nodes in the DAG. 
Specifically, if a view \code{$\vt_i$} takes another view 
or source data \code{$\vt_j$} as input, 
we add an edge: \code{$\vt_j$} $\rightarrow$ \code{$\vt_i$}. 
The \textit{dependents} of \code{$\vt_j$} are defined as 
the views that are reachable from \code{$\vt_j$} in the \viewG. 
For simplicity, we regard the source data as a special type of view 
that performs an identity function over the source data. 
Figure~\ref{fig:graph_example} shows an example of a view graph 
for visualizations in a dashboard, where the source data 
are database tables and each view is defined by a SQL statement. 
There are two base tables: \code{Base~Table~1} and \code{2}, 
also regarded as \code{View~1} and \code{2}, respectively. 
We use $\vt_k$ to represent \code{View~k}.
\code{View~3-6} (denoted $\vt_{3-6}$) 
and \code{View~6-8} (denoted $\vt_{6-8}$) are the 
dependents of $\vt_1$ and $\vt_2$, respectively. 
They define the content for the visualizations in this dashboard. 

\begin{table}[!t]
\fontsize{6.5}{8}\selectfont
\rthree{
\begin{tabular}{ll}
\textbf{Notations} & \textbf{Meanings} \\ \hline
$\txw^{t_i}$             & A \writeTxn that is created at timestamp $t_i$ \\
$\txr^{s_i}$             & A \readTxn that is created at timestamp $s_i$ \\
$G^{t_i}$                & A version of the \viewG created by $\txw^{t_i}$ \\
$\vt_{k}$                & A view in the \viewG \\
$\vt_{k}^{t_i}$          & A \qres for $\vt_{k}$ in $G^{t_i}$ \\
$UC^{t_i}_{k}$          & A state representing the \qres $\vt_{k}^{t_i}$ is under computation  \\ 
$\Vs^{t_i}$              & A set that stores the \qres{s} and UC{s} for the views 
in $G^{t_i}$ \\
$C_f$ & \ConF \\ 
$C_m$ & \ConM \\ 
$C_c$ & \ConC 
\end{tabular}
}
\caption{Notation frequently used in this paper}
\label{tbl:notation}
\vspace{-10mm}
\end{table}

\stitle{\Qres and viewport}
\rtwo{A \emph{\qres} represents the output 
of a view given a \version of the source data and the definition 
of the \viewG. This \qres is rendered on the dashboard as a visualization 
(this includes visualizations of tables or even single values). 
In certain settings, a \vdef may itself be editable and rendered 
as part of the dashboard (e.g., as a filter). 
For the following discussion, 
we assume \vdef{s} are not editable or rendered. 
We discuss reading and modifying a \vdef in Section~\ref{sec:extension}.
}
A dashboard may include many visualizations that 
cannot fit into a single screen. 
The rectangular area on the screen 
a user is currently looking at 
is the \emph{viewport}. In Figure~\ref{fig:graph_example}, 
the viewport includes visualizations for views $\vt_{3-5}$.
A user can change the viewport to explore different parts of a \viewG.


\stitle{Reading and writing a view, and \vstate}
\rtwo{We model the user inspecting a visualization 
in the viewport as reading the corresponding view, 
which returns a {\em \vstate}. 
A \vstate is either a \qres or a state that indicates the \qres 
has not been computed yet (denoted as \emph{\undercompute}) 
and is usually materialized such that 
future reads coming from the user can reuse the materialized \pvstate.}
In Figure~\ref{fig:graph_example}, 
we need to materialize the \vstate{s} for $\vt_{3-8}$ 
to support future reads by the user.


There are two types of writes in \tdb:
\emph{input writes} and \emph{triggered writes}. 
An \emph{input write} is from a user or an external system, 
and modifies the source data 
(e.g., new data inserted to a base table) 
or \viewG definitions. 
In the following, we focus on input writes 
to the source data as it is the most common case of 
refreshing a dashboard. 
\rtwo{We discuss processing 
modifications to the 
\viewG definitions in Section~\ref{sec:extension}.}
The input write will trigger additional writes, 
called \emph{triggered writes}, 
which compute new \pqres{s} 
for the views that depend on the base views which were modified in the input write. 
For example, modifying the base table $\vt_1$ 
in Figure~\ref{fig:graph_example} triggers 
computing new \pqres{s} for $\vt_{3-6}$. 




\begin{figure*}[!t]
\vspace{-15pt}
    \centering
    \includegraphics[width=140mm]{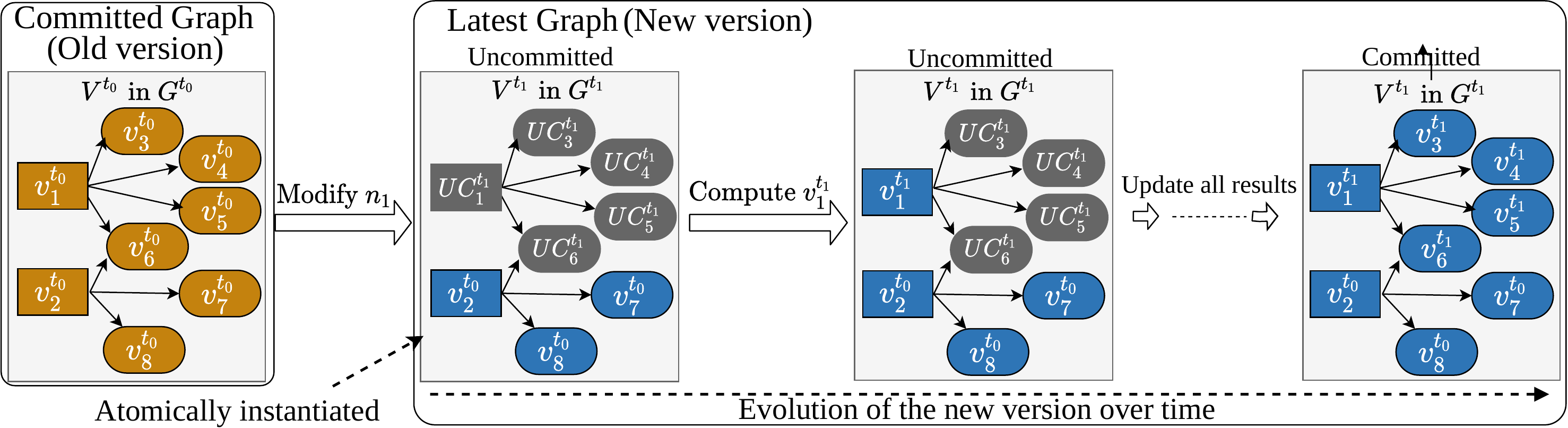}
    \vspace{-4mm}
    \caption{An example of creating a new version of \viewG and computing the \qres for each node}
    \label{fig:multiversion}
    \vspace{-5mm}
\end{figure*}

\subsection{Modeling the Interaction with a \ViewGG}
\label{sec:abstraction}
We model a user's or an external system's interaction 
with a \viewG as \txn{s}, 
and logically associate each \txn 
with a unique timestamp that represents its submission time, 
with \txn{s} being ordered by these timestamps. 
We focus on a single user setting as in 
most user-facing data analysis tools, 
such as Excel~\cite{msexcel}, 
Tableau~\cite{tableau}, and PowerBI~\cite{powerBI}, 
\ptr{and discuss the multiple user setting in Section~\ref{sec:extension}.}
\ppaper{and discuss the multiple user setting 
in the technical report~\cite{tpTR}.}
Our model has two types of \txn{s}:

\stitle{\WriteTxn}
A \writeTxn is issued when a set of input writes on the \viewG 
(e.g., modifications to a set of base tables) 
are submitted together to the system.
A \writeTxn involves processing the input writes  
and recomputing the views that depend on the input writes. 
For example, in Figure~\ref{fig:graph_example} if $\vt_1$ is modified 
the \writeTxn will recompute $\vt_{3-6}$.
The specific algorithm for maintaining the \viewG and recomputing  
\qres{s} is orthogonal to our model, and, for example,  
can employ incremental view maintenance~\cite{Gupta93IVM, DBT, TangSEKF20InQP}. 
We focus on processing one \writeTxn at a time, 
which is typical in existing tools
~\cite{msexcel, tableau, powerBI, superset}, 
\ptr{and discuss the case of multiple, simultaneous \writeTxn{s} in Section~\ref{sec:extension}.}
\ppaper{and discuss the case of multiple, simultaneous \writeTxn{s} in 
the technical report~\cite{tpTR}.}

\stitle{\ReadTxn}
\Tdb models a user inspecting visualizations 
in the viewport as a \readTxn. 
For example, in Figure~\ref{fig:graph_example},  
the \readTxn involves reading views $\vt_{3-5}$. 
A unique property of this \readTxn is that it does not delay 
to wait for the requested \qres{s} to be computed. 
So the \readTxn may return an \undercompute state 
for a visualization if the \qres has not been computed yet. 
If an \undercompute state is returned, 
the user cannot inspect and interact with the 
visualization in the visual interface.
To simulate the effect of the user ``looking at'' 
the viewport, our model assumes the system periodically 
issues new \readTxn{s} to pull \vstate{s}. 
The user may read different parts of the \viewG 
by changing the viewport while the system continues to process writes. 
The location of the user's viewport is 
known to the system throughout.

\subsection{Formalization}
\label{sec:formalize}


We now introduce the aforementioned concepts in more detail 
and more formally. \rthree{Table~\ref{tbl:notation} summarizes 
the notation frequently used in this paper.}
The \viewG is logically multi-versioned, 
where a new \version of \viewG $G^{t_i}= (E, \Vt, \Vs^{t_i})$ 
is instantiated by a \writeTxn with timestamp $t_i$ 
(denoted as $\txw^{t_i}$). 
$\Vt$ represents the set of nodes (i.e., source data and views) 
in the \pG and each edge $e={(\vt_{prec}, \vt_{dep})} \in E$ 
indicates that node $\vt_{dep}$ has another node $\vt_{prec}$ 
as input. $\Vs^{t_i}$ captures the \qres{s} 
and the \undercompute states for the views 
in $G^{t_i}$ and evolves as we process the \writeTxn $\txw^{t_i}$. 
At a given time, $\Vs^{t_i}$ may include: 
i) the \qres for $\vt_{k}$ 
that $\txw^{t_i}$ has already finished computing 
--- this \qres is represented as $\vs^{t_i}_{k}$; 
ii) $\uc^{t_i}_{k}$, which represents 
the state that $\txw^{t_i}$ intends to compute 
the \qres for $\vt_{k}$, but has not done it yet; 
and iii) the \qres for $\vt_{k}$ from the last \version of \viewG 
given that $\vt_{k}$ will not be updated by $\txw^{t_i}$. 
Figure~\ref{fig:multiversion} shows an example of 
computing a new \version of \viewG for a \writeTxn $\txw^{t_1}$ that 
modifies $\vt_1$ in Figure~\ref{fig:graph_example}. 
We see that creating a new \version of \viewG 
logically replicates the \qres{s} of the last \version of the \viewG, 
and marks all of the \qres{s} to be computed as \ua{s} (in gray). 
Each \ua is replaced after the corresponding 
\qres is computed (in blue). 
We guarantee that a new \version of \viewG is atomically 
seen by \readTxn{s} via our concurrency control protocol 
(to be discussed in Section~\ref{sec:system}). 
\rtwo{We call a \version of \viewG \emph{committed} if its  
\writeTxn is committed; 
otherwise, this \version is \emph{uncommitted}. 
A \writeTxn is defined to be committed 
if the system has a) computed all of the new \qres{s} for 
the version of \viewG created by this \writeTxn 
and b) updated a global variable that stores  
the timestamp of the recently \oldG. 
More details about the procedure for processing a \writeTxn 
and the management of the global variable are in Section~\ref{sec:cc}.}
The initial \version of the \viewG is $G^{t_0}$, 
which is modified by a sequence of \writeTxn{s} 
$\TxW = \{\txw^{t_1}, \cdots, \txw^{t_n}\}$, 
where $\txw^{t_i}$ is submitted before $\txw^{t_j}$ if $t_i < t_j$. 

We also have a sequence of \readTxn{s} 
$\TxR = \{\txr^{s_1}, \cdots, \txr^{s_m}\}$, 
where $\txr^{s_i}$ is submitted before $\txr^{s_j}$ if $s_i < s_j$. 
Recall that each \readTxn corresponds to a single viewport 
and all of the views in it. 
We refer to the \vstate{s} returned by a \readTxn $\txr^{s_i}$ 
as $\RR^{s_i}$, which includes \qres{s} 
and/or \ucs for the views in the viewport. 
If a \readTxn returns a \uc for a view, 
its corresponding visualization is marked as 
\emph{\unavailable} in the dashboard. 
On the front end, this can be displayed in various ways: 
grayed out, a progress bar, a loading sign, etc.
We use $\uc$ for $\uc^{t_i}_k$ 
when $k$ and $t_i$ are clear from the context. 

\subsection{\MacPro}
\label{sec:macPro} 

We now formally define the so-called \macPro for \readTxn{s}, 
motivated by user needs in analytical visual interfaces. 
First, the user consumes the \vstate{s} 
returned by \readTxn{s} in order:  
i.e., they consume the \vstate{s} of one \readTxn before the next. 
Therefore, they expect to see \emph{monotonically} newer 
\pvstate{s} for each view, avoiding the confusion that 
the \pvstate{s} seen ``travel back in time''. 
Second, in a user-facing dashboard, 
the notion of \emph{\ava} helps ensure interactivity, 
as it means the user can continuously 
explore the \qres{s} of different visualizations 
without interruption, while the system processes \writeTxn{s}.
Finally, \emph{\con} helps ensure that 
insights derived from multiple visualizations 
on a viewport are computed from the same snapshot of 
source data~\cite{ZhugeGW97Multiview, GolabJ11Consistency, WuCH020InteractionSnapshots}.
We now describe each property in detail. 


\stitle{\Mon} 
\Mon means if a user reads a given version of a \qres or \ua 
for a view, any successive reads on the same view will  
return the same or more recent version of the \qres or \ua. 
Formally, \mon is defined as: 
\vspace{-2mm}
\begin{definition}[\Mon]
\label{def:mon} 
A sequence of \readTxn{s} $\TxR = \{\txr^{s_1}, \cdots, \txr^{s_m}\}$ 
maintains \mon if the following holds: 
for any view $\vt_k$ read by any two \txn{s} 
$\txr^{s_i}$ and $\txr^{s_j}$, the timestamps of the returned 
\pvstate{s} are $t_p$ and $t_q$, respectively: 
$t_p \leq t_q$ if $s_i < s_j$. 
\vspace{-2mm}
\end{definition}

\ptr{\noindent To maintain \mon, the system needs to track the timestamps 
of the \res{s} for each view returned by the recent read operations, 
as will be discussed in Section~\ref{sec:system}.}


\stitle{\Ava}
This property says that for any view that is read by any \readTxn, 
the system should not return an \undercompute state, \ua. 
Formally, \ava is defined as:
\vspace{-2mm}
\begin{definition}[\Ava]
\label{def:ava}
A sequence of \readTxn{s} $\TxR = \{\txr^{s_1}, \cdots, \txr^{s_m}\}$ 
maintains \ava if for the \pvstate{s} $\RR^{s_i}$ that are 
returned by any \txn $\txr^{s_i}$, we have $\ua \not\in \RR^{s_i}$.
\vspace{-2mm}
\end{definition}
\noindent \rone{The user may also sacrifice \ava 
by opting for partial \ava, where they accept a controlled number of \ua{s} 
as a trade-off for reading fresher \qres{s}, discussed next.}






\stitle{\Con} 
In our setting, \con means that the \vstate{s} 
returned by each \readTxn belongs to a single \version of the \viewG.  
\Con is formally defined as: 
\vspace{-2mm}
\begin{definition}[\Con]
\label{def:con}
Let $\RR^{s_i}$ be the \vstate{s} returned by $\txr^{s_i}$. 
We say $\txr^{s_i}$ maintains \con if 
there exists a \version of \viewG $G^{t_j}=(E, \Vt, \Vs^{t_j})$ 
such that $t_j \leq s_i$ and $\RR^{s_i}
\subseteq \Vs^{t_j}$. 
\vspace{-2mm}
\end{definition}
\noindent Intuitively, $t_j \leq s_i$ requires that a user read a \version 
created by the \writeTxn{s} that happen before $\txr^{s_i}$. 
The condition $\RR^{s_i} \subseteq \Vs^{t_j}$ guarantees that 
the \vstate{s} returned belong to a single \version of \viewG. 
Consider a \readTxn $\txr^{s_1}$ that reads $\vt_{3-5}$. 
Say for $G^{t_1}$ in Figure~\ref{fig:graph_example}, 
we have computed $\vs_3^{t_1}$ but not $\vs_{4-5}^{t_1}$. 
If the returned \pvstate{s} for $\txr^{s_1}$ is 
$\RR^{s_1} = \{\vs_3^{t_1}, \ua_{4-5}^{t_1}\}$, 
then $\txr^{s_1}$ maintains \con because 
$\RR^{s_1}$ belongs to $V^{t_1}$. 

Note that \con in \tdb is different from traditional 
Consistency (C) in ACID for database \txn{s}. 
C in ACID refers to the property that 
each \txn correctly brings the database 
from one valid state to another. 
In our context, \con  
is more closely related to Isolation (I) in ACID, 
which defines when \qres{s} created by one \txn can be read by others. 
Our notion of \con allows a \readTxn to read uncommitted 
\res{s} from a concurrently running \writeTxn 
(e.g., reading the uncommitted $G^{t_1}$), 
but additionally maintains the semantics that the returned 
\pvstate{s} correspond to a single version. 



A follow-up question about preserving \con is 
which \version of \viewG a \readTxn should read. 
Specifically, since we process one \writeTxn at a time, 
a \readTxn can choose between reading the last committed 
\version of \viewG, which we call the \emph{\oldG}, 
and the \version that the latest \writeTxn is computing, 
which we call the \emph{\newG}. 
Depending on the \version that is read, we define three types of \con, 
which opt for different trade-offs between \uaMetric and \snMetric.
(Recall that \uaMetric refers to the time during which 
the views in the viewport are \unavailable, 
while \snMetric refers to the time during which the returned \qres{s} 
are not consistent with the \newG; both will be defined 
in Section~\ref{sec:metrics}.)

The first type of \con is \textbf{\ConF} or \cf, 
which always reads the \newG. 
\cf returns fresh \pqres{s} but suffers high \uaMetric. 
For the example in Figure~\ref{fig:multiversion}, 
with \cf, \readTxn{s} always 
read $G^{t_1}$ while we are processing the \writeTxn. 

Another type of \con, called \textbf{\ConC} or \cc, 
always reads the most recently \oldG. 
\cc does not have \unavailable views, but the \snMetric 
of the returned \qres{s} could be high.
In Figure~\ref{fig:multiversion}, 
if \cc is used, \readTxn{s} cannot read $G^{t_1}$ 
until we have computed all of the \qres{s} for $G^{t_1}$ 
(i.e., $\vs_1^{t_1}$ and $\vs_{3-6}^{t_1}$). 

We additionally introduce a type of \con 
that lies between \cf and \cc. 
This type of \con requires 
that a \readTxn read the most recent version of the \viewG 
that returns the minimum number of \uas for this \txn, 
which we call \textbf{\ConM} or \cm for short.
With \cm, we would typically read the \oldG 
to avoid returning \uas when the new \qres{s} 
in the viewport are not yet computed. 
Once they are computed, 
we can read the \newG to return fresh \qres{s}.
Consider reading $\vt_{3-5}$ in Figure~\ref{fig:multiversion}. 
Initially, the \readTxn{s} will read $G^{t_0}$ because 
$G^{t_0}$ does not include \ua{s}. 
After the new \pqres{s} for $\vt_{3-5}$ are computed, 
we will read $G^{t_1}$ because 
reading $G^{t_1}$ for $\vt_{3-5}$ 
does not return $\ua{s}$, and $G^{t_1}$ is more recent 
than $G^{t_0}$. 
\rone{Note that \cm is different from \cc 
because for \cc, \readTxn{s} will read the \newG 
$G^{t_1}$ after the \writeTxn is committed, 
while for \cm, \readTxn{s} will read $G^{t_1}$ after 
all of the new \pqres{s} in the viewport are computed, 
which could be sooner.}
In addition, when the user changes the viewport, 
the minimum number of \uas returned by a \readTxn 
may not always be zero for \cm.
As we prove next, adopting \cm may sacrifice \ava 
if we need to additionally maintain \mon. 

\begin{reviewone}
\stitle{Correctness and Performance}
Among the \macPro, both \mon and \con impact correctness, 
i.e., maintaining one or both of these properties 
may be essential to guaranteeing correct insights 
from the visual interface, depending on the application. 
The former ensures that users are not deceived by updates 
that get ``undone'', while the latter ensures that users 
viewing multiple visualizations on a screen can draw correct 
joint inferences based on the same snapshot of the source data. 
On the other hand, all of the \macPro impact performance 
since maintaining these properties will increase 
\snMetric and/or \uaMetric 
as we will show in Section~\ref{sec:order}. 
Users can further make 
performance trade-offs between \uaMetric and \snMetric 
by choosing which \version of the \viewG to read for \con. 
\end{reviewone}


\subsection{Feasibility of Property Combinations}
\label{sec:feasibility}


We now develop a series of theorems, that we call the \macThe, 
to characterize the complete subsets of \macPro 
that can be maintained together. 
\ppaper{
\rone{We omit proofs due to space limitations, which
can be found in the technical report~\cite{tpTR}.}}
Specifically, we define the possible property combinations 
involving each type of \con (i.e., \cc, \cf, and \cm). 
The first two straightforward theorems establish the 
fact that always reading the \oldG provides \mon 
and \ava for free, while always reading the \newG 
maintains \mon, but sacrifices \ava. 

\begin{theorem}[\cc]
\vspace{-2mm}
Maintaining \conC will also maintain \mon and \ava for \readTxn{s}. 
\label{thr:f_cc}
\end{theorem}
\ptr{\begin{proof}
\vspace{-2mm}
(Sketch) If \conC is adopted, 
\readTxn{s} read a committed \version of \viewG, 
which does not include \ua{s}. So \ava is preserved. 
Since \conC requires reading the most recently committed \version, 
whose timestamp advances monotonically, \mon is preserved. 
\end{proof}
}

\begin{theorem}[\cf]
\vspace{-2mm}
Maintaining \conF will also maintain \mon for \readTxn{s}, 
but \conF and \ava cannot always be maintained. 
\label{thr:f_cf}
\end{theorem}
\ptr{\begin{proof}
\vspace{-2mm}
\ConF (or \cf) requires always reading the \newG, 
whose timestamp monotonically advances. 
So both \cf and \mon are maintained. 
In addition, always reading the \newG can return \ua{s}, 
violating \ava.
\end{proof}}

\begin{figure*}[!t]
\vspace{-15pt}
    \centering
    \includegraphics[width=140mm]{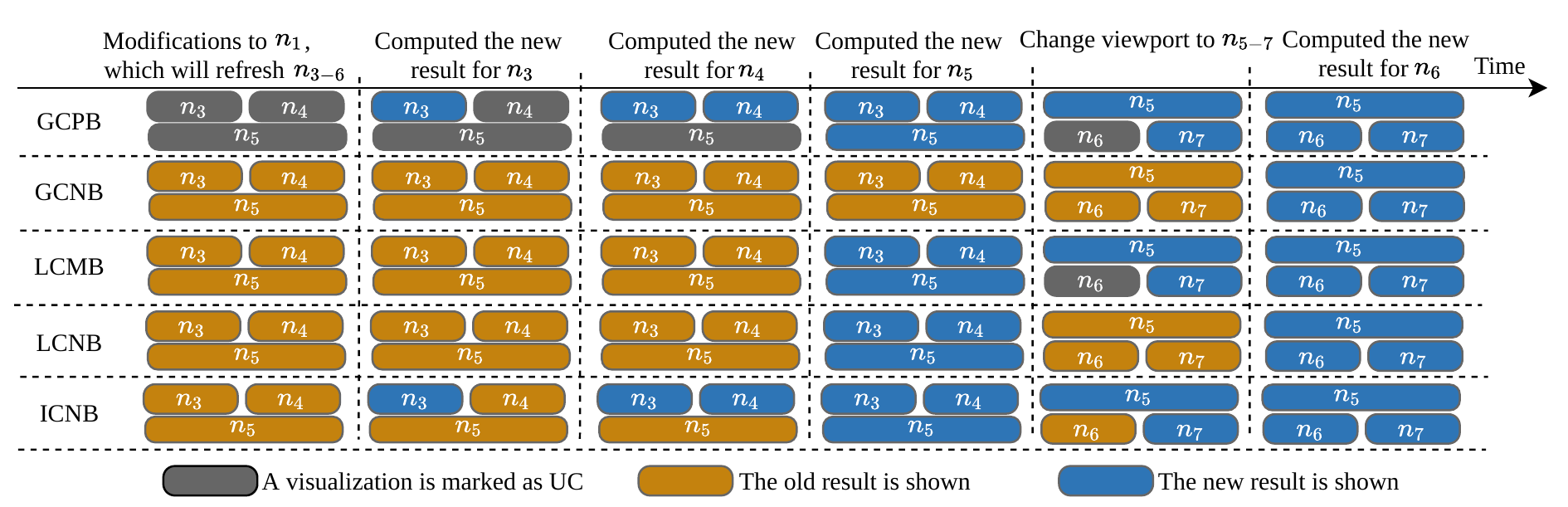}
    \vspace{-4mm}
    \caption{An example timeline for each \approach 
    presenting \pqres{s} to the user.}
    \label{fig:timeline}
    \vspace{-3mm}
\end{figure*}

\vspace{-1mm}
\noindent Unfortunately, with \cm, we cannot have both \mon and \ava: if two consecutive \readTxn{s} 
involve overlapping views, 
the latter \txn needs to read the same 
or more recent \version of \viewG 
compared to the former one to maintain \mon. 
Therefore, the latter \txn may read the \newG, 
which may include \ua{s}, thereby violating \ava. 
\begin{theorem}[\cm-impossibility]
\vspace{-1mm}
We cannot always simultaneously maintain \mon, \ava, 
and \conM for \readTxn{s}. 
\label{thr:f_cm1}
\end{theorem}
\ptr{
\begin{proof}
\vspace{-3mm}
(Sketch) We construct a counterexample 
where the three properties cannot be met together. 
We assume the initial graph is $G^{t_0}$ 
and a user modifies the base table $\vt_1$ 
in Figure~\ref{fig:multiversion}.  
This modification creates a \writeTxn $\txw^{t_1}$ 
that updates $\vt_1$ and $\vt_{3-6}$, 
and generates a new \version $G^{t_1}$. 
We further assume we have computed the new \res{s} 
$\vs_1^{t_1}$ and $\vs_{3-5}^{t_1}$, but not $\vs_{6}^{t_1}$.

Based on this setup, consider two consecutive 
\readTxn{s} $\txr^{1}$ (reading $\vt_{3-5}$) 
and $\txr^{2}$ (reading $\vt_{5-7}$), which 
correspond to the case that the user moves the viewport. 
To maintain \conM, 
$\txr^{1}$ will read $G^{t_1}$ and return $\vs_{3-5}^{t_1}$ 
since $G^{t_1}$ does not include \ua{s} for $\txr^{1}$. 
Now we show the subsequent \txn $\txr^{2}$ cannot maintain the 
three aforementioned properties simultaneously. 
To maintain \mon and \conM, $\txr^{2}$ 
has to read $G^{t_1}$. 
This is because both $\txr^{2}$ and $\txr^{1}$ need to read $\vt_5$, 
and $\txr^{1}$ has already read $\vs_5^{t_1}$ in $G^{t_1}$. 
However, reading $G^{t_1}$ violates \ava because 
$\txr^{2}$ needs to read $\vt_6$ 
but $\vs_{6}^{t_1}$ has not yet been computed for $G^{t_1}$. 
This example proves that \mon, \ava, and \conM 
cannot always be met together.
\end{proof}
}

\vspace{-1mm}
\noindent Interestingly, if we sacrifice one property among the three 
properties, we can always maintain the other two. 
\begin{theorem}[\cm-possibility]
\vspace{-2mm}
\Tdb can always maintain any two properties 
out of \mon, \ava, and \conM for \readTxn{s}. 
\label{thr:f_cm2}
\end{theorem}
\ptr{
\begin{proof} 
\vspace{-2mm}
(Sketch) First, we can maintain \mon and \ava together
for any \readTxn because for each view read by this \txn, 
we can always return the most recent \qres, 
without considering \con. 
Second, we can also maintain \ava and \conM (or \cm). 
To achieve this, each \readTxn reads the most recent \version 
of \viewG that has zero \uas for this \txn. 
Finally, we can maintain \mon and \cm together 
because for a \readTxn we could always find a \version of \viewG 
that meets \mon (e.g., the latest one). 
Among all of the \versions that maintain \mon, 
we choose the one that minimizes the number of \uas for this \txn 
and thus achieves \cm. 
\end{proof}
}



\begin{reviewone}
\subsection{Property Combinations and \Approaches}
\label{sec:new_property} 
Given the feasible property combinations, we now 
define different ways of presenting \pqres{s}
to the user while preserving a given property combination, 
which we call \emph{\approach{es}}. 
We use Figure~\ref{fig:timeline} to show illustrate 
how each \approach presents \res{s} to the user 
for the example in Figure~\ref{fig:multiversion}. 
In this example, the base table $\vt_{1}$ is modified, 
which will refresh $\vt_{3-6}$, but not $\vt_{7}$; 
the viewport initially includes $\vt_{3-5}$ 
and is modified to $\vt_{5-7}$ after we have 
computed the new \qres{s} for $\vt_{3-5}$. 
\ptr{Then, we discuss the \approach{es} that are covered  
in existing tools and the newly discovered ones. 
Finally, we introduce three new variants 
that allow users to make better trade-offs 
between \snMetric and \uaMetric.}
For simplicity, we use \MonS for \Mon and \AvaS for \Ava.

\stitle{\Approaches from Theorems~\ref{thr:f_cc}-\ref{thr:f_cf}}
We first define the \approach{es} derived from Theorems~\ref{thr:f_cc}-\ref{thr:f_cf}:
\begin{tightitemize}
\item \pGCNB: \GCNB
\item \pGCPB: \GCPB
\end{tightitemize}
Theorem~\ref{thr:f_cc} shows that it is possible to 
preserve \MonS-\AvaS-\cc together. 
We denote the \approach for this property combination \pGCNB, 
which always reads and presents the \qres{s} 
from the recently \oldG to the user. 
On the other hand, \approach \pGCPB preserves 
\MonS-\cf based on Theorem~\ref{thr:f_cf}. 
It always reads the \newG, 
presents new \qres{s} that are consistent 
with the newly modified source data, 
and marks a view \unavailable if its \qres 
has not been computed yet. 
Figure~\ref{fig:timeline} has shown the 
examples for the two \approach{es} 
presenting \pqres{s} in the visual interface. 
We include ``Globally Consistent (GC)'' 
in the names of the \approach{es} \pGCNB and \pGCPB 
to indicate that for these two \approach{es} all of the \res{s} 
(in the viewport or otherwise) 
are consistent with a single \version of the \viewG. 

\stitle{\Approaches from Theorems~\ref{thr:f_cm1}-\ref{thr:f_cm2}}
Next, we define the \approach{es} derived from Theorems~\ref{thr:f_cm1}-\ref{thr:f_cm2}:
\begin{tightitemize}
\item \pLCNB: \LCNB
\item \pLCMB: \LCMB
\item \pICNB: \ICNB
\end{tightitemize}
\Approach \pLCNB adopts \AvaS-\cm from Theorem~\ref{thr:f_cm2}. 
\rtwo{Between the \oldAndNew \viewG{s}, it 
reads the most recent \version that 
does not have \ua{s} for any \readTxn, 
ensuring that each viewport does not have \unavailable views 
and that the \res{s} within a viewport are consistent.} 
\Approach \pLCMB adopts \MonS-\cm from Theorem~\ref{thr:f_cm2}. 
Between the \oldAndNew \viewG{s}, 
it reads the most recent \version that 
returns the minimum number of \uas 
for each \readTxn and preserves \mon. 
Specifically, if reading either the \oldOrNew \pG preserves \mon, 
\pLCMB chooses the version that has the minimum number of \ua{s} 
for the \readTxn, where the minimum number of \ua{s} is zero 
since the \oldG includes zero \ua{s}. 
Otherwise, \pLCMB reads the \newG to preserve \mon, 
which may include \ua{s}. 
Finally, \approach \pICNB preserves \MonS-\AvaS. 
It allows a user to always inspect \pqres{s} of any views and 
refreshes each view independently, which sacrifices \con. 
Figure~\ref{fig:timeline} has shown the 
examples for the three \approach{es} 
presenting \pqres{s} in the visual interface. 
Note that all \approach{es} except for \pICNB 
provide consistent \pqres{s} for the user (this include 
locally and globally consistent \approach{es}). 
However, users will perceive different levels of \snMetric and 
\uaMetric for different \approaches. 
In Section~\ref{sec:order}, we formally characterize  
the relative orders across \approach{es} 
in terms of \snMetric and \uaMetric. 

\begin{figure}[!t]
    \centering
    \includegraphics[width=80mm]{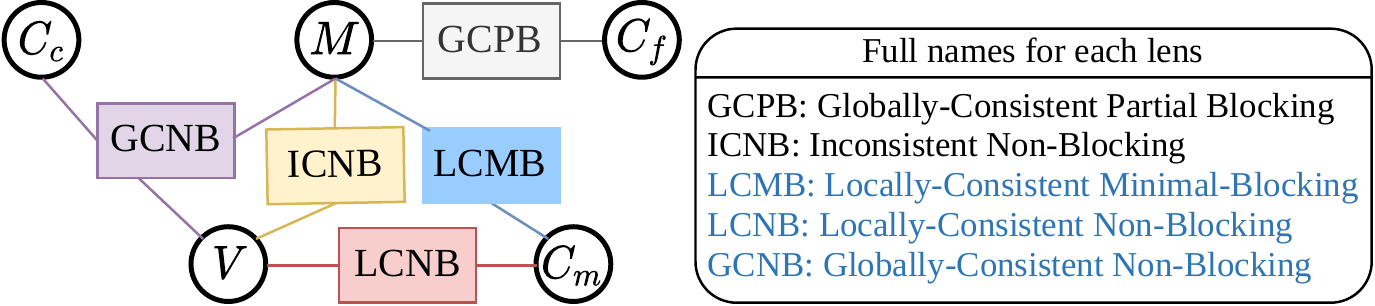}
    \vspace{-4mm}
    \caption{The possible property combinations and the corresponding \baseApproach{es} covered in \tdb}
    \label{fig:approach_summary}
    \vspace{-5mm}
\end{figure}

\stitle{Discovering new \approach{es}}
We call the above \approach{es} as \emph{\baseApproach{es}}, 
and their names and corresponding properties are summarized in Figure~\ref{fig:approach_summary}.
\pGCPB is adopted by Power~BI~\cite{powerBI}, Superset~\cite{superset}, and Dataspread~\cite{Antifreeze}, 
and \pICNB is adopted by Google~Sheets~\cite{sheets}. 
Recall that there is an existing \approach, \GCFB (\pGCFB), 
that is adopted by Excel~\cite{msexcel}, Calc~\cite{libreCalc}, 
and Tableau~\cite{tableau}. 
It marks all of the views \unavailable 
until the system has computed all of the new \qres{s}. 
We don't consider it henceforth since it is dominated by \pGCPB---\pGCPB maintains the same properties, 
and has lower \uaMetric and equal \snMetric compared to \pGCFB. 
\pGCNB, \pLCNB, and \pLCMB are newly discovered \approach{es} 
in our model.
\end{reviewone}

\begin{figure}[!t]
    \centering
    \includegraphics[width=60mm]{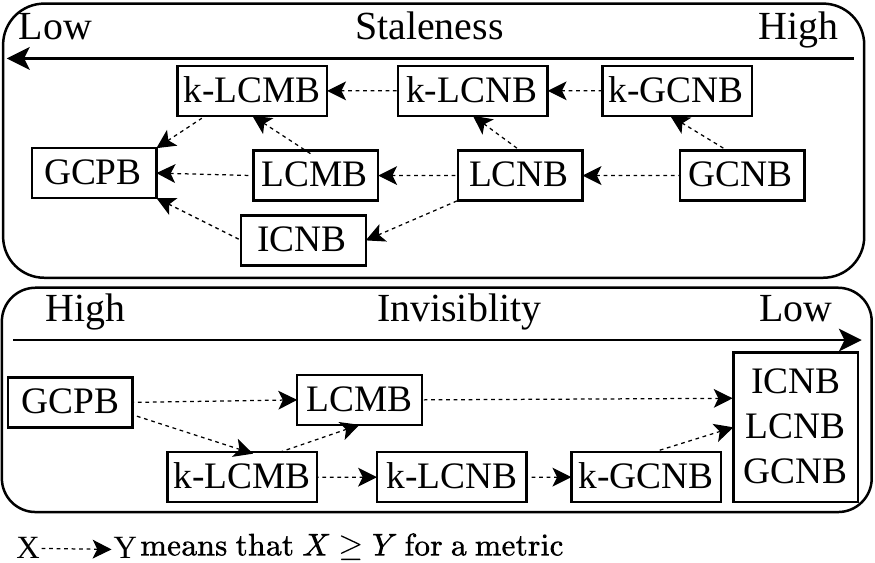}
    \vspace{-4mm}
    \caption{Summary of the orders across different \approaches based on \snMetric and \uaMetric}
    \label{fig:order}
    \vspace{-5mm}
\end{figure}

\stitle{$k$-\relaxed variants}
As we will show in Figure~\ref{fig:order}, 
the three new \approaches above 
are optimized to achieve low \uaMetric, 
but have high \snMetric. 
Therefore, we introduce their $k$-\relaxed variants 
to allow for more \unavailable views to reduce \snMetric 
such that a user can gracefully explore 
the performance trade-off between \uaMetric and \snMetric, 
as visualized in Figure~\ref{fig:trade-off_full}. 
Specifically, \kGCNB, the variant of \pGCNB, 
will read the \newG if this \pG involves $k$ or fewer \uas, 
while \pGCNB reads the \newG only when it is committed. 
Similarly, \kLCNB, corresponding to \pLCNB, 
reads the most recent version of \viewG 
that has $k$ or fewer \uas for the \readTxn, 
ensuring the viewport has $k$ or fewer \unavailable views. 
\kLCMB, the variant of \pLCMB, 
needs to maintain \mon and \con, and works as follows. 
If reading either the \oldOrNew \pG preserves  
\mon, \kLCMB reads the recent \version 
that has $k$ or fewer \uas for the \readTxn, similar to \kLCNB. 
Otherwise, \kLCMB reads the \newG to preserve \mon. 

\subsection{Performance Metrics}
\label{sec:metrics}


With the different \approach{es} defined, 
we now formally define the performance metrics: 
\uaMetric and \snMetric, for these \approach{es}, 
and study their relative 
performance in Section~\ref{sec:order}. 

\UaMetric represents the total time 
when the views read by a user are \unavailable. 
We adapt the metric from previous work~\cite{Antifreeze} 
to our scenario of modeling reading the \viewG as \readTxn{s}. 
We define $\uaFunc$, the \uaMetric 
for a set of \readTxn{s} $\TxR = \{\txr^{s_1}, \cdots, \txr^{s_m}\}$, 
as: 
\vspace{-1.5mm}
\[
\uaFunc(\TxR) = \sum_{i=1}^{m-1} |\RR_{\ua}^{s_i}| \times (\TxnT(\txr^{s_{i+1}}) - \TxnT(\txr^{s_{i}}))
\vspace{-1.5mm}
\]
$\TxnT(\txr^{s_i})$ is the time during which $\txr^{s_i}$  
returns and $\RR_{\ua}^{s_i}$ is the set of \uas in the 
returned \vstate{s}. 
So $|\RR_{\ua}^{s_i}| \times (\TxnT(\txr^{s_{i+1}}) 
- \TxnT(\txr^{s_{i}}))$ represents the time 
when the views read by $\txr^{s_{i}}$ 
stay \unavailable between two consecutive \readTxn{s}. 

\SnMetric represents the total time 
during which \readTxn{s}' returned \qres{s} 
are not consistent with the latest \version of the \viewG. 
We use $\snFunc$ to denote \snMetric for 
a set of \readTxn{s} $\TxR = \{\txr^{s_1}, \cdots, \txr^{s_m}\}$. 
Say $G^{t_i}$ is the latest \version of the \viewG before 
the \readTxn $\txr^{s_i}$ starts, 
and say the returned \qres by $\txr^{s_i}$ for view $\vt_k$ 
is $\vs_k^{t_j}$. 
$\snFunc$ is defined as:
\vspace{-1.5mm}
\[
\snFunc(\TxR) = \sum_{i=1}^{m-1} \sum_{\vs_k^{t_j}\in \RR^{s_i}_{qr}}
\mathtt I[\vs_k^{t_j} \notin \Vs^{t_i}] \times (\TxnT(\txr^{s_{i+1}}) - \TxnT(\txr^{s_i}) )
\vspace{-1.5mm}
\]
Here, $\RR^{s_i}_{qr}$ represents 
the \qres{s} that are returned by $\txr^{s_i}$. 
$\mathtt I[\vs_k^{t_j} \notin \Vs^{t_i}]$ is 1 
if the \qres is stale (i.e., $\vs_k^{t_j}$ does not belong to 
the latest \version of the \viewG); 
otherwise, it is 0. 
So the inner summation represents the total time 
when the \qres{s} returned by $\txr^{s_i}$ stay stale 
between two consecutive \readTxn{s}. 


\subsection{Performance Metrics: Guarantees}
\label{sec:order} 
\ppaper{Figure~\ref{fig:order} plots 
the ordering across different \approach{es} 
with respect to \uaMetric and \snMetric.}
\ptr{We now study the trade-off between 
\uaMetric and \snMetric by ordering 
different \approaches by these metrics, 
as is summarized in Figure~\ref{fig:order}.}
\ppaper{Our analysis assumes the same \writeTxn, 
the same order of computing the new \qres{s}, 
and the same sequence of \readTxn{s} for all \approaches.
\rone{The theorems and proofs for guaranteeing the ordering can be found 
in the technical report~\cite{tpTR}.}}
\ptr{Recall that we already 
assume the system processes one \writeTxn 
at a time, so the \readTxn{s} either read the \oldOrNew \pG. 
We use the acronyms for the respective \approaches; 
the full names can be found in 
Figure~\ref{fig:approach_summary}. 
For simplicity, we use $\pSN(A)$ and $\pIV(A)$ 
to represent the \snMetric and \uaMetric for \approach $A$, 
respectively.  

We start with comparing \pGCPB and \pGCNB with other \approaches. 
These two represent extreme points across the trade-off between 
\snMetric and \uaMetric since \pGCPB always reads the \newG, 
while \pGCNB always reads the \oldG, and therefore sandwich 
other \approach{es}. 
\begin{theorem}
\vspace{-2mm}
$\pSN(\pGCPB) \leq \pSN(\text{all other \approaches}) \leq \pSN(\pGCNB)$,
and $\pIV(\pGCPB) \geq \pIV(\text{all other \approaches}) \geq 
\pIV(\pGCNB)$. 
\label{thr:o1}
\end{theorem}
\ptr{
\begin{proof}
\vspace{-2mm}
(Sketch) \pGCPB always reads the \newG, so its \snMetric 
is zero and no larger than the other \approaches, 
and its \uaMetric is no smaller than the other \approaches 
(except for \pGCFB, which is excluded).
Symmetrically, \pGCNB reads the \newG only after all of the 
new \qres{s} are computed, 
so its \snMetric is no smaller than the other \approaches, 
and its \uaMetric is no larger than the other \approaches. 
\end{proof}}

Next, we introduce two theorems that help
identify a relative ordering between 
\pLCMB, \pLCNB, and \pICNB. 
Intuitively, \pLCMB always reads the same 
or more recent version than \pLCNB because 
if \pLCMB needs to read the \newG to maintain \mon, 
\pLCNB may still read the \oldG; 
otherwise, both \approach{es} 
read the same version of the \viewG. 
So, \pLCMB has no smaller \uaMetric and no larger \snMetric than \pLCNB. 
In terms of \pLCNB and \pICNB, 
\pICNB reads the new \qres for a view whenever it is computed, 
but \pLCNB needs to wait for all the new \qres{s} 
in the viewport to be computed before reading them. 
So, \pLCNB has no smaller \snMetric than \pICNB, and 
they have the same zero \uaMetric.  
Finally, we find that 
the order between \pLCMB and \pICNB for \snMetric
depends on the workload, but \pLCMB has no smaller \uaMetric 
since \pLCMB may return \uas while \pICNB will not.
\begin{theorem}
\vspace{-2mm}
$\pSN(\pLCMB) \leq \pSN(\pLCNB)$, 
$\pSN(\pICNB) \leq \pSN(\pLCNB)$, 
and the order between \pSN(\pLCMB) 
and \pSN(\pICNB) depends on the workload. 
\label{thr:o2}
\end{theorem}
\ptr{
\begin{proof}
\vspace{-2mm}
(Sketch)  We prove $\pSN(\pLCMB) \leq \pSN(\pLCNB)$ 
by showing that 
\pLCMB always reads the same or more recent \version than \pLCNB. 
Specifically, we show the two cases are true: 
1) if \pLCMB reads the \oldG, then \pLCNB also reads 
the \oldG, and 2) if \pLCMB reads the \newG, 
\pLCNB may not read the \newG. 
The first case is true because if \pLCMB reads the \oldG, 
it means that the \newG includes \uas for the \readTxn 
and \pLCNB also reads the \oldG. 
The second case is also true 
because when \pLCMB reads the \newG, 
\pLCNB may still read the \oldG, sacrificing \mon in the process. \par 
$\pSN(\pICNB) \leq \pSN(\pLCNB)$ because 
\pICNB reads the new \qres for a view whenever it is computed, 
but \pLCNB needs to wait for all the new \qres{s}
in the viewport to be computed before reading them. 
The order between $\pSN(\pICNB)$ and $\pSN(\pLCMB)$ 
depends on the configuration of the \viewG, viewport, 
and read/\writeTxn{s}.
Consider when a viewport involves multiple views to update 
and all of the \readTxn{s} read this viewport. 
In this case, \pLCMB has higher \snMetric than \pICNB 
because \pICNB can read the new \res for a view whenever it is computed 
but \pLCMB needs to wait for all of the new \qres{s} 
in the viewport to be computed. 
\pLCMB can also have lower \snMetric than \pICNB. 
Say a user first waits for 
the views in the viewport to be up-to-date 
and then proceeds to examine other views. 
If any two consecutive \readTxn{s} have overlapping views, 
then after \pLCMB reads the \newG for the 
first viewport, it will always read the \newG for   
the rest of the viewports to preserve \mon. 
In this case \pLCMB has high \uaMetric, but low \snMetric. 
\pICNB, on the other hand, does not always 
read the \qres{s} in the \newG after the initial viewport 
and can have higher \snMetric than \pLCMB.
\end{proof}}

\begin{theorem}
\vspace{-2mm}
$\pIV(\pLCMB) \geq \pIV(\pLCNB) = \pIV(\pICNB)$ for \uaMetric.
\label{thr:o3}
\end{theorem}
\ptr{\begin{proof}
\vspace{-2mm}
(Sketch) \pLCNB and \pICNB have the same zero \uaMetric 
since they are non-blocking. 
\pLCMB has larger or equal \uaMetric compared to \pLCNB and \pICNB 
because \pLCMB may return \ua{s} and 
does not always have zero \uaMetric. 
\end{proof}}

\noindent Our final two theorems order each $k$-\relaxed variant 
and their corresponding \baseApproach, 
and the three $k$-\relaxed variants relative to each other. 
Intuitively, a $k$-\relaxed variant 
always reads the same or more recent \version than the 
corresponding \baseApproach due to the $k$ additional 
\ua{s} allowed, so it has no greater \snMetric 
and no smaller \uaMetric 
than the corresponding \baseApproach. 
In addition, the ordering across $k$-\relaxed variants 
is the same as their corresponding \baseApproach{es} 
since they allow the same number of \ua{s}. 
\begin{theorem}
\vspace{-2mm}
A $k$-\relaxed variant has no greater \snMetric 
and no smaller \uaMetric than the corresponding \baseApproach.
\label{thr:o4}
\end{theorem}
\ptr{
\begin{proof}
\vspace{-2mm}
(Sketch) The $k$-\relaxed variant 
always reads the same or more recent \version than the 
corresponding \baseApproach due to the $k$ additional \ua{s} allowed, 
and thus has no larger \snMetric and no smaller \uaMetric 
than the corresponding \baseApproach.
\end{proof}}

\begin{theorem}
\vspace{-2mm}
$\pSN(\kLCMB) \leq \pSN(\kLCNB) \leq \pSN(\kGCNB)$ 
and $\pIV(\kLCMB) \geq \pIV(\kLCNB) \geq \pIV(\kGCNB)$.
\label{thr:o5}
\end{theorem}
\ptr{
\begin{proof}
\vspace{-2mm}
(Sketch) To compare \kLCMB and \kLCNB, 
we show that \kLCMB will consistently read the same or 
more recent \version of the \viewG than \kLCNB. 
Specifically, if \kLCMB reads the \oldG,  
it means the \newG includes more than $k$ \uas 
for the viewport, so \kLCNB also reads the \oldG by definition. 
In addition, if \kLCMB reads the \newG, 
\kLCNB may read the \oldG 
because \kLCNB does not need to preserve \mon. 
So we have $\pSN(\kLCMB) \leq \pSN(\kLCNB)$  
and $\pIV(\kLCMB) \geq \pIV(\kLCNB)$.
In terms of \kLCNB and \kGCNB, \kLCNB reads the \newG 
when this \pG has $k$ or fewer \uas for the \readTxn, 
but \kGCNB can read the \newG only 
when the \newG has $k$ or fewer \uas overall. 
So we have $\pSN(\kLCNB) \leq \pSN(\kGCNB)$  
and $\pIV(\kLCNB) \geq \pIV(\kGCNB)$. 
\end{proof}}
}

\ptr{
\begin{figure}[!t]
    \centering
    \includegraphics[width=70mm]{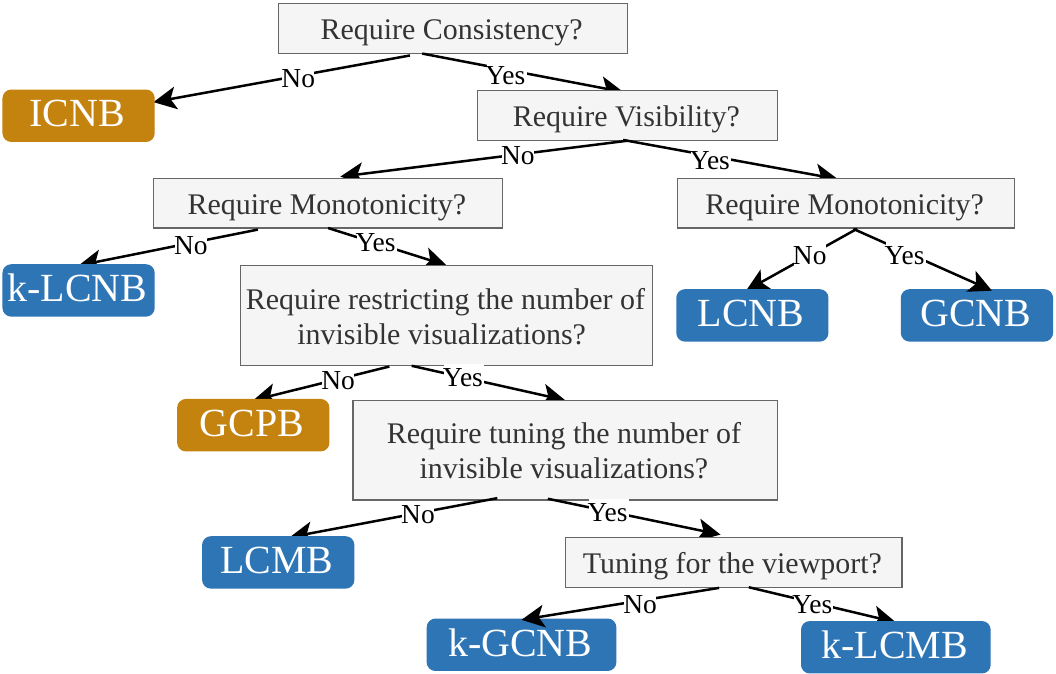}
    \vspace{-4mm}
    \caption{Selection of the appropriate \approach{es}}
    \label{fig:use_cases}
    \vspace{-5mm}
\end{figure}
}

\ptr{
\subsection{Selecting Appropriate \Approach{es}}
\label{sec:use_cases}
We've introduced a number of \approaches so far, many
of which do not strictly dominate each other. Selecting 
the right \approach depends on user needs and
use cases. We now provide guidance for selecting 
the appropriate \approach, summarized as a decision tree 
in Figure~\ref{fig:use_cases}, where the existing \approach{es} 
are marked as gold boxes and the new \approach{es} 
are marked as blue.
\ppaper{We discuss this diagram in detail in our technical report~\cite{tpTR}, as well as
how we may design appropriate user interfaces
to help
users make sense of the \res{s} and guarantees
as part of
a given \approach.}
\ptr{If the user does not require \con across multiple visualizations, 
\pICNB is the best choice since it maintains 
both \mon and \ava, and has low \snMetric. 
Otherwise, we check whether \ava is required. 
If yes, we choose between \pLCNB and \pGCNB, depending 
on the whether the user wants to additionally maintain \mon. 
If \ava is not required, we additionally check 
whether \mon is required. If not, we use \kLCNB 
to reduce \snMetric and \uaMetric. Note that \kLCNB 
subsumes \pLCNB. If the user does not want to tune the $k$ value, 
they can adopt \pLCNB. 
On the other hand, if \mon is needed, 
we check whether the user needs to restrict the number of \unavailable 
visualizations in the dashboard. If not, which means 
the user wants to see the new \qres{s} without worrying 
about \uaMetric, then \pGCPB is adopted. 
Otherwise, we further check whether the user wants to tune 
the $k$ value. If not, \pLCMB is adopted. 
If yes, \kGCNB or \kLCMB is adopted depending on 
whether the user wants to tune the $k$ value 
with respect to the viewport or the dashboard overall.}
}

\begin{reviewtwo}
\subsection{Extensions}
\label{sec:extension}
We now discuss extending our \tdb \model to support 
reading and modifying the definition of a \viewG, 
and user actions beyond modifying the location of the viewport
\ptr{,concurrent \writeTxn{s}, 
and multiple users, and various UI designs}.

\stitle{Supporting reading and modifying \vdef{s}}
\ptr{We now consider the settings where 
the user can read and modify \vdef{s}.}
The user can read a \vdef if it is rendered 
alongside the corresponding visualization 
in the viewport. 
Here, a \readTxn is extended to read \qres{s} and/or \vdef{s} 
in the viewport, depending on which of them 
are rendered. 
To support modifying a \vdef, 
the \viewG's edges and \vdef{s} are multi-versioned; 
that is, a \version of \viewG is defined to be 
$G^{t_i}= (E^{t_i}, \Vt^{t_i}, \Vs^{t_i})$. 
A modification to a \vdef is modeled as a \writeTxn 
that modifies the \vdef and refreshes the visualizations 
that depend on it 
(e.g., modifying a filter on a base table will 
refresh the visualizations derived from this base table). 
The system and scheduler processes such a \writeTxn 
in the same way as they process the \writeTxn 
stemming from modifying the source data. 


\stitle{Supporting user actions beyond modifying the viewport}
In addition to moving the viewport, the user can also 
modify the visualizations in a dashbaord, 
by filtering, panning and shifting, brushing and linking, 
and drilling down/up/across charts. 
We break such actions into two broad categories. 
If the user action needs to be processed at the back-end 
(e.g., a web server), where \tdb operates, 
then this action is 
interpreted as a \writeTxn that involves 
modifying the \vdef 
(e.g., modifying a filter). 
Otherwise, if the user action is performed at the front-end 
(e.g., shifting a visualization), 
\tdb does not need to intervene. 

All that said, while supporting the aforementioned actions, 
the visual interface may present the old \vdef 
for some \approach{es} even after  
the user modifies the \vdef  
(e.g., presenting the old value of a filter after 
it is modified for \pGCNB to preserve \con), 
which introduces additional complications 
regarding the desired user experience. 
If, for example, the dashboard designer determines that 
the dashboard should not show an old filter value to the user 
after they have updated it --- since that might be confusing to the user, 
then the selected \approach can be overridden to 
be either \pGCPB or \pICNB to process 
the case for modifying and reading a filter. 
For all other cases, the selected \approach will be used. 

\end{reviewtwo}


\ptr{
\stitle{Supporting concurrent \writeTxn{s}}
We also support concurrent \writeTxn{s} in a single user setting, 
which means a user or an external system can submit a \writeTxn 
while the previous one is not finished. 
Here, each new \writeTxn creates a new \version of \viewG 
as discussed in Section~\ref{sec:formalize} 
and \writeTxn{s} are committed with 
respect to the order they are submitted. 
The definitions of \macPro for \readTxn{s} do not change 
in this scenario. For example, for \conM, 
a \readTxn returns the \res{s} of the \version 
that has the minimal \ua{s} for this \txn.}

\ptr{
\stitle{Supporting multiple users}
\tdb can be easily extended to support multiple users reading 
the same dashboard. Here, each user can choose different \macPro 
and we process each user's \readTxn{s} 
based on the selected properties separately. 
We leave the case of supporting concurrent writes from multiple users 
to future work.}

\ptr{
\stitle{UI designs}
Instantiating \tdb requires new UI designs in the dashboard, 
such as helping the user identify 
the version of \viewG they are looking at 
and effectively demonstrate 
the multiple choices of \approaches to the user in the dashboard. 
For example, to differentiate two versions 
(for the case of processing one \writeTxn at a time), 
we can annotate the visualization that belongs to the \oldG 
and requires to be updated with a special indicator, 
such as a progress bar on the side. 
Prior work also studies UI designs for differentiating 
multiple versions~\cite{WuCH020InteractionSnapshots}, 
which can be leveraged in \tdb if 
multiple concurrent \writeTxn{s} exist. 
However, the UI designs are not 
the focus of this paper and left for future work.}

%% file: system.tex
\section{Maintaining \macPro}
\label{sec:system}
We now discuss how to maintain different property combinations 
for different \approach{es} defined in \tdb. 
Specifically, we discuss the design of the \viewG and 
auxiliary data structures (Section~\ref{sec:viewG}), 
and the algorithms for maintaining \macPro separately (Section~\ref{sec:cc}-\ref{sec:mv_properties})
and maintaining property combinations for each \approach  (Section~\ref{sec:property_combinations}). 
We assume a \emph{\RTxnMng} responsible for processing \readTxn{s}, 
and another \emph{\WTxnMng} responsible for processing \writeTxn{s}. 
The \pRTxn and \WTxnMng{s} are assumed to run on separate  
threads to enable concurrent execution of the two types of \txn{s}. 
\rtwo{In Section~\ref{sec:prototype}, we discuss the strategy 
for triggering \writeTxn{s}.}

\subsection{\ViewGG and Auxiliary Data Structures}
\label{sec:viewG} 
We maintain a multi-versioned \viewG. 
Each node stores a list of \pitem{s}, called \emph{\itemlist}, 
where a \pitem could be a \qres or \ua, and created by a new \writeTxn. 
Recall that a \ua is a place-holder for the corresponding \qres. 
Each node in the \viewG is associated with a latch to 
synchronize concurrent reads/writes to its \itemlist.  

We additionally maintain an auxiliary table, \metaInfo, 
to store the timestamps of the last \oldAndNew \viewG{s} 
(denoted as $\tc$ and $\ts$, respectively), 
and the number of \ua{s} for the \newG (denoted as $\civ$). 
The quantities $\tc$, $\ts$, and $\civ$ are maintained by the \WTxnMng 
and will be used by the \RTxnMng to 
preserve the properties specified by the user. 
We also include a latch to synchronize concurrent accesses to 
the \metaInfo table, which means any access to \metaInfo 
needs to acquire this latch.

\subsection{Maintaining \Con}
\label{sec:cc}
We now discuss preserving three types of \con: \cc, \cf, and \cm, 
and will discuss preserving \mon and \ava separately 
in the next subsection. 
Intuitively, maintaining \con for a \readTxn 
means this \txn can read the recently \oldAndNew \viewG{s}. 
However, traditional concurrency control protocols, such as 2PL
or OCC~\cite{BernsteinHG87CCBook}, do not apply here 
because they do not support the type of \con that 
reads an uncommitted version of the \viewG (i.e., \cf and \cm). 
To maintain \con, we process a \readTxn in two steps: 
\begin{enumerate}[1)]
\vspace{-1.5mm}
    \item Atomically find the timestamps for the last \oldAndNew \viewG{s} (i.e., $\tc$ and $\ts$ in \metaInfo)
    \item Read the \versions of \viewG for $\tc$ and $\ts$.
\vspace{-1.5mm}
\end{enumerate}
Step 1) is done correctly via the latch on the \metaInfo table. 
Step 2) requires that the \viewG{s} for $\tc$ and $\ts$ exist, 
which is done by the \WTxnMng. 
Step 2) additionally requires 
an algorithm for reading a \version of \viewG for given a timestamp, 
which is done in the \RTxnMng. 
We now discuss the designs of \pRTxn and \WTxnMng{s} 
for maintaining \con. 

\stitle{\WTxnMng}
The \WTxnMng processes a \writeTxn $\txw^{t_i}$ in three steps: 
\begin{enumerate}[1)]
\vspace{-1.5mm}
    \item Create a new \version of \viewG for $\txw^{t_i}$
    \item Compute the \qres{s} for the views involved in $\txw^{t_i}$ 
    and update the \viewG with the new \pqres{s}
    \item Update $\tc$ with $t_i$
\vspace{-1.5mm}
\end{enumerate}
Step 1) guarantees that the timestamp of the latest \viewG, $\ts$, 
exists. Specifically, the \WTxnMng creates a new \version of \viewG 
by appending \ua{s} to the \itemlist{s} of the nodes 
that $\txw^{t_i}$ needs to update\footnote{For \pGCNB and \pICNB, 
which do not need to read the uncommitted \version, 
we can skip generating \ua{s} as an optimization}.
Then it atomically updates $\ts$ with the timestamp of 
the running \writeTxn $\txw^{t_i}$, and $\civ$, 
the number of \ua{s} for the \newG, in \metaInfo. 
Step~2) computes the \qres and replaces the corresponding $\ua$ 
for each view, and updates $\civ$. 
It leverages a scheduler to 
decide the order of computing the \qres{s} 
to reduce \uaMetric and/or \snMetric, 
which we discuss in Section~\ref{sec:scheduler}. 
Step~3) updates the timestamp of the last committed \version (i.e., $\tc$) 
with $t_i$, which guarantees that the version of \viewG for $\tc$ exists. 
\rtwo{We say $\txw^{t_i}$ is committed if we have successfully performed 
the aforementioned three steps for $\txw^{t_i}$.}

\stitle{\RTxnMng} 
The \RTxnMng uses timestamps $\tc$ and $\ts$ 
to read the last \oldAndNew \viewG{s}. 
Depending on the properties that need to be maintained, 
the \RTxnMng decides the \version to read, 
which is discussed in Section~\ref{sec:property_combinations}. 
Here, we present the algorithm for reading a version of \viewG. 
Assuming a \txn $\txr^{s_j}$ needs to read $G^{t_i}$, 
the intuition is that for each view read by $\txr^{s_j}$, 
we read the recent \qres/\ua 
whose timestamp is no larger than $t_i$. 
The reason is that we have two possible cases 
if a \qres/\ua for a node $\vt_k$ belongs to $G^{t_i}$: 
1) $\vt_k$ is or will be modified by $\txw^{t_i}$, in which case 
we have a \qres/\ua whose timestamp is $t_i$; 
2) $\vt_k$ is not modified by $\txw^{t_i}$, in which case 
the most recent \qres whose timestamp is smaller than $t_i$ 
belongs to $G^{t_i}$. 

\subsection{Maintaining \Mon and \Ava}
\label{sec:mv_properties}

To guarantee \mon, in the \RTxnMng we maintain 
a table (denoted as \emph{\lastread}) 
that stores the timestamps of the \qres{s} or \uas 
that are last read. 
\Mon requires that a \readTxn reads the \qres{s} or \ua{s} 
whose timestamps are no smaller than the corresponding timestamps 
in the \lastread table. 
To maintain \ava, we guarantee that the \vstate{s} returned by 
a \readTxn do not include \uas. 

\subsection{Maintaining Property Combinations}
\label{sec:property_combinations}

For the \approach{es} that need to maintain \con, 
we read the \metaInfo and \lastread table to decide 
which \versions of the \viewG to read. 
Specifically, to maintain \MonS-\cf for \pGCPB 
we use $\ts$ in \metaInfo to read the \newG. 
Similarly, to preserve \MonS-\AvaS-\cc for \pGCNB, 
we use $\tc$ to read the last \oldG. 
For \kGCNB, we need to check 
whether the number of \ua{s} for the \newG 
(i.e., $\civ$ in \metaInfo) is no larger than $k$. 
If so, we read the \version for $\ts$, otherwise, $\tc$ is used. 

Maintaining \AvaS-\cm for \pLCNB will read both 
the last \oldAndNew \viewG{s}. 
Among the two sets of returned \vstate{s}, 
we choose to return the set  
that does not have \ua{s} and corresponds to the more recent \version. 
Maintaining \MonS-\cm for \pLCMB requires preserving \mon. 
This is done by checking the \lastread table to 
see whether reading the committed \version violates \mon. 
If not, \pLCMB follows the same procedure of \pLCNB. 
Otherwise, \pLCMB will read the \newG. 
\kLCNB and \kLCMB are processed similarly with $k$ \ua{s} relaxed. 

For the property combination \MonS-\AvaS (adopted by \pICNB), 
which sacrifices \con, we do not need to read \metaInfo. 
Instead, we directly read the \viewG and 
return the most recent \qres for each node involved in the \readTxn. 

\section{\writeTxn Scheduler}
\label{sec:scheduler}


\ptr{
\begin{figure}[!t]
    \centering
    \includegraphics[width=75mm]{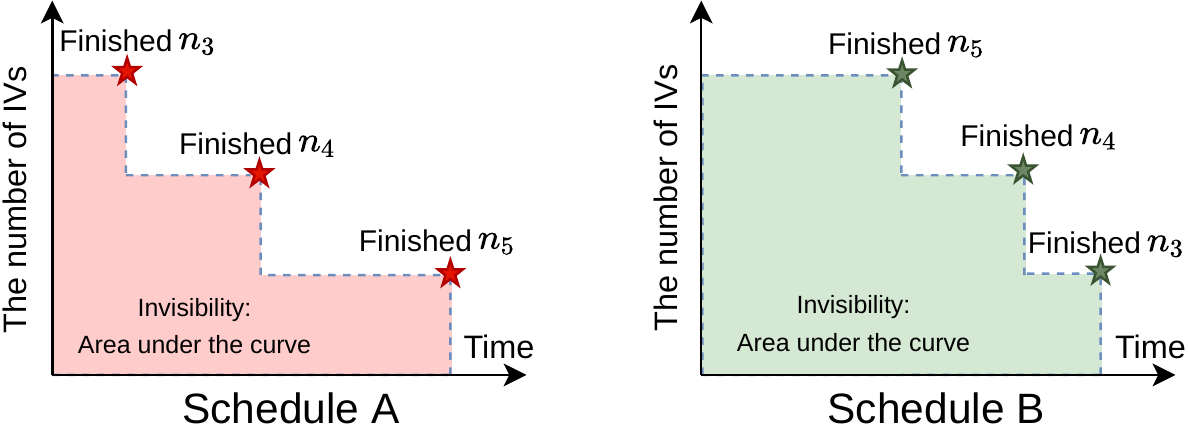}
    \vspace{-3mm}
    \caption{The impact of different schedules on \uaMetric}
    \label{fig:unavailability}
    \vspace{-5mm}
\end{figure}}

\begin{reviewone}
As mentioned in Section~\ref{sec:system}, 
the scheduler, which decides the order 
of computing the new \qres{s} for a \writeTxn, 
impacts \uaMetric and \snMetric for all \approach{es} except \pGCNB. 
\pGCNB will only read the \newG after all of 
the views are updated, so its \snMetric and \uaMetric are 
independent of the scheduler. 
We analyze two factors that 
impact the performance of the scheduler 
and design a scheduling algorithm that considers 
these factors to reduce \uaMetric and \snMetric. 
Our discussion assumes processing a \writeTxn $\txw^{t_i}$ 
that updates a set of nodes $\WS_w^{t_i}$. 

\stitle{Factors that impact \uaMetric and \snMetric} 
\SnMetric (or \uaMetric) is increased if a view that 
is read by a user is stale (or \unavailable), respectively. 
Therefore, prioritizing computing new \pqres{s} 
for views that a user will spend more time reading 
will best reduce the values of the two metrics. 
Since it is impossible to exactly predict how long 
the user will spend reading a 
view, we use $\RTime_{k}^{t_i}$, the total time 
during which a view $\vt_k$ has been read in the viewport since 
the \writeTxn $\txw^{t_i}$ started as a proxy. 
Using $\RTime_{k}^{t_i}$ is based on 
the assumption that the user will spend more time on a view 
in the future if they spent more time on this view in the past. 
However, as this assumption weakens, 
$\RTime_{k}^{t_i}$ will be less effective in 
reducing \uaMetric and \snMetric, as we will see 
in Section~\ref{sec:exp_optimizations}. 
For a set of \readTxn{s} $\TxR = \{\txr^{s_1}, 
\cdots, \txr^{s_m}\}$ after $\txw^{t_i}$ is started, 
$\RTime_{k}^{t_i}$ is defined as 
\vspace{-3mm}
\[
\RTime_{k}^{t_i} = \sum_{j=1}^{m-1} \mathtt I[\vt_k \text{ is read 
by } \txr^{s_{j}}] \times (\TxnT^\prime(\txr^{s_{j+1}}) - \TxnT^\prime(\txr^{s_{j}}))
\]
$\TxnT^\prime(\txr^{s_{j}})$ is the time when the system receives 
the \readTxn $\txr^{s_{j}}$. 
$\mathtt I[\vt_k \text{ is read by } \txr^{s_{j}}]$ is 1 
if the view $\vt_k$ is in the viewport 
when $\txr^{s_{j}}$ is issued, otherwise 0. 
Therefore, $\mathtt I[\vt_k \text{ is read 
by } \txr^{s_{j}}] \times (\TxnT^\prime(\txr^{s_{j+1}}) - \TxnT^\prime(\txr^{s_{j}}))$ represents the duration 
when the view $\vt_k$ stays in the viewport 
between two consecutive \readTxn{s}. 
The system tracks the arrival time of each \readTxn and 
the views that are read by this \txn 
to calculate $\RTime_{k}^{t_i}$. 
In our scheduling algorithm, 
we prioritize scheduling the view that has 
higher $\RTime_{k}^{t_i}$.
\end{reviewone}

In addition, there is another factor that 
impacts the \snMetric and \uaMetric: 
the different amounts of time for computing the \pqres{s} 
for different views. Intuitively, prioritizing computing 
the new \pqres for the view 
that has the least execution time will allow the user 
to read a fresh view earlier, which is also observed 
by previous work~\cite{Antifreeze}. 
\ptr{We use an example to illustrate this. 
Here, say the \writeTxn $\txw^{t_i}$ 
updates the base table $\vt_1$ 
and $\vt_{3-6}$ in Figure~\ref{fig:multiversion}, 
the views $\vt_{3-5}$ are in the current viewport, 
and the viewport does not change. 
We additionally assume we need to maintain properties \MonS-\cf 
(i.e., always reading the last created \version) 
and want to minimize \uaMetric. 
Figure~\ref{fig:unavailability} shows two different 
schedules for $\vt_{3-5}$, where 
the x-axis plots the execution time while the y-axis plots 
the number of \uas. 
The \uaMetric is essentially the area under the curve: 
\emph{the sum of the times of each view being \unavailable}.
We see that while both schedules have the same execution time, 
Schedule~A has lower \uaMetric
than Schedule~B because it always prioritizes 
computing the \qres that has the least execution time. 
This heuristic can similarly reduce \snMetric 
for other property combinations (e.g., \MonS-\cm). }
We use $\ExecTime_{k}^{t_i}$ to represent the amount of 
time for computing the new \pqres for the view $\vt_{k}$ 
while processing $\txw^{t_i}$. 
The view that has a smaller $\ExecTime_{k}^{t_i}$ 
should have a higher priority.

\ptr{
\begin{algorithm}[!t]
    \SetAlgoLined\DontPrintSemicolon
    \SetArgSty{textrm}
    $\WS_w^{t_i} \gets$ Topologically sort $\WS_w^{t_i}$ \;
    $\varname{\RG}$ $\gets$ Break $\WS_w^{t_i}$ into topologically independent groups\;
    \For{$\RG^l \in \RG$}{
        \While{$\RG^l$ is not empty} {
            $\vt_{max} \gets \displaystyle \argmax_{\vt_k \in \RG_l}
            ~\Metric_{k}^{t_i} = \frac{\RTime_{k}^{t_i}}{\ExecTime_{k}^{t_i}}$\;
            Update the view $\vt_{max}$\;
            $\RG^l \gets \RG^l \backslash \{\vt_{max}\}$\;
        }
    }
  \caption{\small Scheduling algorithm in the \WTxnMng}
  \label{alg:schedule}
\end{algorithm}
}

\begin{reviewone}
\stitle{Scheduling algorithm}
We design a metric $\Metric_{k}^{t_i} = 
\RTime_{k}^{t_i}/\ExecTime_{k}^{t_i}$ 
to decide the priority of a view $\vt_k$. 
$\Metric_{k}^{t_i}$ captures the characteristics 
of the two aforementioned factors. 
If two views have the same $\RTime_{k}^{t_i}$, 
a lower $\ExecTime_{k}^{t_i}$ yields a higher $\Metric_{k}^{t_i}$. 
Similarly, for the same $\ExecTime_{k}^{t_i}$, 
a higher $\RTime_{k}^{t_i}$ results in a higher $\Metric_{k}^{t_i}$. 
\end{reviewone}

\ptr{Algorithm~\ref{alg:schedule} shows the scheduling algorithm.}
\ppaper{The scheduling algorithm works as follows.}
We first sort $\WS_w^{t_i}$, the set of nodes $w^{t_i}$ will update, topologically, 
break them into topologically independent groups, 
and compute each group with respect to the topological order. 
That is, we should only compute \qres{s} for views 
whose precedents are updated. 
To schedule a view to be updated within a group, 
we compute $\Metric_{k}^{t_i}$ for each yet 
computed view $\vt_k$ in this group 
and choose the view with the highest $\Metric_{k}^{t_i}$.

%% file: prototype.tex
\section{Prototype Implementation}
\label{sec:prototype}

\ptr{
\begin{figure}[!t]
    \centering
    \includegraphics[width=80mm]{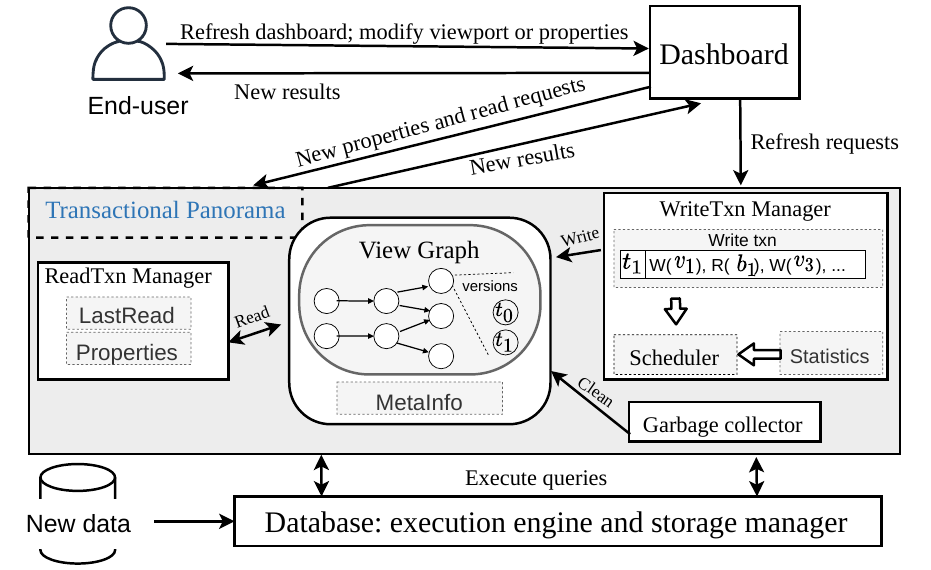}
    \vspace{-4mm}
    \caption{Overview of our implementation in Superset}
    \label{fig:system_overview}
    \vspace{-7mm}
\end{figure}
}

We now discuss implementing the \tdb \framework
in Superset~\cite{superset}, 
a widely used open-source BI tool\ppaper
\ptr{that has 47.3k stars on Github and similar functionality to 
Tableau~\cite{tableau} and PowerBI~\cite{powerBI}}.
Superset provides a web-based client interface, 
where a user can define 
visualizations and organize them as part of a dashboard.
\ppaper{\rtwo{The dashboard can be refreshed manually or configured 
to refresh periodically.}} 
\ppaper{\rtwo{Each refresh is interpreted as a \writeTxn.}} 
Superset adopts a web server to process front-end requests 
and employs a database to store the base tables 
and compute new \qres{s} for visualizations. 
\ppaper{\rone{The details of the prototype can be found in 
the extended version~\cite{tpTR}}.}

\ptr{
Figure~\ref{fig:system_overview} shows an overview of the system. 
The user interacts with a dashboard by 
changing the viewport to explore different 
subsets of visualizations, selecting desired properties,  
and refreshing the dashboard with 
respect to changes to the underlying data. 
As in today's many BI tools, the dashboard 
can be refreshed manually or configured to 
refresh periodically (e.g., every 1 min).
When the dashboard needs to be refreshed, 
it sends a refresh request to the \WTxnMng, 
where each refresh request is interpreted as a \writeTxn. 
After a refresh request is sent, 
the dashboard will regularly send read requests to the \RTxnMng 
to pull refreshed visualizations. 
Each read request includes the viewport information, 
which is leveraged by the \RTxnMng to construct \readTxn{s}. 
When the system has finished the \writeTxn, 
the \RTxnMng returns the new \qres{s} 
that have not yet been read by the user 
to the dashboard to refresh all of its visualizations, 
after which, the dashboard stops sending new read requests. 
When the system is not processing a \writeTxn, 
the user can change the desired properties. 
Changing the properties on the fly is left for future work. 

We store the base tables in the database and 
maintain the \res{s} of the derived visualizations along 
with the \viewG in memory in the web server. 
We also include a garbage collector to clean up 
\qres{s} in the \viewG that are guaranteed to not be read 
in the future. To safely clean up the useless \qres{s}, 
we maintain a timestamp, $\tr$, 
representing the oldest version of \viewG 
that a \readTxn is reading or will read in the future. 
With $\tr$, the garbage collector can safely remove the \qres{s} 
that belong to the older versions of \viewG than $\tr$. 
We maintain $\tr$ by updating it with $\tc$, 
the timestamp for most recently committed version of \viewG, 
before each \readTxn starts, 
because the oldest possible version a \readTxn will read is the \oldG. }

%


%% file: experiment.tex
\section{Experiments}
\label{sec:experiment}
The high-level goal of our experiments is to 
characterize the relative benefits of different \approaches
for various workloads---to help users 
select the right \approach for their needs 
and make appropriate performance trade-offs (Section~\ref{sec:exp_discrete}). 
Our experiments also seek to demonstrate the value of 
the new \approach{es}, 
which provide new trade-off points for the user to select (Section~\ref{sec:exp_discrete}-\ref{sec:exp_k_relaxed}), 
and evaluate the performance benefit and overhead 
of the optimizations for the \writeTxn scheduler (Section~\ref{sec:exp_optimizations}). 
\ptr{
Our experiments address the following research questions:
\begin{enumerate}[1)]
    \item \text{[Workload impact on performance]} How do different user behaviors 
    and dashboard configurations impact \uaMetric and \snMetric for the \baseApproach{es} in \tdb? (Section~\ref{sec:exp_discrete})
    \item \text[Performance benefit of new \approach{es}] 
    How much do the new \approach{es} reduce \uaMetric 
    while maintaining \con, compared to existing \approach{es}? (Section~\ref{sec:exp_discrete})
    \item \text{[Trade-offs provided by $k$-\relaxed variants]} How do different $k$ values impact the \uaMetric and \snMetric for the $k$-\relaxed variants? (Section~\ref{sec:exp_k_relaxed})
    \item \text{[Impact of optimized scheduling]} How much does our \writeTxn scheduler decrease or increase 
    \uaMetric and \snMetric? (Section~\ref{sec:exp_optimizations})
\end{enumerate}
}
%

\begin{table}[!t]
\begin{minipage}[b]{0.3\linewidth}
\centering
\includegraphics[width=18mm]{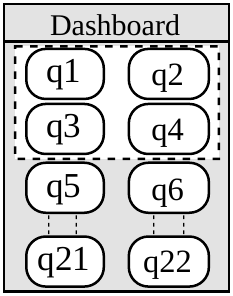}
\vspace{-3mm}
\captionof{figure}{TPC-H dashboard}
\label{fig:tpch_dashboard}
\end{minipage}\hfill
\begin{minipage}[b]{0.7\linewidth}
\centering
\fontsize{7}{9}\selectfont
\begin{tabular}{lll}
Configurations & Options & Default Value \\ \hline
\Readb  & \begin{tabular}[c]{@{}l@{}}\{\regMove,\\  \waitMove,\\ \ranMove\}\end{tabular} & \regMove \\ \hline
\Explorerange  & \{22, 16, 10, 4\} & 22  \\ \hline
\Viewportsize  & \{4, 10, 16, 22\} & 4   \\ \hline
\end{tabular}
\caption{Experiment configurations}
\label{tbl:configuration}
\end{minipage}
\vspace{-7mm}
\end{table}


\begin{figure}[!t]
  \begin{tabular}{cc}
      \begin{minipage}[b]{0.48\linewidth}
        \centering
        \includegraphics[width=\linewidth]{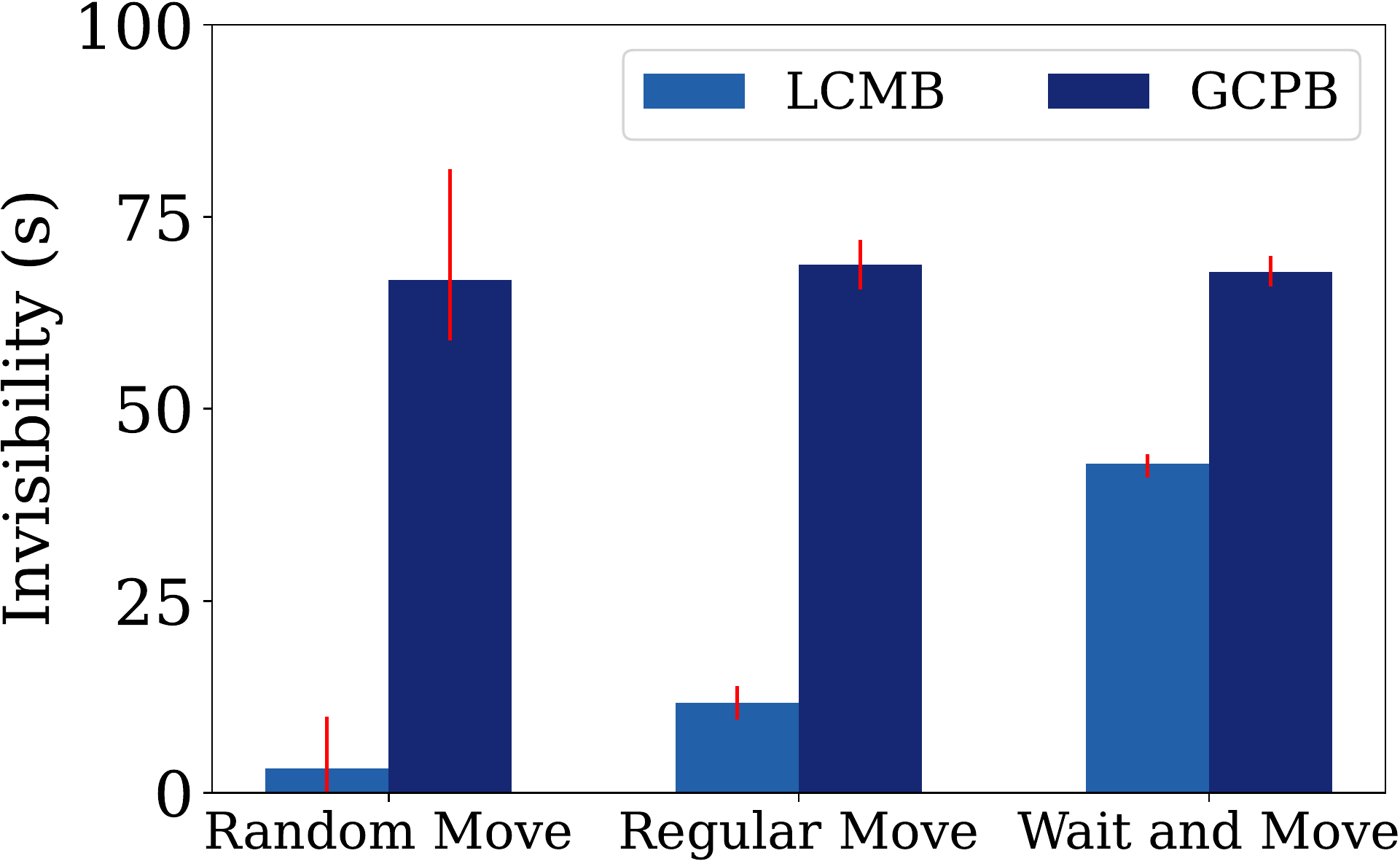}
        \vspace{-5mm}
        \subcaption{\UaMetric}
        \label{fig:exp_read_behavior_iv}
      \end{minipage} 

      \begin{minipage}[b]{0.48\linewidth}
        \centering
        \includegraphics[width=\linewidth]{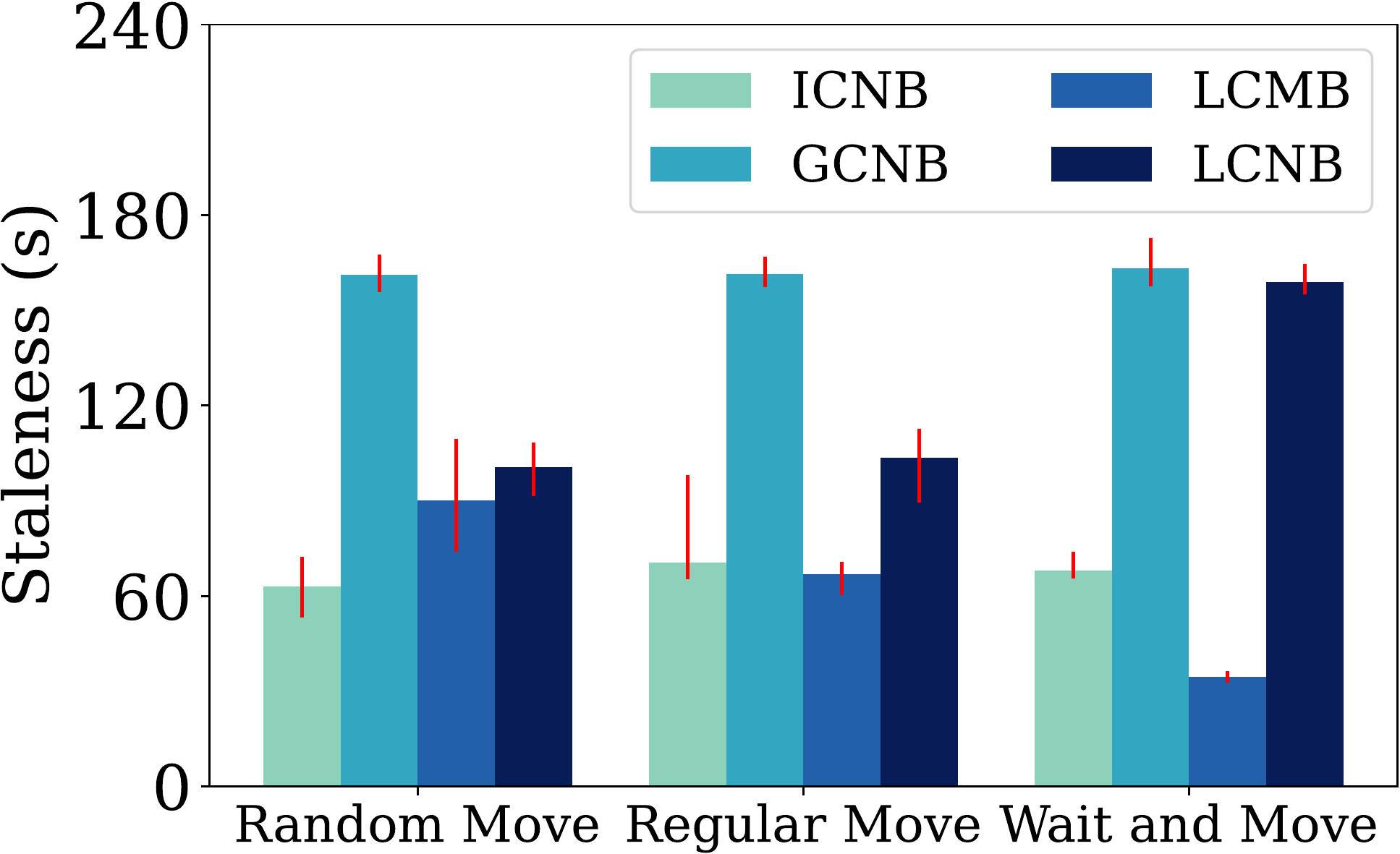}
        \vspace{-5mm}
        \subcaption{\SnMetric}
        \label{fig:exp_read_behavior_sl}
      \end{minipage} 
  \end{tabular}
  \vspace{-6mm}
  \caption{Evaluations of different read behaviors}
  \label{fig:exp_read_behavior}
  \vspace{-7mm}
\end{figure}

\begin{figure*}[!t]
\centering
\begin{tabular}{cc}
    \begin{minipage}[b]{0.24\linewidth}
    \begin{tabular}{c}
    \begin{minipage}{\linewidth}
    \includegraphics[width=\linewidth]{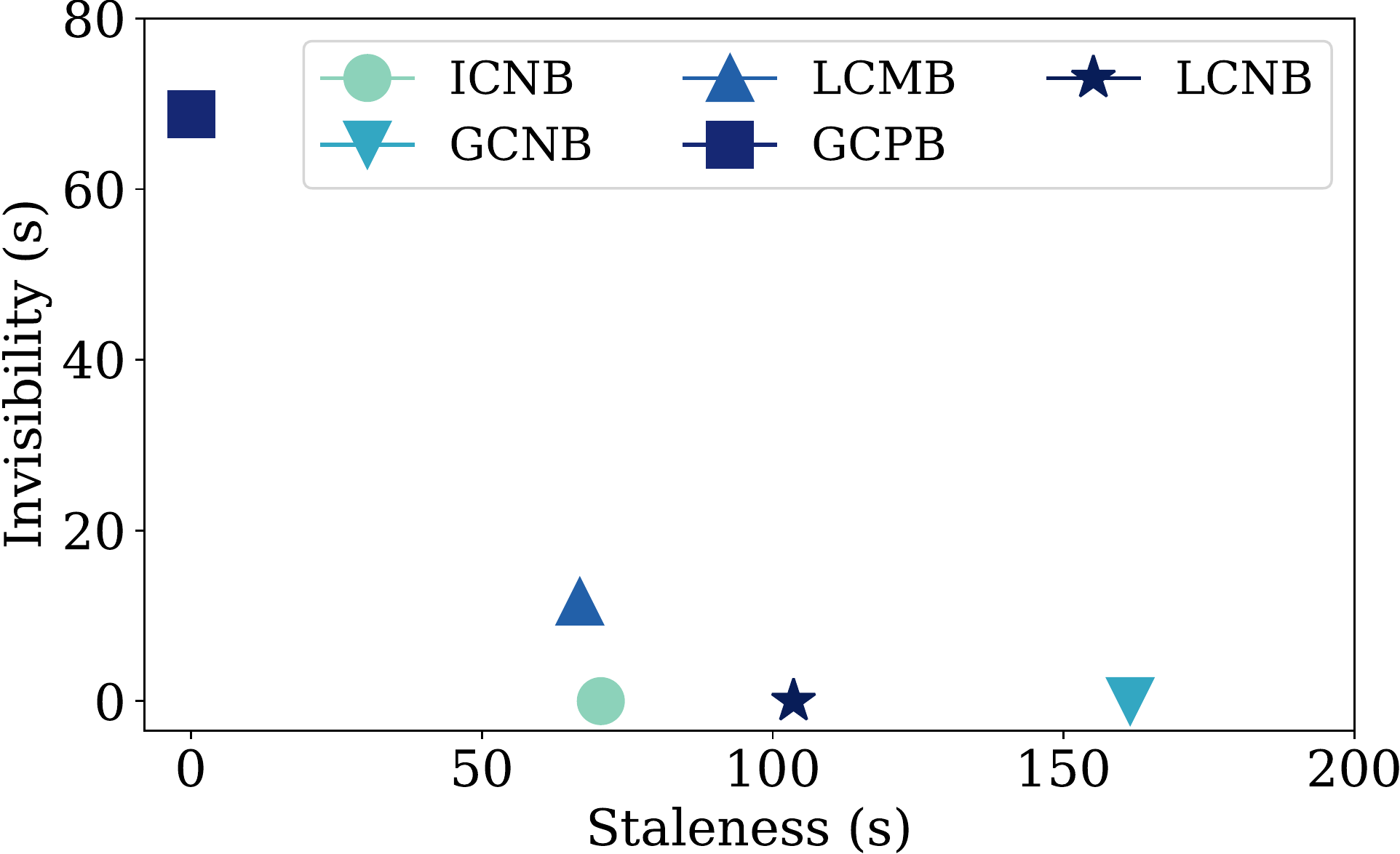}
    \vspace{-5mm}
    \caption{Performance trade-off for \regMove}
    \vspace{-3mm}
    \label{fig:exp_tradeoff_regular_move}
    \end{minipage} 
    \end{tabular}
    \end{minipage}

   \begin{minipage}{0.74\textwidth}
        \begin{tabular}{cc}
      \begin{minipage}[b]{0.5\linewidth}
        \centering
        \includegraphics[width=\linewidth]{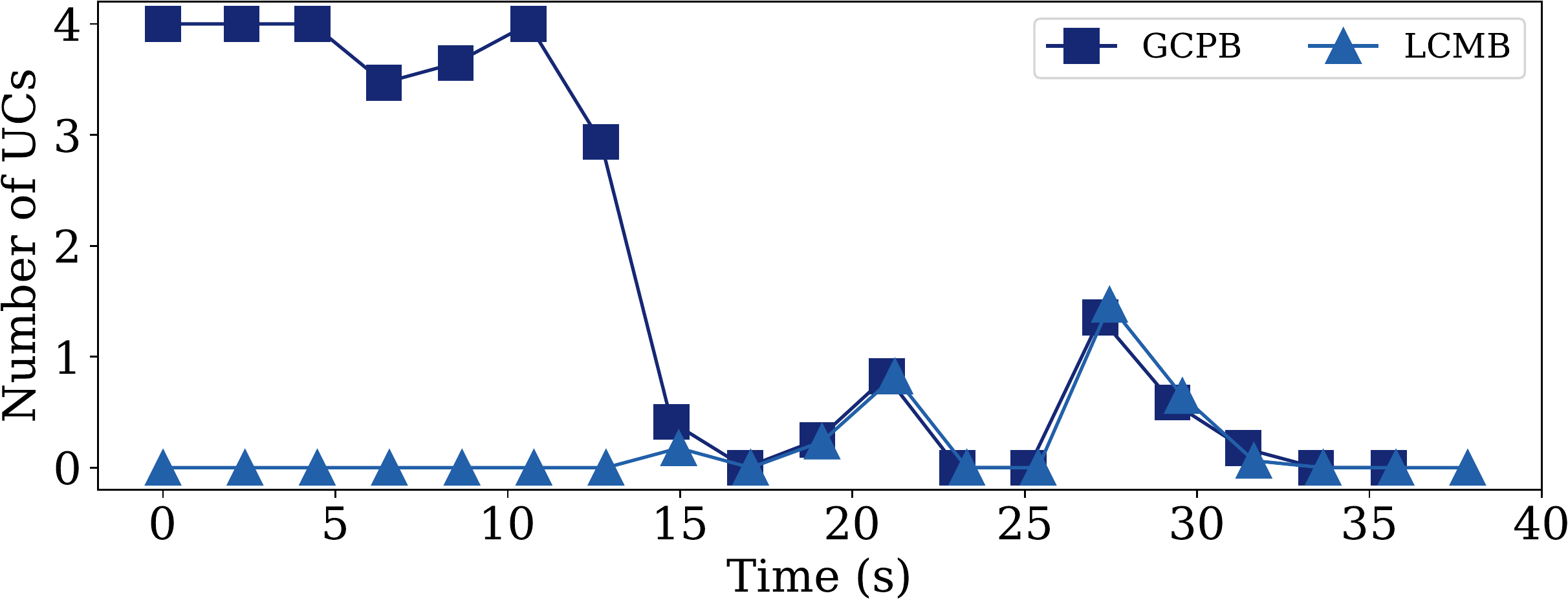}
        \vspace{-5mm}
        \subcaption{\UaMetric}
        \label{fig:exp_trace_iv}
      \end{minipage}
             \begin{minipage}[b]{0.5\linewidth}
        \centering
        \includegraphics[width=\linewidth]{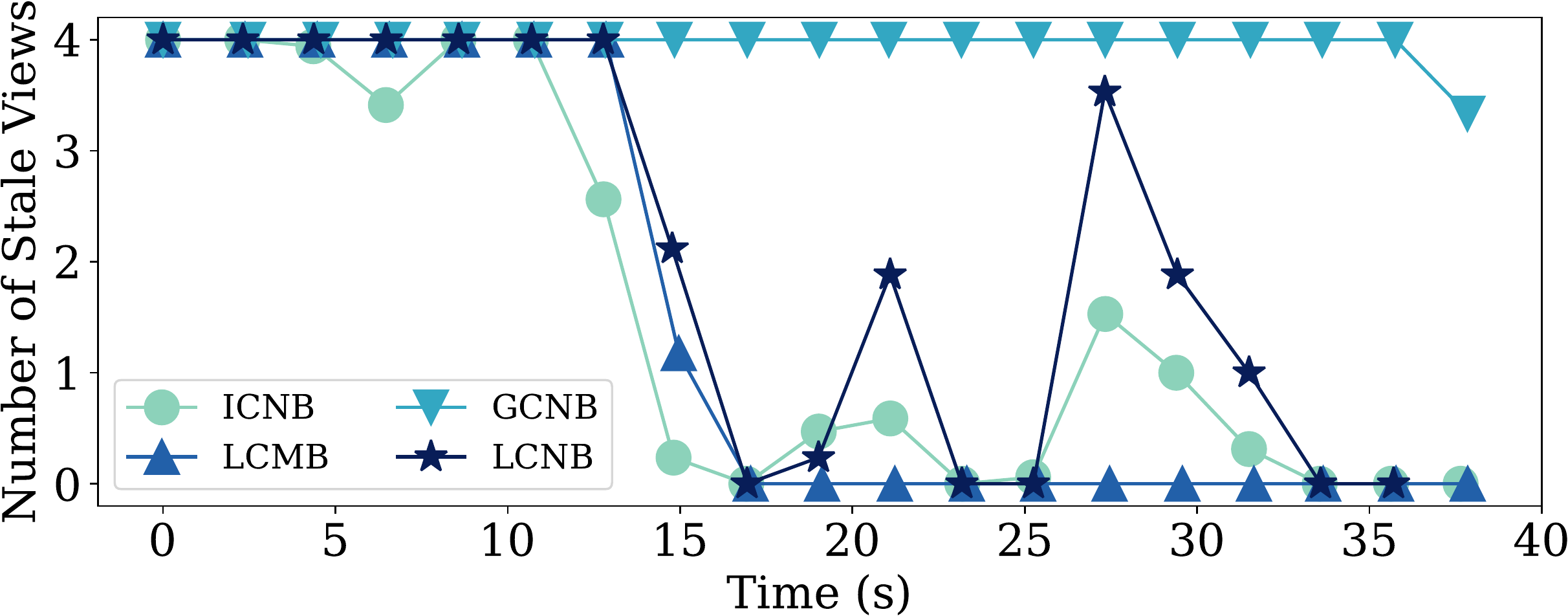}
        \vspace{-5mm}
        \subcaption{\SnMetric}
        \label{fig:exp_trace_sl}
      \end{minipage} 
        \end{tabular}
        
          \vspace{-5mm}
  \caption{The number of \unavailable and stale views in the viewport during the refresh \\ \hfill}
  \label{fig:exp_trace}
  \vspace{-3mm}
   \end{minipage}
\end{tabular}
\vspace{-5mm}
\end{figure*}

\stitle{Benchmark} 
We build a dashboard based on the TPC-H benchmark. 
This dashboard includes 22 visualizations 
for all of the 22 TPC-H queries and runs on 1 GB of data stored in PostgreSQL. 
This dashboard places two visualizations in a row, 
as in Figure~\ref{fig:tpch_dashboard}. 
\rtwo{We test one refresh of the dashboard with respect 
to modifications to the base tables unless otherwise specified, 
since the main focus of this paper is on the scenario 
where there is only one write transaction in the system at a time. 
This case happens when the period for triggering a refresh 
is longer than the time for executing the refresh, or if 
the system only processes one refresh at a time. 
For the test of one refresh, 
we insert 0.1\% new data to the tables Lineitem, 
Orders, and PartSupp and then refresh  
all of the 22 visualizations.}
One test ends when we have 
computed the new \qres{s} for all visualizations. 

We build a test client to simulate different user behaviors 
and dashboard configurations. 
Similar to the web client of Superset, this client sends  
web requests to the web server to trigger a refresh 
(i.e., start a \writeTxn), 
configure the \approach used for processing a refresh, 
and regularly pull refreshed results of visualizations in the viewport
(i.e., start \readTxn{s}). 
We simulate three types of user behaviors in moving the viewport 
to read different visualizations, which we call the \emph{\readb}:
1)~\textit{\regMove}: regularly moving the viewport 
downward or upward and reversing the direction 
if we reach the boundary of the dashboard; 
2)~\textit{\waitMove}: similar to the first one  
with the difference that it only moves the viewport 
after all of visualizations in the viewport are refreshed; 
and \rtwo{3)~\textit{\ranMove}: randomly chooses a viewport, 
which simulates the behavior where the user moves around a lot 
in the dashboard.  
For the three behaviors, 
the viewport is placed at the top of the dashboard 
at the beginning of each test and moved every 1 second.}
For the first two behaviors, each move 
changes the viewport by a row of visualizations. 
The test client can additionally vary 
the number of visualizations the user will inspect
in the dashboard (denoted as \emph{\exploreRange}) during a test. 
We assume these visualizations are at the top of the dashboard.
For example, \exploreRange 4 means that the user will 
explore the visualizations for $q_{1-4}$ in 
Figure~\ref{fig:tpch_dashboard} during a refresh. 
Our experiments also vary the number of visualizations 
in the viewport (denoted as \emph{\viewportSize}) to evaluate 
how the relative sizes of the viewport and the dashboard 
impact \uaMetric and \snMetric. 
The experiment configurations are summarized in Table~\ref{tbl:configuration} and we use default configurations 
unless otherwise specified. 

\stitle{Configurations, and measuring \uaMetric and \snMetric}
The experiments are run on a t3.2xlarge instance of AWS EC2, 
which has 16 GB of memory and 8 vCPUs, 
and uses Ubuntu 20.04 as the OS. 
Our experiments use PostgreSQL~10.5 with default configurations. 
The time interval between two consecutive \readTxn{s} is 
set to be 100 ms to avoid overwhelming the web server. 
that is, the test client sends requests to 
pull refreshed \res{s} every 100 ms. 
We run each test three times and report the mean number 
except for the tests that involve \ranMove. 
For those tests, we run each 
10 times and report the min, max, and mean. 

\rthree{
To measure \uaMetric and \snMetric, the test 
client tracks the timestamp when each \readTxn returns 
and the content of the returned \vstate{s}, 
which include information for whether 
each returned \vstate is a \ua or a stale \qres. 
Using this information and the definitions in Section~\ref{sec:metrics}, 
we can compute \uaMetric and \snMetric. 
For example, the \uaMetric for one test is initialized to 0. 
During the test, if a \readTxn has returned a \ua for a view, 
then the time difference between when this and 
the next \readTxn return will be added to the \uaMetric. }

%
%

\subsection{Performance of \bBaseApproach{es}}
\label{sec:exp_discrete}
 
\ptr{
\noindent\fbox{\begin{minipage}{80mm}\small
\textit{Takeaways: 1) The new \approaches significantly reduce \uaMetric while preserving \con, 
compared to the existing 
\approach \pGCPB; 2) A larger \exploreRange increases the \uaMetric 
and \snMetric for all \approaches except \pGCNB; 
3) The \snMetric of all \approaches except \pGCPB increases with 
a larger \viewportSize and converges to \pGCNB when the \viewportSize equals the dashboard size.}
\end{minipage}}
}

\vspace{1mm}
\noindent We evaluate the configurations 
in Table~\ref{tbl:configuration} for the \baseApproach{es} 
using one refresh. 
\rtwo{Afterwards, we test the impact of varied \prefreshinterval{s} 
for multiple refreshes.}

\stitle{\Readb} 
Figure~\ref{fig:exp_read_behavior} reports the \uaMetric and 
\snMetric for the \baseApproach{es} under different \readb{s}. 
Each test reports the mean
with the min/max as the error bar (i.e., the red line). 
We note that if the \uaMetric or \snMetric for a \approach 
is zero, then that \approach is not shown in the figure. 
To better see the trade-off between \uaMetric and \snMetric, 
we also plot the two metrics together 
in Figure~\ref{fig:exp_tradeoff_regular_move} for \regMove. 
We observe significant differences in \uaMetric and \snMetric 
for different \approaches in Figure~\ref{fig:exp_read_behavior}---the new \approaches 
(i.e., \pLCMB, \pLCNB, and \pGCNB) can significantly 
reduce \uaMetric while maintaining \con compared 
to the existing \approach \pGCPB. 
Specifically, \pGCPB has the highest \uaMetric 
because it always reads the \newG, 
and \pLCMB, the \approach that first reads the 
\oldG and switches to read the \newG, 
can reduce \uaMetric by up to 95.2\% compared to \pGCPB (i.e., \ranMove). 
\pLCMB reduces less \uaMetric in the case of \waitMove because 
it spends a longer time on reading the \newG. 
That is, after the first viewport, \pLCMB always reads the \newG  
for the rest of the viewports\ptr{ due to the overlapping views 
across consecutive viewports}. 
\pLCNB and \pGCNB have zero \uaMetric. 
Recall that \pLCNB always reads the \version of \viewG 
with zero \ua{s} for the viewport to 
maintain \con and \ava, but sacrifices \mon, 
and \pGCNB reads the last \oldG until all of the new \qres{s} 
are computed for the \newG. 
\pICNB also has zero \uaMetric, but sacrifices \con. 

On the other hand \pLCMB, \pLCNB, and \pICNB can significantly reduce \snMetric
relative to \pGCNB. 
For example, \pLCMB reduces \snMetric 
by up to 78.9\% compared to \pGCNB (i.e., for \waitMove). 
\pLCMB reduces the \snMetric by sacrificing \ava 
while \pICNB and \pLCNB need to sacrifices \con and \mon, respectively. 
Overall, these results show that it is valuable to enable a user 
to have these options to make appropriate trade-offs. 
In addition, Figure~\ref{fig:exp_read_behavior_sl} verifies 
the result that the order between \pICNB and \pLCMB 
for \snMetric is undecided in Section~\ref{sec:order}---the \snMetric of \pICNB can be either higher 
or lower than \pLCMB in Figure~\ref{fig:exp_read_behavior_sl}. 

\begin{figure}[t]
  \begin{tabular}{cc}
      \begin{minipage}[b]{0.48\linewidth}
        \centering
        \includegraphics[width=\linewidth]{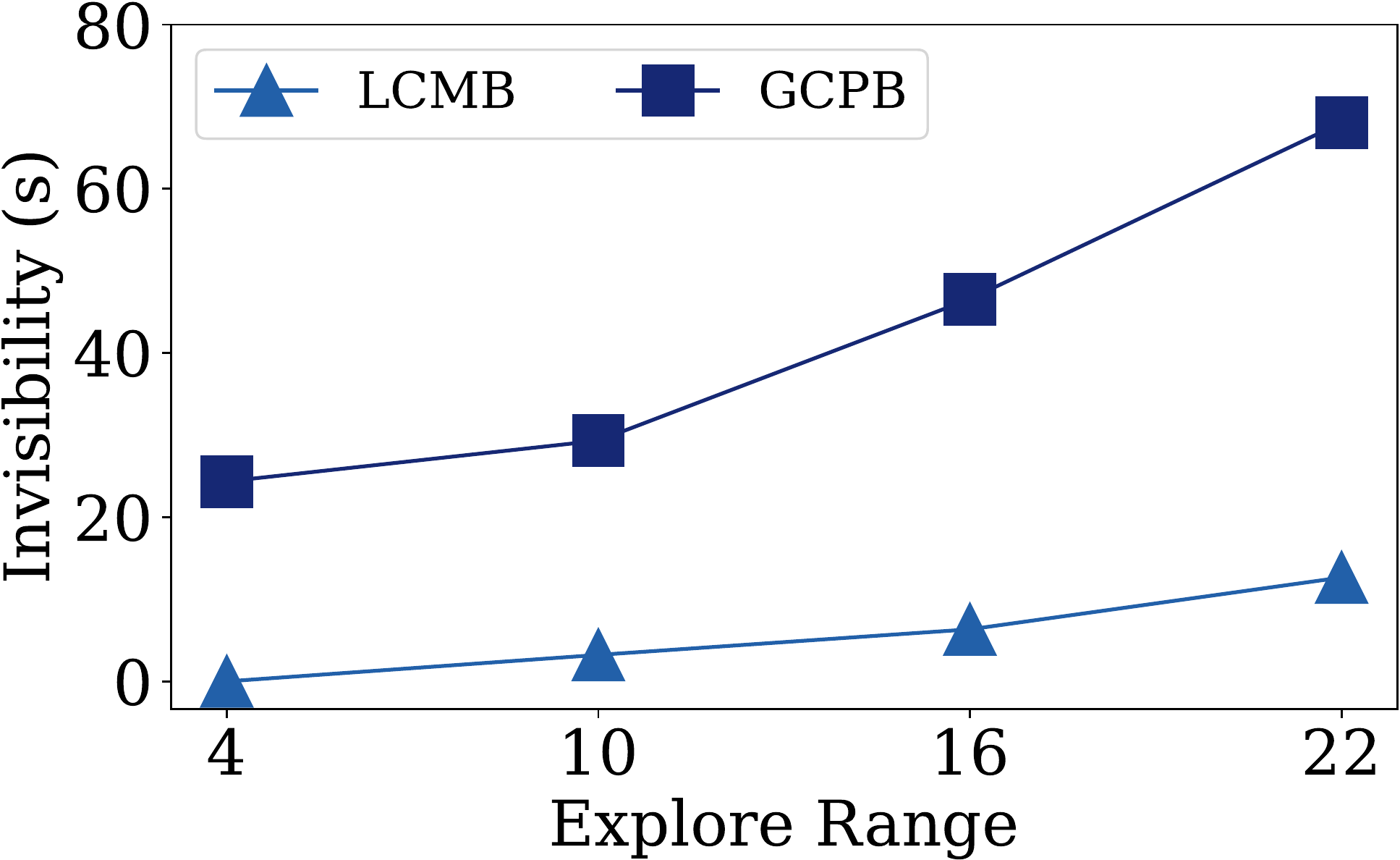}
        \vspace{-5mm}
        \subcaption{\UaMetric}
        \label{fig:exp_explore_range_iv}
      \end{minipage}
      
       \begin{minipage}[b]{0.48\linewidth}
        \centering
        \includegraphics[width=\linewidth]{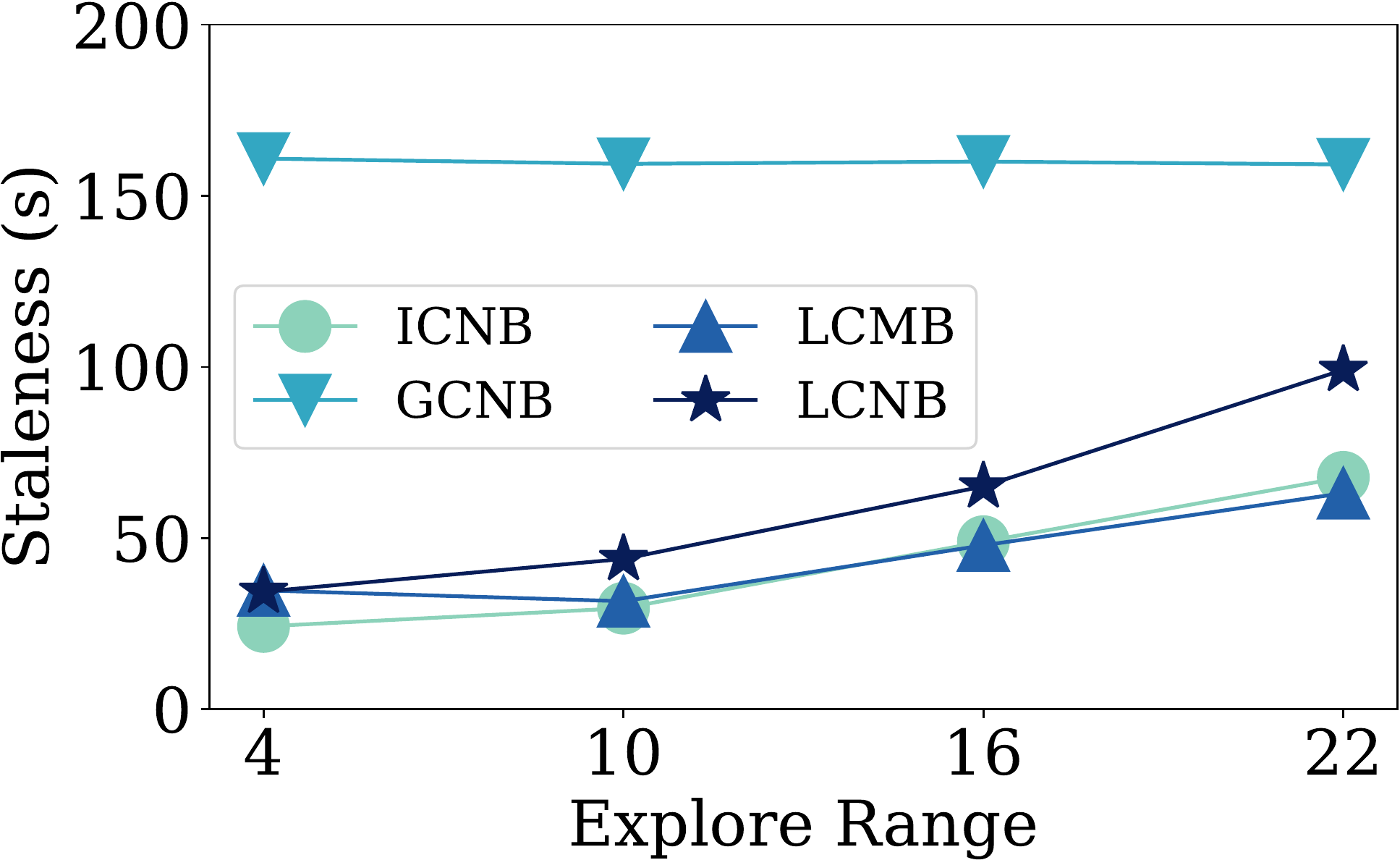}
        \vspace{-5mm}
        \subcaption{\SnMetric}
        \label{fig:exp_explore_range_sl}
      \end{minipage} 
  \end{tabular}
  \vspace{-6mm}
  \caption{Evaluations of different \exploreRange{s}} 
  \label{fig:exp_explore_range}
  \vspace{-5mm}
\end{figure}

\begin{figure}[t]
  \begin{tabular}{cc}
  
      \begin{minipage}[b]{0.48\linewidth}
        \centering
        \includegraphics[width=\linewidth]{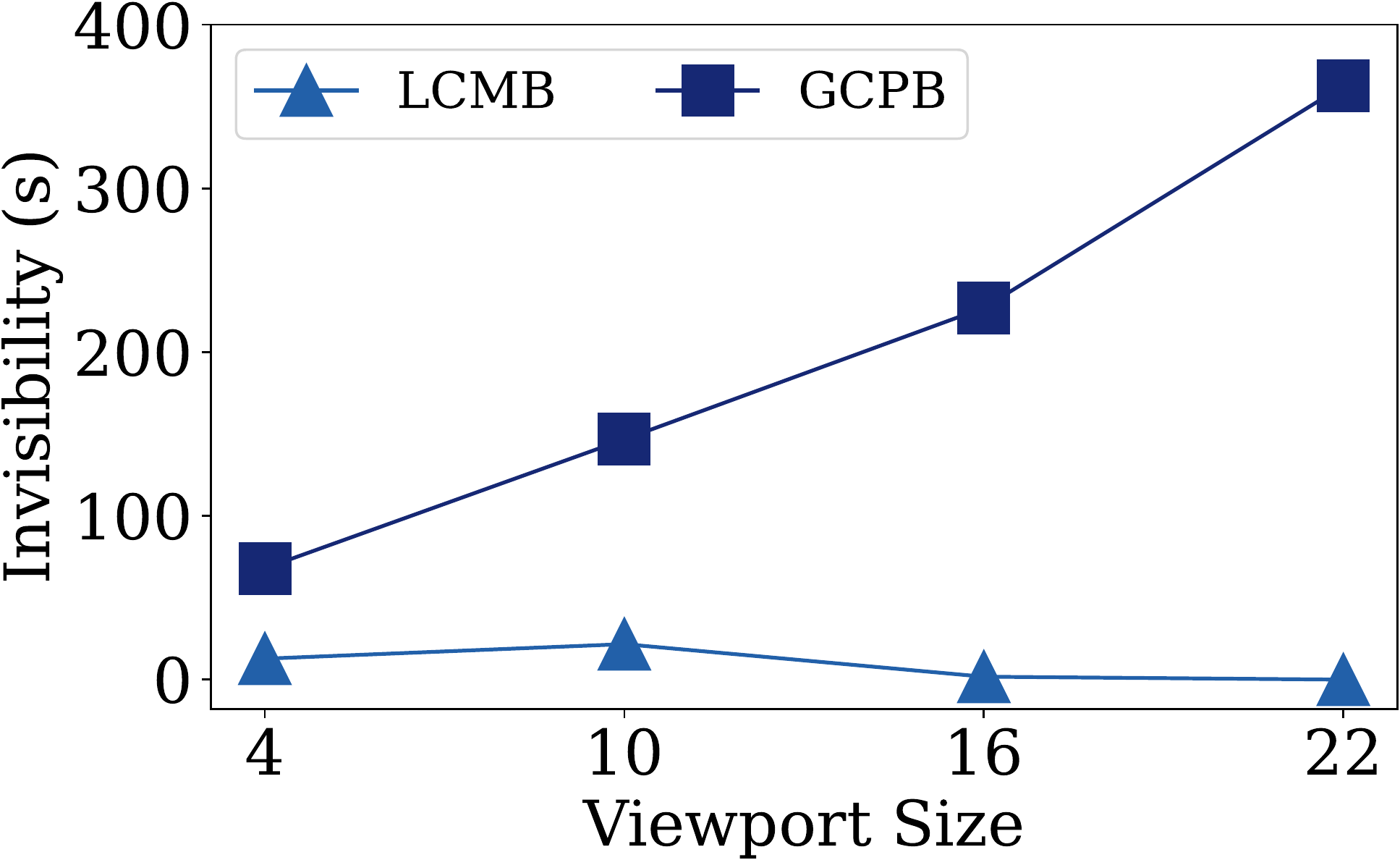}
        \vspace{-5mm}
        \subcaption{\UaMetric}
        \label{fig:exp_viewport_size_iv}
      \end{minipage} 

      \begin{minipage}[b]{0.48\linewidth}
        \centering
        \includegraphics[width=\linewidth]{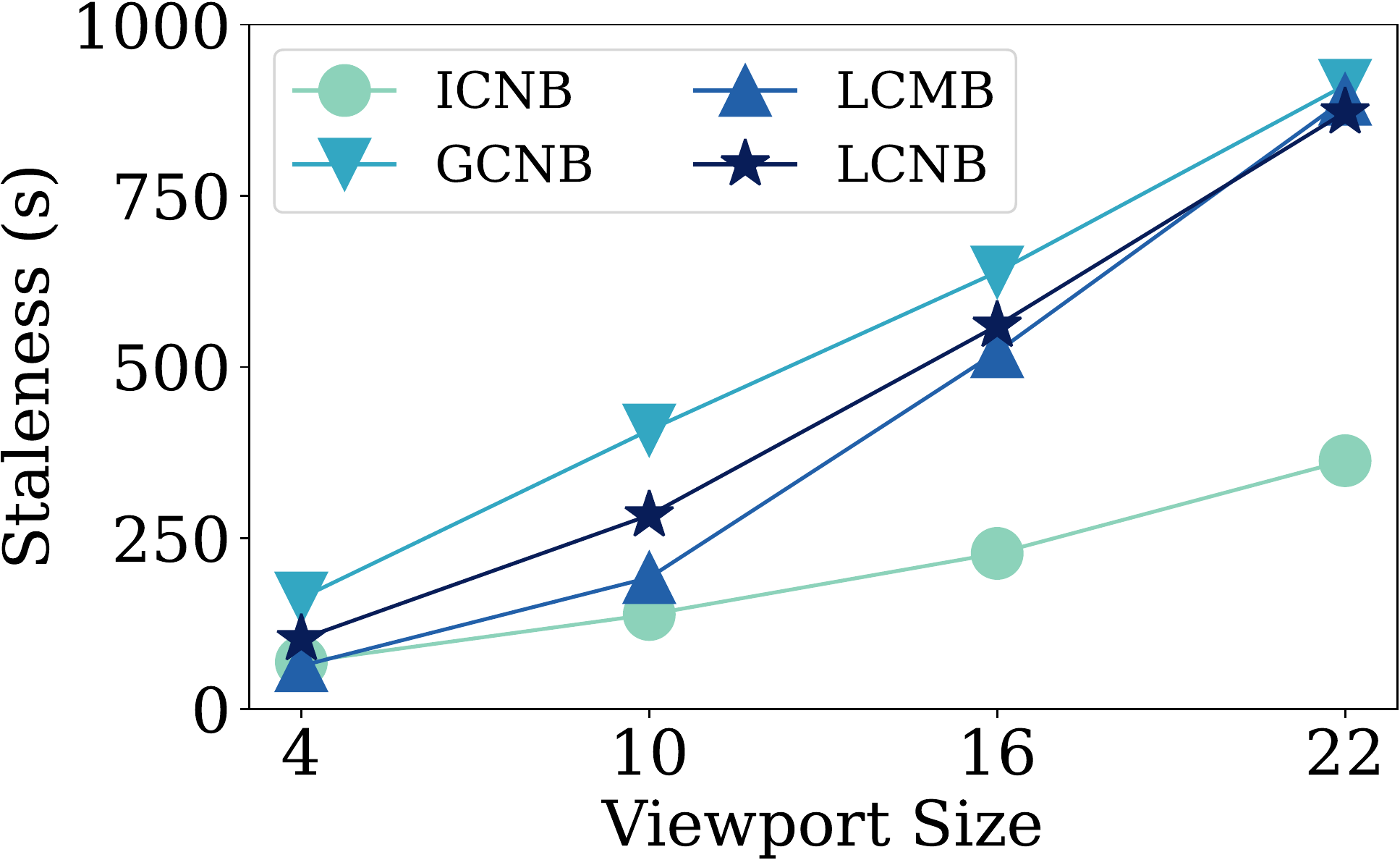}
        \vspace{-5mm}
        \subcaption{\SnMetric}
        \label{fig:exp_viewport_size_sl}
      \end{minipage}     
  \end{tabular}
  \vspace{-6mm}
  \caption{Evaluations of different \viewportSize{s}} 
  \label{fig:exp_viewport_size}
  \vspace{-5mm}
\end{figure}

\begin{figure}[t]
  \begin{tabular}{cc}
      \begin{minipage}[b]{0.48\linewidth}
        \centering
        \includegraphics[width=\linewidth]{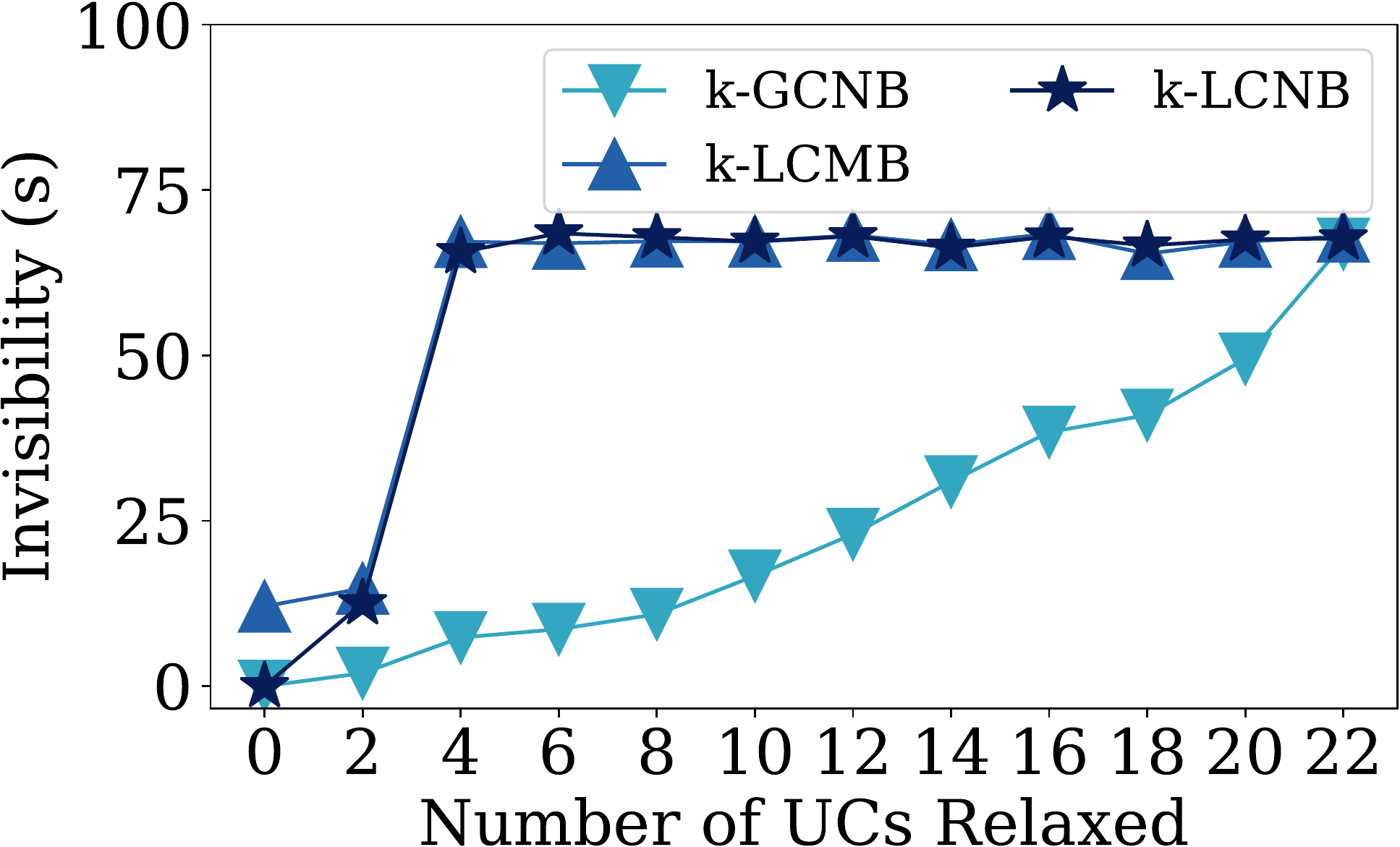}
        \vspace{-5mm}
        \subcaption{\UaMetric}
        \label{fig:exp_k_relaxed_iv}
      \end{minipage} 

      \begin{minipage}[b]{0.48\linewidth}
        \centering
        \includegraphics[width=\linewidth]{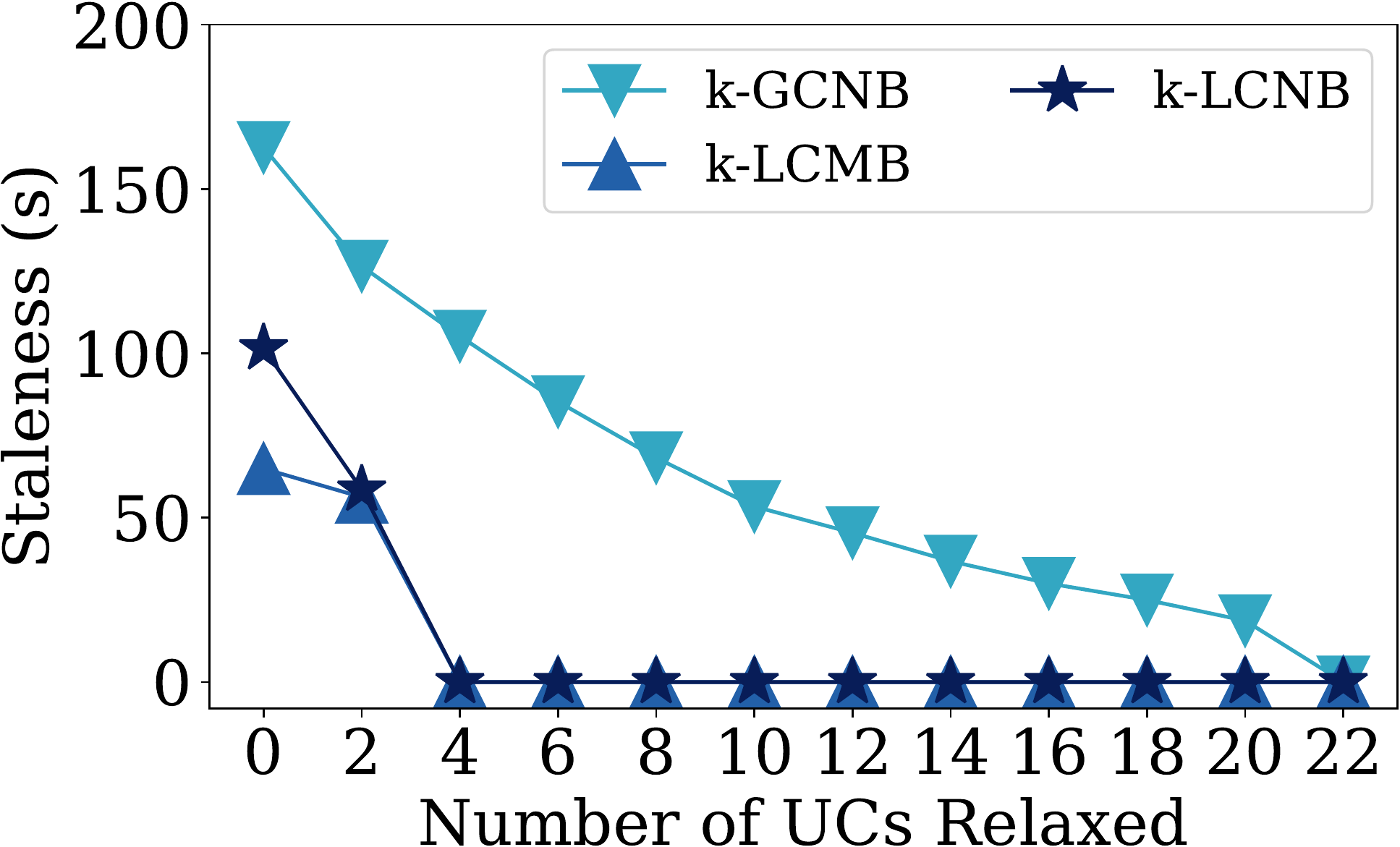}
        \vspace{-5mm}
        \subcaption{\SnMetric}
        \label{fig:exp_k_relaxed_sl}
      \end{minipage}
      
  \end{tabular}
  \vspace{-6mm}
  \caption{Evaluations of varied $k$ values} 
  \label{fig:exp_k_relaxed}
  \vspace{-5mm}
\end{figure} 

\begin{figure}[t]
  \begin{tabular}{cc}
      \begin{minipage}[b]{0.48\linewidth}
        \centering
        \includegraphics[width=\linewidth]{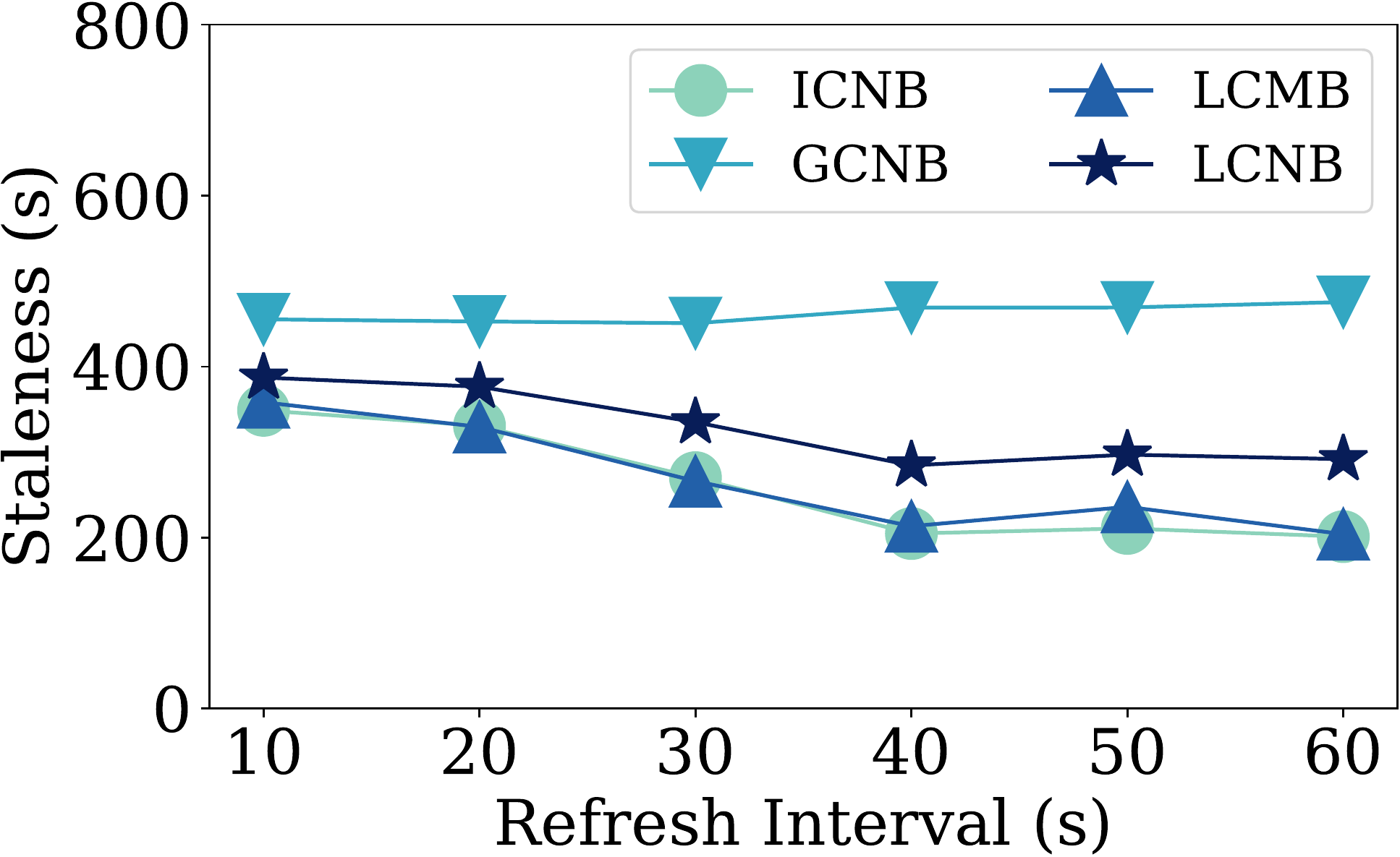}
        \vspace{-5mm}
        \subcaption{\SnMetric}
        \label{fig:exp_refresh_sl}
      \end{minipage}
      
      \begin{minipage}[b]{0.48\linewidth}
        \centering
        \includegraphics[width=\linewidth]{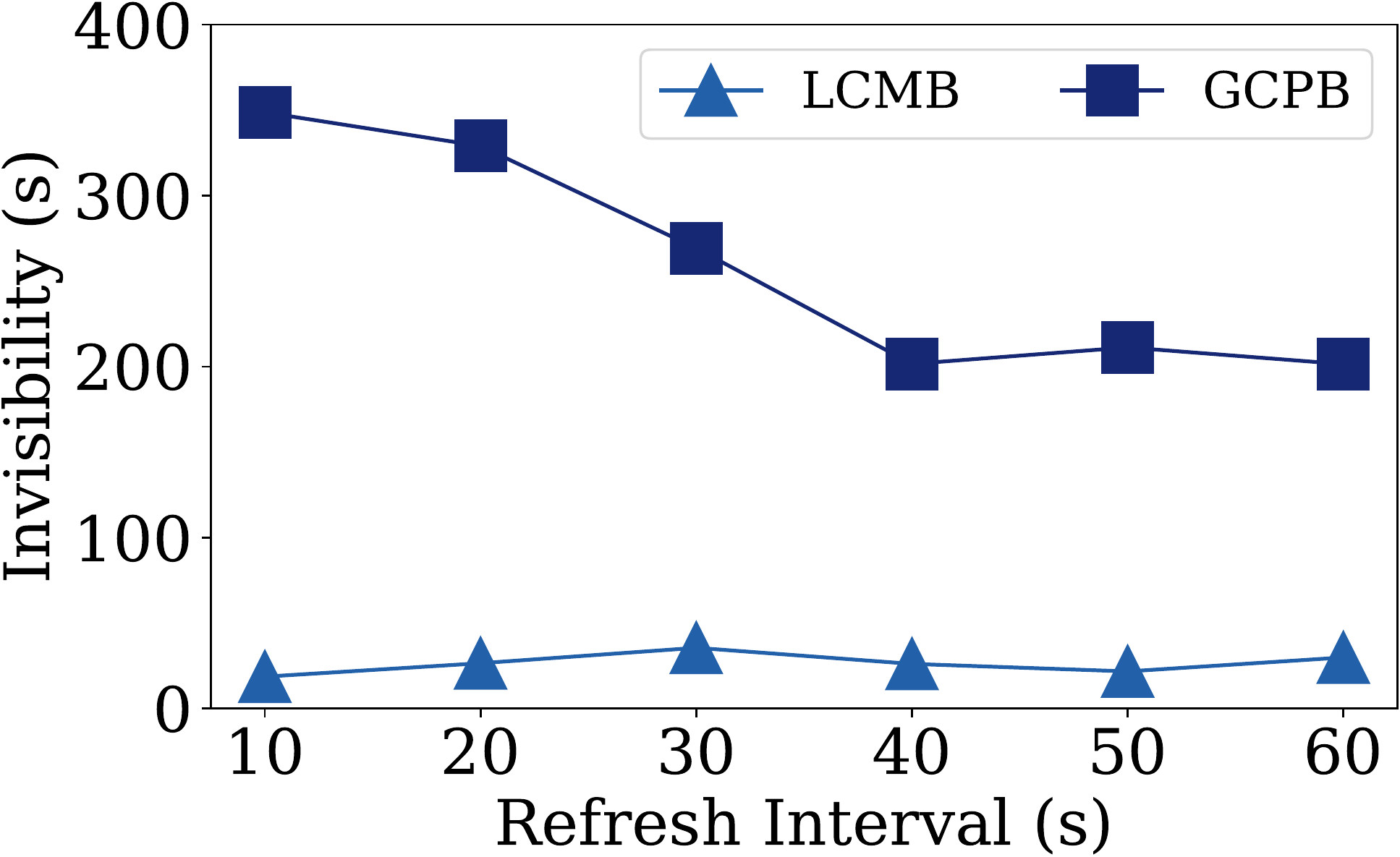}
        \vspace{-5mm}
        \subcaption{\UaMetric}
        \label{fig:exp_refresh_iv}
      \end{minipage} 
  \end{tabular}
  \vspace{-6mm}
  \caption{\rone{Evaluation of varied refresh intervals}}
  \label{fig:exp_refresh_interval}
  \vspace{-5mm}
\end{figure}

\begin{figure*}[t]
\vspace{-15pt}
  \begin{tabular}{ccc}
      \begin{minipage}[b]{0.28\linewidth}
        \centering
        \includegraphics[width=\linewidth]{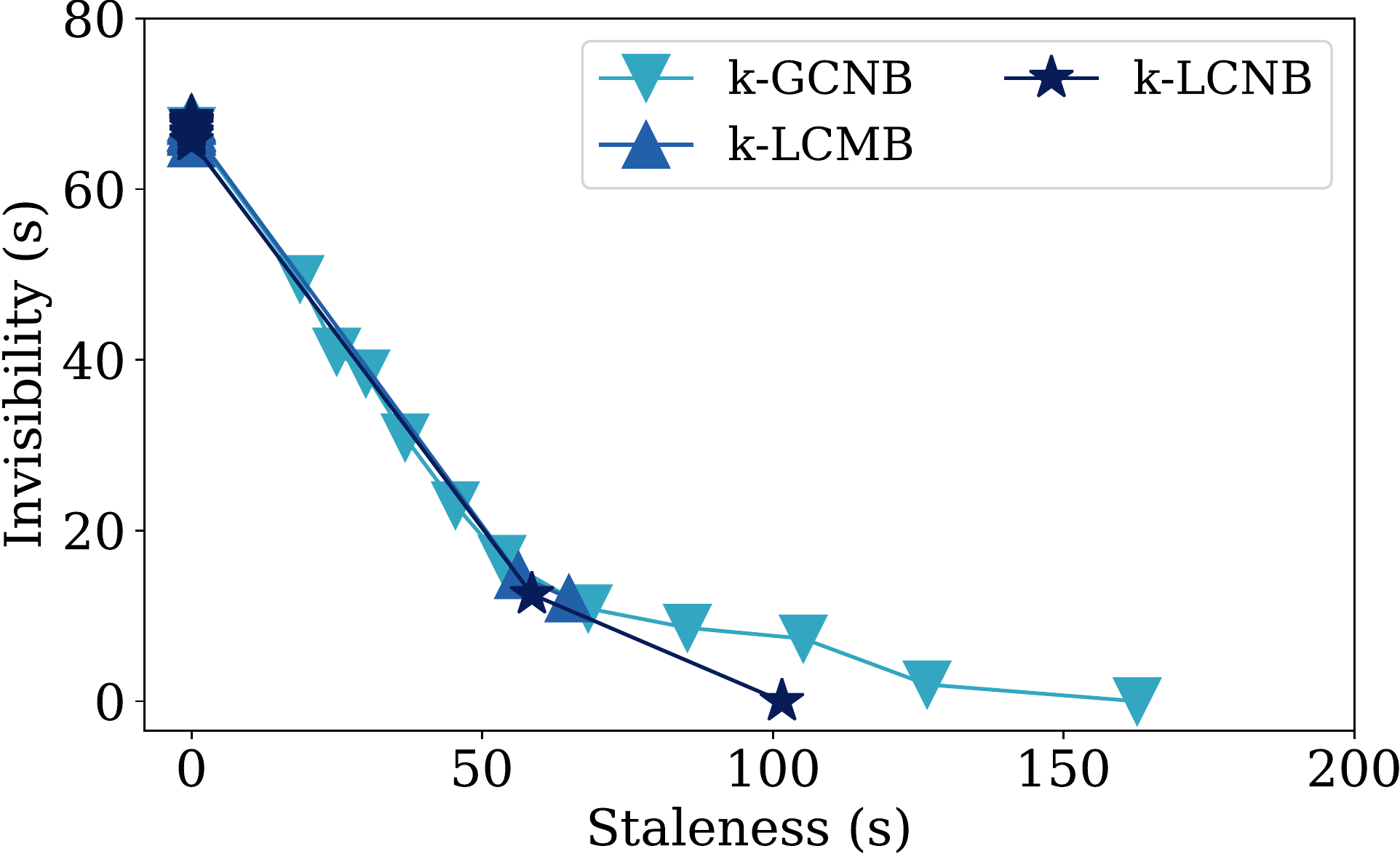}
        \vspace{-5mm}
        \subcaption{\regMove}
        \label{fig:exp_k_regular}
      \end{minipage} 

      \begin{minipage}[b]{0.28\linewidth}
        \centering
        \includegraphics[width=\linewidth]{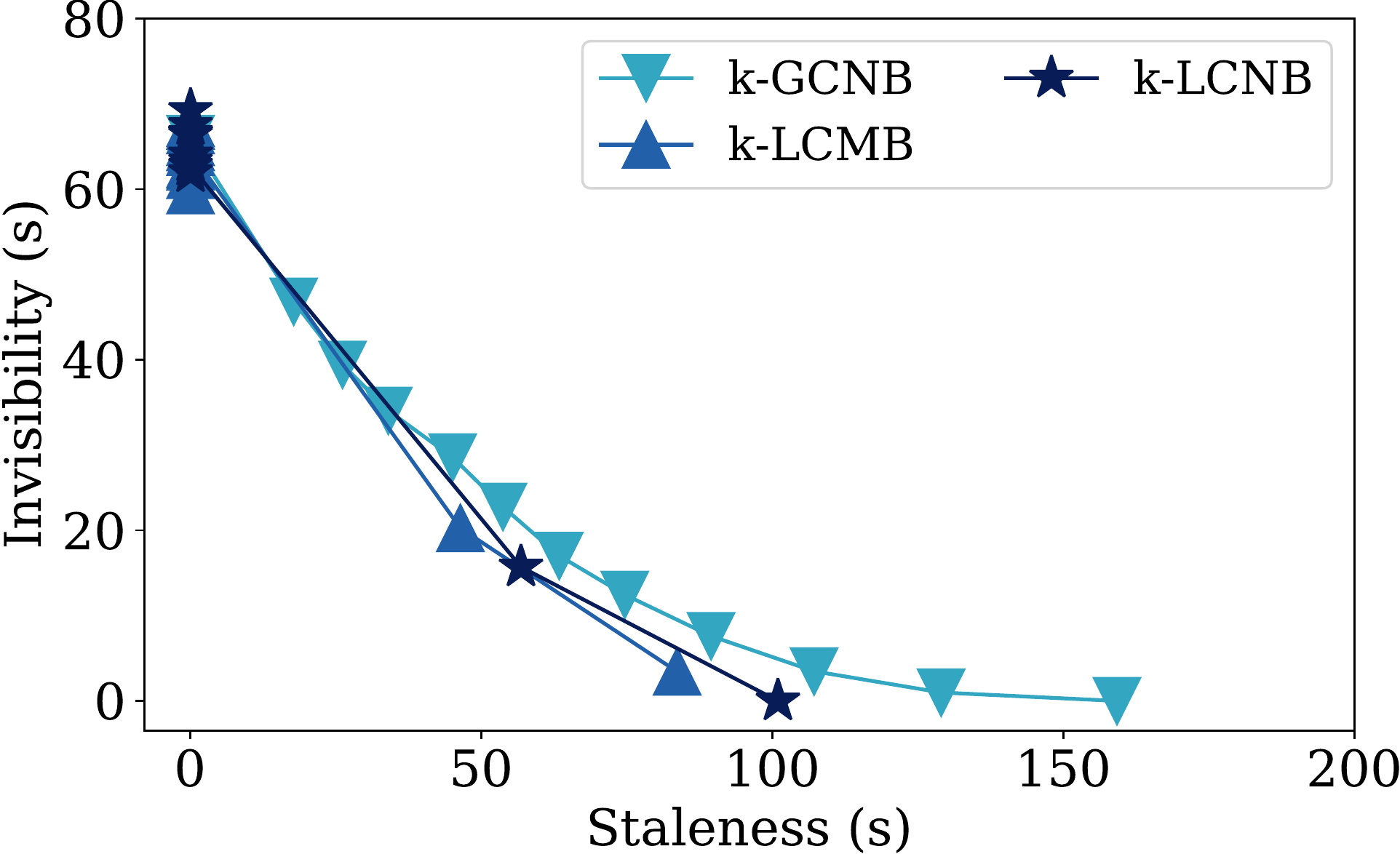}
        \vspace{-5mm}
        \subcaption{\ranMove}
        \label{fig:exp_k_random}
      \end{minipage}
      
      \begin{minipage}[b]{0.28\linewidth}
        \centering
        \includegraphics[width=\linewidth]{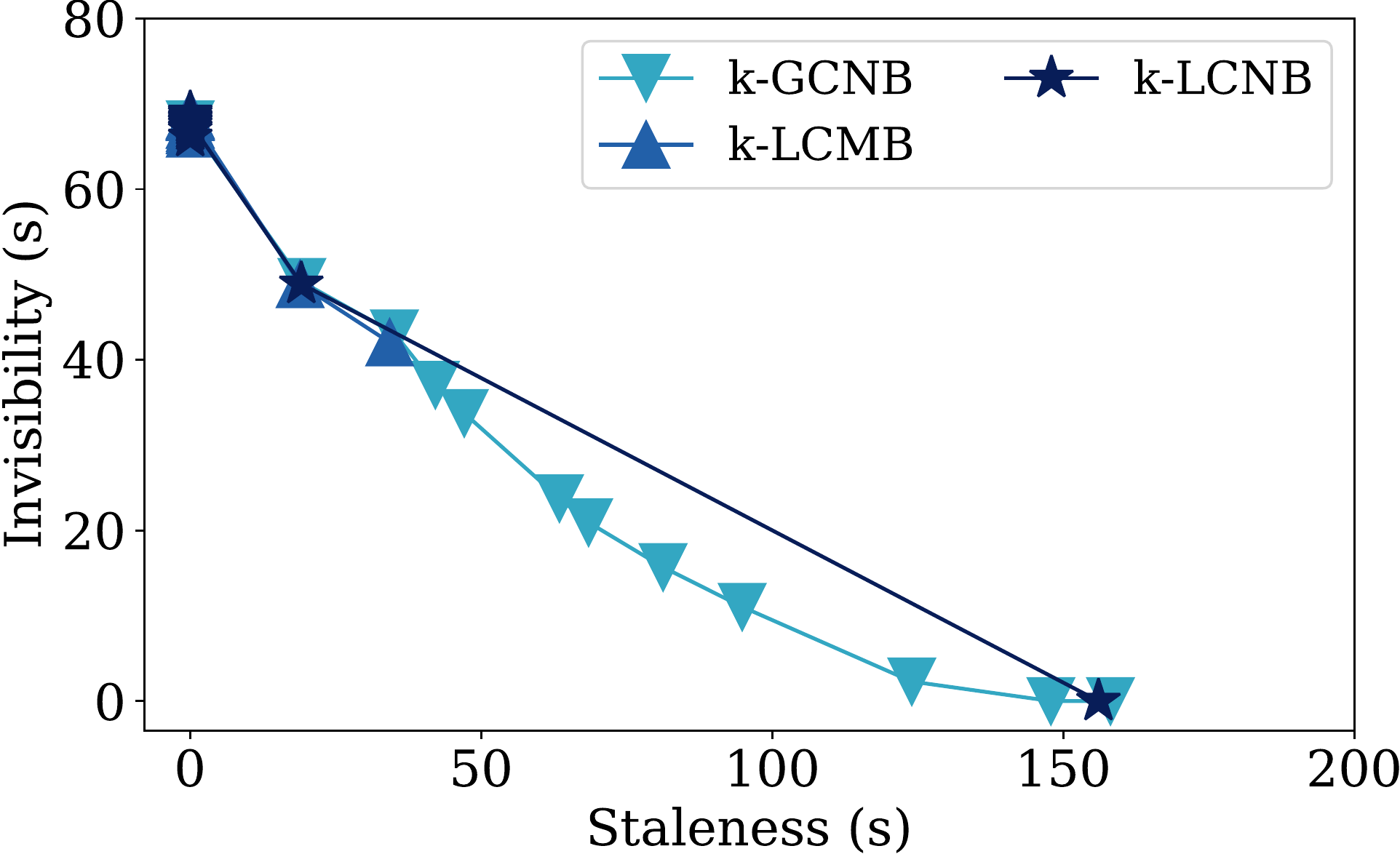}
        \vspace{-5mm}
        \subcaption{\waitMove}
        \label{fig:exp_k_see}
      \end{minipage} 
  \end{tabular}
  \vspace{-6mm}
  \caption{Trade-off between \uaMetric and \snMetric with varied $k$ values}
  \label{fig:exp_k_trade_off}
  \vspace{-4mm}
\end{figure*}

\begin{figure*}[t]
  \begin{tabular}{cccc}
      \begin{minipage}[b]{0.24\linewidth}
        \centering
        \includegraphics[width=\linewidth]{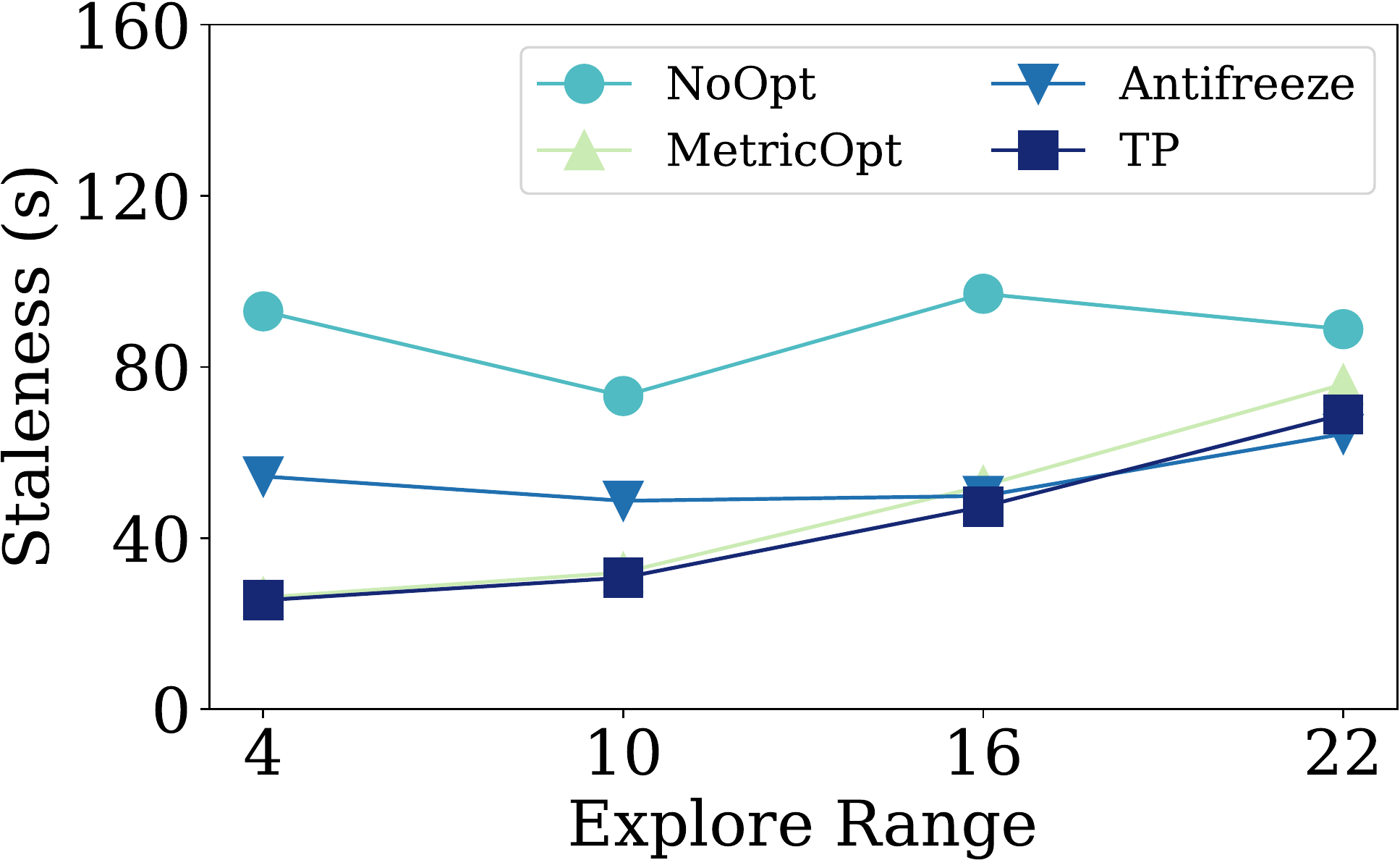}
        \vspace{-5mm}
        \subcaption{\pICNB}
        \label{fig:exp_opt_sl_icnb}
      \end{minipage}
      
      \begin{minipage}[b]{0.24\linewidth}
        \centering
        \includegraphics[width=\linewidth]{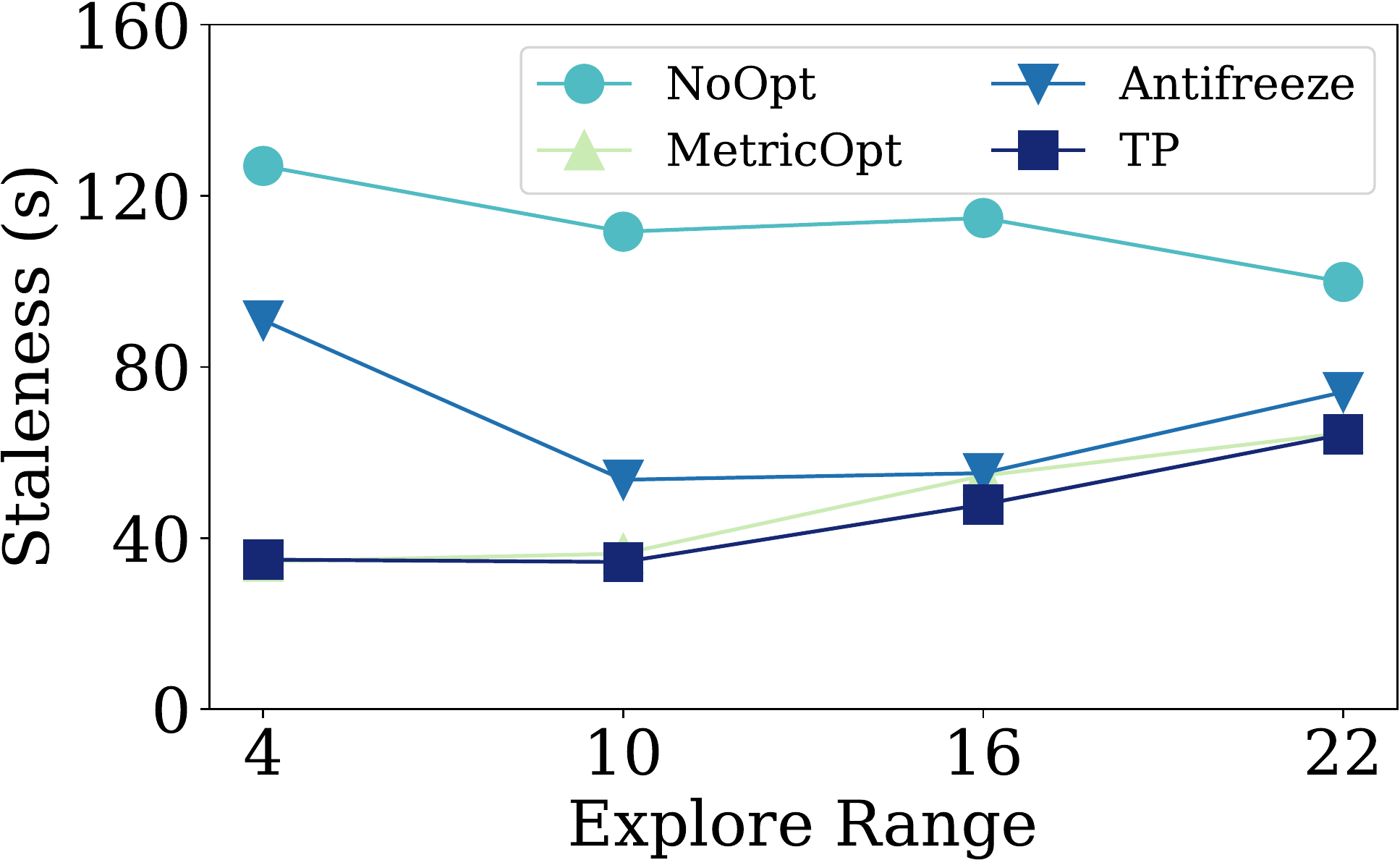}
        \vspace{-5mm}
        \subcaption{\pLCMB}
        \label{fig:exp_opt_sl_lcmb}
      \end{minipage} 

      \begin{minipage}[b]{0.24\linewidth}
        \centering
        \includegraphics[width=\linewidth]{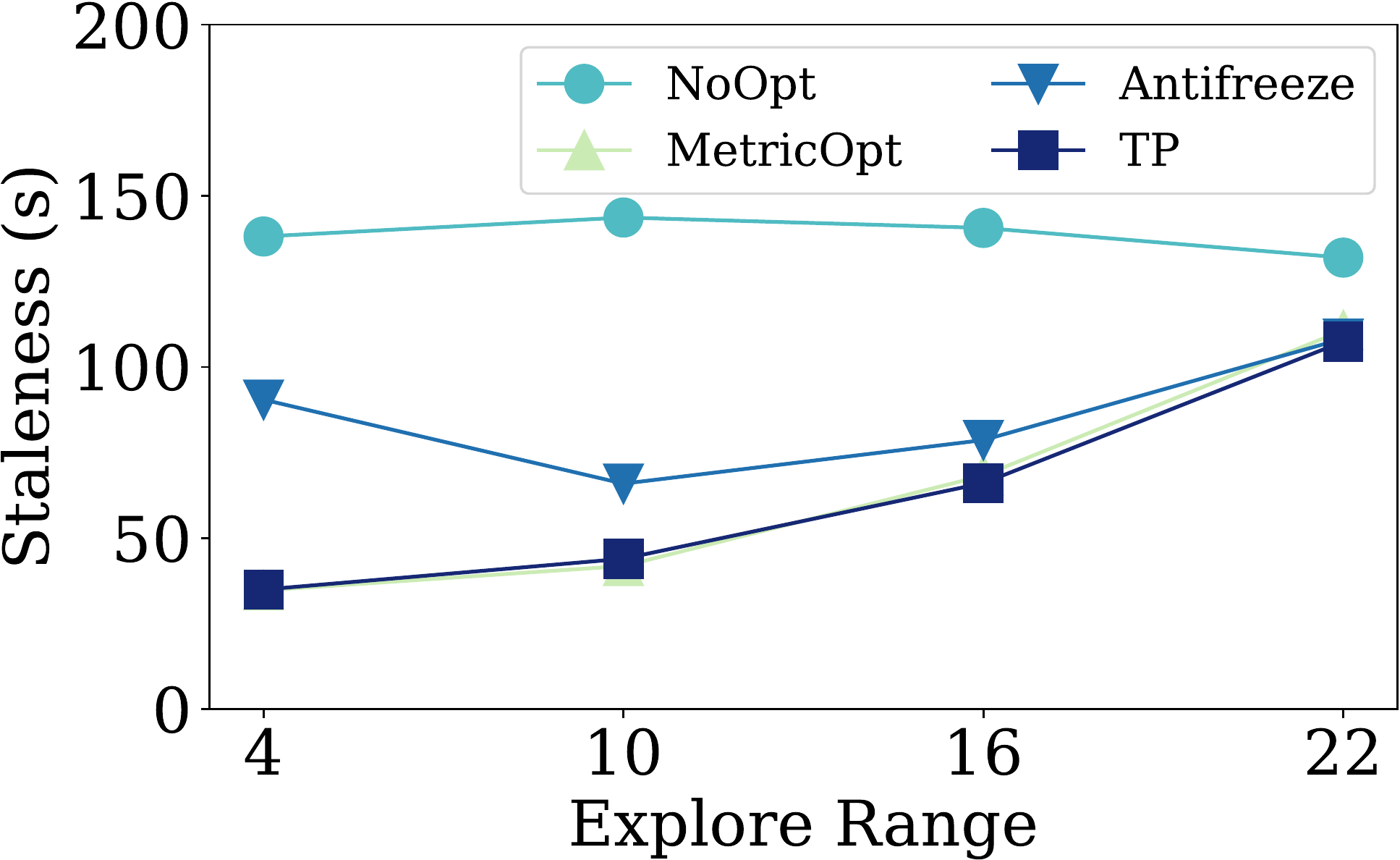}
        \vspace{-5mm}
        \subcaption{\pLCNB}
        \label{fig:exp_opt_sl_lcnb}
      \end{minipage}
      
      \begin{minipage}[b]{0.24\linewidth}
        \centering
        \includegraphics[width=\linewidth]{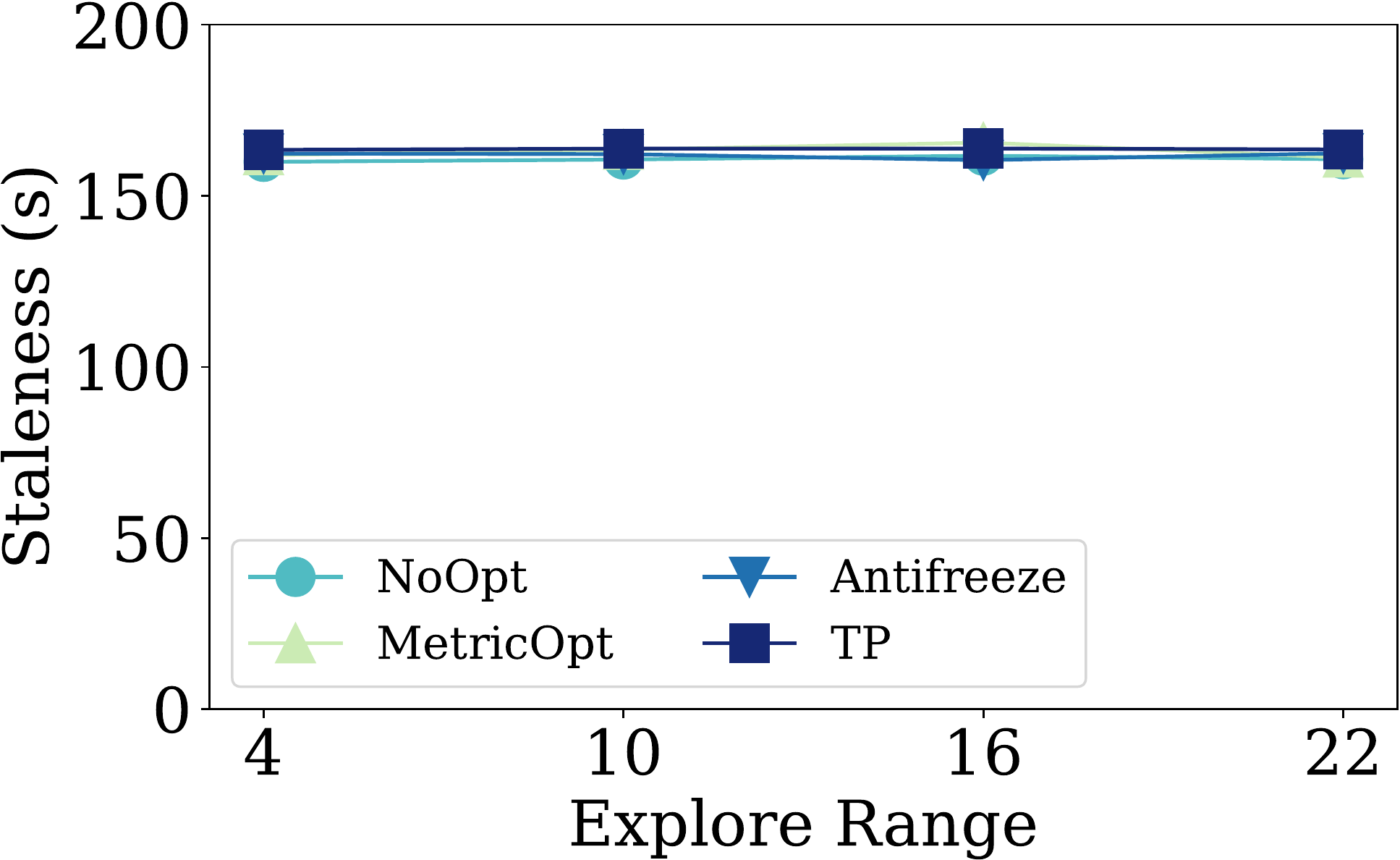}
        \vspace{-5mm}
        \subcaption{\pGCNB}
        \label{fig:exp_opt_sl_gcnb}
      \end{minipage} 
  \end{tabular}
  \vspace{-6mm}
  \caption{\rone{\small Evaluation of scheduler optimizations (\snMetric)}}
  \label{fig:exp_opt_sl}
  \vspace{-5mm}
\end{figure*}

\begin{figure}[t]
  \begin{tabular}{cc}
      \begin{minipage}[b]{0.48\linewidth}
        \centering
        \includegraphics[width=\linewidth]{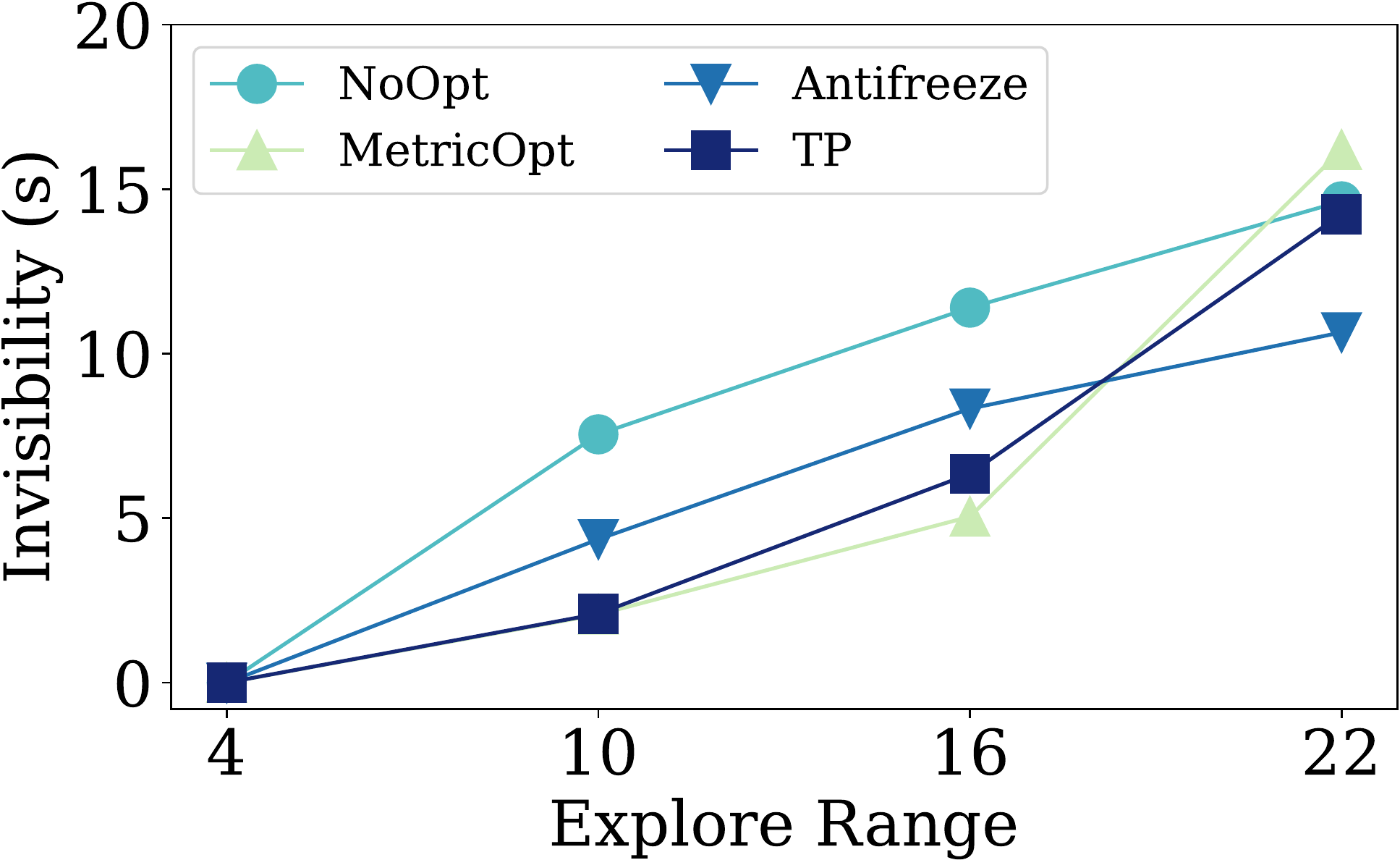}
        \vspace{-5mm}
        \subcaption{\pLCMB}
        \label{fig:exp_opt_iv_lcmb}
      \end{minipage}
      
      \begin{minipage}[b]{0.48\linewidth}
        \centering
        \includegraphics[width=\linewidth]{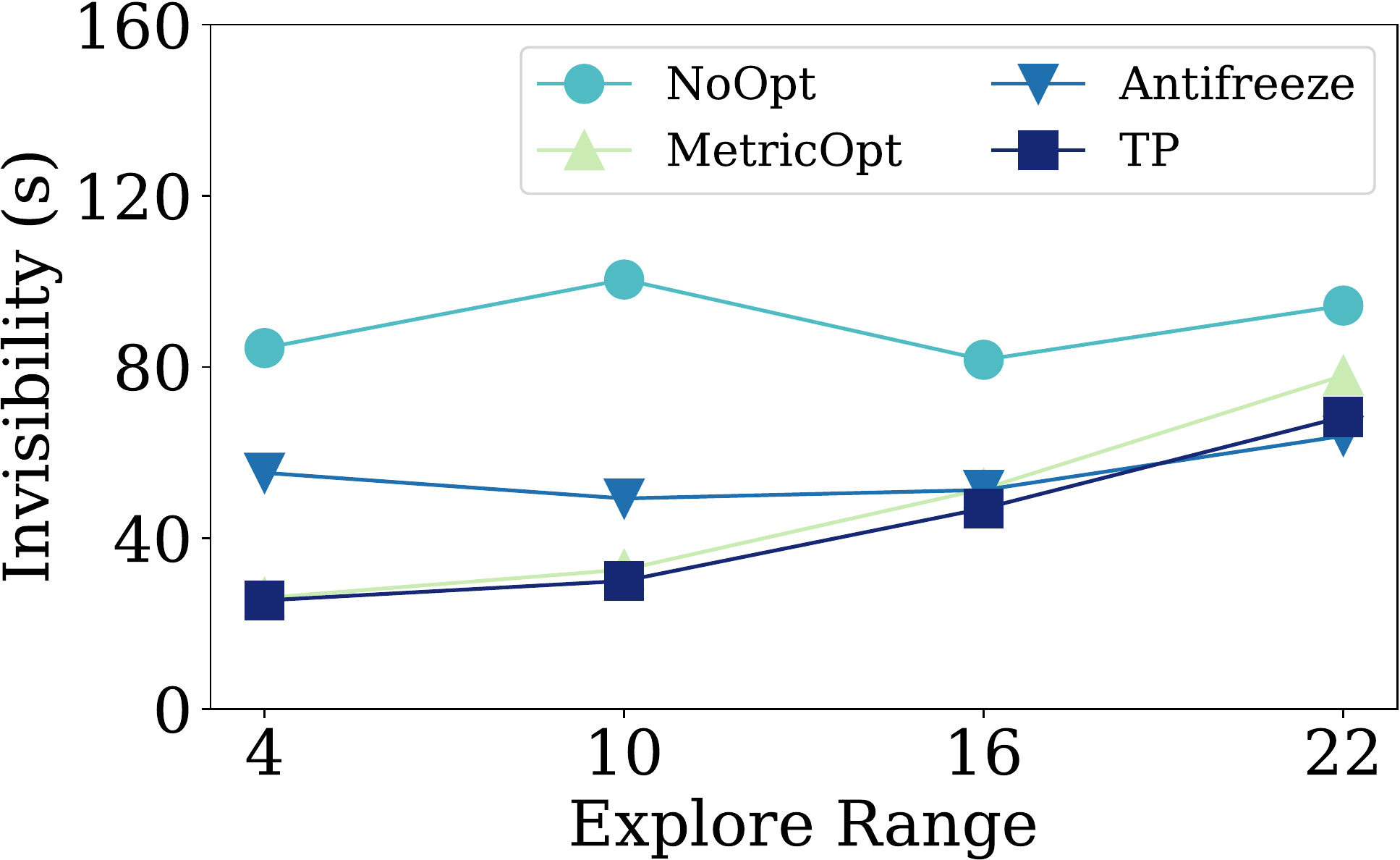}
        \vspace{-5mm}
        \subcaption{\pGCPB}
        \label{fig:exp_opt_iv_gcpb}
      \end{minipage} 
  \end{tabular}
  \vspace{-5mm}
  \caption{\rone{\small Evaluation of scheduler optimizations (\uaMetric)}}
  \label{fig:exp_opt_iv}
  \vspace{-5mm}
\end{figure}

%

To better understand the behavior of different \approaches, 
we further report the returned 
number of \unavailable and stale views by \readTxn{s} 
for \regMove while we are processing the \writeTxn. 
We aggregate the \readTxn{s} 
that finish for every 2s 
and report the mean in Figure~\ref{fig:exp_trace}. 
The areas under the curve represent 
the \uaMetric/\snMetric in the respective figures. 
In Figure~\ref{fig:exp_trace_iv}, 
\pGCPB initially returns many \ua{s} 
as it reads the new \version, 
and the number of \ua{s} decreases as we compute more \qres{s}. 
Specifically, for the first 10s, 
a user sees more than 3 \ua{s} on average, 
out of the 4 visualizations in the viewport. 
Therefore, this user cannot interact  
for the first 10s, significantly diminishing interactivity. 
\pLCMB, on the other hand, initially reads the \oldG  
to avoid \unavailable views. 
Then, it reads the \newG (i.e., after 15s) 
to present fresh \res{s} to the user as \pGCPB does. 
Figure~\ref{fig:exp_trace_sl} shows the number of stale views over time. 
\pGCNB reads the same number of stale views as the \viewportSize 
(i.e., 4 in our test) until the last \txn. 
\pLCMB, \pLCNB, and \pICNB can read the new \version 
during refresh, which reduces  \snMetric. 

\stitle{\Explorerange} 
This experiment evaluates the impact of varied \exploreRange{s} 
on different \approaches. 
The results in Figure~\ref{fig:exp_explore_range} 
show that we have smaller \uaMetric/\snMetric for \pGCPB, 
\pLCMB, \pLCNB, and \pICNB when the user explores a smaller 
number of visualizations because these \approaches 
are more likely to read 
the new results for smaller \exploreRange{s}. 
The \snMetric for \pGCNB is independent of the value of \exploreRange 
since it only refreshes the visualizations after all of the \qres{s} 
for the new \version are computed. 

\stitle{\Viewportsize} 
Figure~\ref{fig:exp_viewport_size} reports the results 
of varying the \viewportSize{s}. 
The \uaMetric for \pGCPB increases as 
 \viewportSize increases because the user will 
read more \unavailable views for each \readTxn. 
However, the \uaMetric for \pLCMB slightly increases 
and then decreases to zero. 
The reason for decreasing \uaMetric 
is that a larger \viewportSize pushes \pLCMB to 
wait longer to read the \newG, which decreases \uaMetric. 
In an extreme case, when the \viewportSize covers the whole 
dashboard, \pLCMB's performance converges to \pGCNB, 
which has zero \uaMetric but the highest \snMetric, 
as shown in Figure~\ref{fig:exp_viewport_size_sl}. 
The \snMetric for \pGCNB, \pLCMB, \pLCNB, and \pICNB 
increases as they will read more stale views in one \readTxn. 
\rthree{Overall, when the user explores more visualizations, 
the difference on \uaMetric and \snMetric will become more 
significant across different \approach{es}.}

\begin{reviewone}
\stitle{\Refreshinterval}
We test three refreshes triggered periodically, where the interval 
between two succeeding refreshes (i.e., \prefreshinterval) 
is varied from 10s to 60s. 
Same as the test for one refresh, before starting each refresh, 
we insert 0.1\% new data to the database. 
We report the \snMetric and \uaMetric 
for different \approach{es}. 
Figure~\ref{fig:exp_refresh_interval} shows 
that a smaller \prefreshinterval introduces 
higher \snMetric and \uaMetric 
when the \prefreshinterval is less than the 
execution time for processing a refresh (i.e., 37s in our test) 
for all \approach{es} except \pGCNB. 
This is because while a \version of 
the \viewG is being computed, 
a smaller \prefreshinterval creates 
a new \version of the \viewG earlier. 
This leads the \qres{s} of the under-computation 
\version of the \viewG to become stale earlier, 
and, in turn the \snMetric and \uaMetric are increased. 
The \snMetric of \pGCNB is 
not impacted by the varied \prefreshinterval 
because \pGCNB presents the up-to-date \qres{s} 
to the user when it finishes all of the refreshes in the system 
and its \snMetric is determined by the 
execution time for finishing multiple refreshes, 
which is independent of the \prefreshinterval.
\end{reviewone}



\subsection{Performance of $K$-Relaxed Variants}
\label{sec:exp_k_relaxed}

\ptr{
\noindent\fbox{\begin{minipage}{80mm} \small
\textit{Takeaway: The $k$-\relaxed variants allow the user 
to gracefully explore the trade-off between \uaMetric and \snMetric, 
and enable more trade-off points that are not covered 
in \baseApproach{es}.}
\end{minipage}}}

\vspace{1mm}
\noindent We evaluate the impact of $k$  
for the $k$-\relaxed variants; 
Recall that $k$ represents the additional $\ua{s}$ 
permitted while reading the \newG  
for \pLCMB, \pLCNB, and \pGCNB. 
Here, we vary the $k$ from 0 to 22 with an interval of 2. 
Our results in Figure~\ref{fig:exp_k_relaxed}-\ref{fig:exp_k_trade_off}  
show that the $k$-\relaxed variants gracefully 
explore the trade-off between \uaMetric and \snMetric, 
and enable more trade-off points that are not covered 
in \baseApproach{es}.  
Figure~\ref{fig:exp_k_relaxed_iv} shows 
that as we admit more \ua{s}, the \uaMetric increases  
for the $k$-\relaxed variants. 
However, when $k$ becomes the same as or larger than 
the \viewportSize (i.e., 4 in our test), 
the \uaMetric does not change for \kLCNB and \kLCMB 
since they have converged to \pGCPB. 
However, \snMetric decreases 
as we have a larger $k$ as shown in Figure~\ref{fig:exp_k_relaxed_sl}. 

Figure~\ref{fig:exp_k_trade_off} shows the 
trade-offs between \uaMetric and \snMetric under three \readb{s}. 
We see that the $k$-\relaxed variants have different trade-offs 
for different \readb{s}. 
For example, for \regMove in Figure~\ref{fig:exp_k_regular} 
\kLCNB has better 
trade-offs than \kGCNB when the \snMetric is larger than 100s, 
meaning that for the same \uaMetric, \kLCNB has smaller 
\snMetric than \kGCNB. 
When the \snMetric is smaller than 100s, 
all of the three $k$-\relaxed variants stay on the same trade-off curve. 
For \ranMove, \kLCMB has the best trade-offs compared to the other two variants, and for \waitMove, \kGCNB has the best trade-offs.

\subsection{Effectiveness of Scheduler Optimizations}
\label{sec:exp_optimizations}

\ptr{
\noindent\fbox{\begin{minipage}{80mm}\small
\textit{Takeaway: The optimized scheduler in \tdb 
reduces \snMetric and \uaMetric in most cases.}
\end{minipage}}
}

\vspace{1mm}
\noindent 
\rone{This experiment evaluates the benefit 
and overhead of the scheduler optimizations
in \bothOpt (\pBothOpt for short). 
We compare \pBothOpt with three baselines: 
1) \noOpt, after updating base tables 
randomly picking a view to compute, which is from Superset; 
2) \execOpt, from existing work 
that prioritizes computing the view with 
the least execution time~\cite{Antifreeze}, 
which is the second factor in \pBothOpt's scheduling metric; 
3) \metricOpt, which prioritizes computing the 
view that introduces the most \uaMetric plus \snMetric. 
Since the \uaMetric and \snMetric is increased 
only when a view is read, 
\metricOpt effectively prioritizes computing 
the view that the user spent the most time reading, 
which corresponds to the first factor in \pBothOpt's 
scheduling metric (i.e., $\RTime_{k}^{t_i}$). 
Recall that the effectiveness of $\RTime_{k}^{t_i}$ 
depends on the property that 
a view that was read more in the past is more likely 
to be read in the future, which we call 
{\em temporal locality}. 
To study impact of temporal locality, we vary \exploreRange{s} 
and report \snMetric and \uaMetric for \baseApproach{es} 
under different scheduling metrics. }


%

Figure~\ref{fig:exp_opt_sl} shows that 
\pBothOpt has smaller \snMetric compared to \noOpt and \execOpt 
for all \approaches. 
The performance benefit of \pBothOpt over the baselines 
is larger when we have smaller \exploreRange{s}. 
Specifically, \pBothOpt reduces \snMetric by up to 75\%
and 62\% compared to \noOpt and \execOpt, respectively. 
\pBothOpt and the baselines have the same \snMetric for \pGCNB 
because \pGCNB refreshes the views after all of the new \res{s} 
are computed and its \snMetric is independent of a scheduler policy. 
Figure~\ref{fig:exp_opt_iv} shows that \pBothOpt has smaller 
\uaMetric compared to \noOpt and \execOpt in most cases. 
Similar to the results of \snMetric, \pBothOpt has greater
benefit when  \exploreRange is smaller except for \pLCMB 
with \exploreRange 4. Here, the \exploreRange equals the 
\viewportSize, so \pLCMB does not have \uaMetric. 
Overall, \pBothOpt reduces \uaMetric by up to 70\%
and 54\% compared to \noOpt and \execOpt, respectively. 
However, \pBothOpt may have higher \uaMetric 
than \execOpt when the locality of reading the \viewG weakens, 
such as for \pLCMB with \exploreRange being 22. 
In this case, \pBothOpt increases \uaMetric by 33\%. 

\rone{\metricOpt has similar \res{s} compared to \pBothOpt 
since \metricOpt also prevails when there is strong 
temporal locality. However, when the locality weakens 
(e.g., the \exploreRange is 22), 
\pBothOpt has lower \snMetric and \uaMetric 
because \pBothOpt additionally considers the different 
execution time for refreshing different views 
(i.e., the factor from \execOpt). 
Specifically, \pBothOpt reduces \snMetric and \uaMetric 
by up to 12\% and 13\%, respectively, compared to \metricOpt.} 

%% file: related_conclusion.tex
\section{Related work}
\label{sec:related}
Our work is related to work in transaction processing, view maintenance and 
stream processing, and rendering results 
in interfaces. 

\stitle{Transaction processing} 
There is a long line of work on improving the 
performance of transaction processing 
while maintaining guarantees such as serializability or snapshot isolation~\cite{CCYuBPDS14, TangJE17ACC, CahillRF09SSI, Newmann15MultiVersion, WangJFP17Serializable, ChenYKAAS22Plor, ZhangH0L22Skeena, LuHNML16SNOW}. 
For example, the 
SNOW Theorem~\cite{LuHNML16SNOW} studies fundamental trade-offs 
between power (e.g., consistency level) 
and latency of read-only transactions, and defines 
the properties read-only transactions 
can maintain simultaneously. 
On the other hand, SAGAS~\cite{Garcia-Molina87Sagas} 
breaks up a long-lived transaction 
into sub-transactions, which can interleave with other 
concurrent transactions to improve performance. 
However, none of these projects consider maintaining \con 
while reading uncommitted results or other
desired user properties in visual interfaces, such as \ava and \mon. 
SafeHome~\cite{AhsanYNG21Safehome} adapts \txn{s} 
to define atomicity and serializability for concurrent routines 
in smart homes, and includes a series of visibility models 
to trade off between performance and user visibility 
(i.e., what intermediate states of smart devices are visible to users). 
\Tdb is different from SafeHome because we focus on 
an end-user data analysis scenario; 
so the \macPro are not considered in SafeHome. 
\ptr{The definition of ``\ava'' in SafeHome is also different ours.} 

\stitle{View maintenance and stream processing}
Many papers propose various efficient incremental view maintenance algorithms~\cite{Gupta93IVM, LazyIVM,DeferredIVM, MVPolicy, MVBook, nikolic2016win, DBT, TangSEKF20InQP}. 
These techniques are orthogonal to our model and can be used 
to improve performance. 
S-Store~\cite{MeehanTZACDKMMP15SStore} and 
transactional stream processing~\cite{BotanFKT12Transaction} 
integrate transactions into stream processing to guarantee 
consistency for shared states. 
In a related vein, 
Golab and Johnson~\cite{GolabJ11Consistency} study different
consistency levels for materialized views in a stream warehouse 
with respect to the source data, 
while Zhuge et al.~\cite{ZhugeGW97Multiview} allow users to 
define multiple views to be refreshed consistently. 
\Tdb is different from these work because they do not consider 
the user's semantics of consuming the results in a visual interface 
along with properties such as \mon and \ava. 

\stitle{Rendering analysis results in a visual interface}
As summarized in Table~\ref{fig:existing_tools}, 
many existing data analysis tools\ptr{, including Excel~\cite{msexcel}, 
Google Sheets~\cite{sheets}, Dataspread~\cite{Antifreeze}, 
Libre~Calc~\cite{libreCalc}, Superset~\cite{superset}, Power ~BI~\cite{powerBI}, and Tableau~\cite{tableau}} 
make fixed choices on the properties maintained 
while rendering analysis results with respect to an update. 
Interaction Snapshots~\cite{WuCH020InteractionSnapshots} 
additionally presents
a scaled-down display of the dashboard for each interaction 
(e.g., cross-filter), 
where this scaled-down version serves as the new snapshot,
with an indicator for whether the new snapshot is computed. 
This way, the user can interact with the old snapshot and 
replace it with the new snapshot later, similar to \pGCNB. 
However, Interaction Snapshot does not 
allow a user to read uncommitted results 
and choose the different properties they desire. 
Another line of research renders approximate 
results~\cite{raman1999online, MoritzFD017:Trust,rahman2017ve,zgraggen2016progressive, SynopsisZhao0L20} 
and refines them later;
we don't use approximation. 

\section{Conclusion\ptr{ and Future Work}}
\label{sec:conclusion} 
We introduced \tdb, a \framework that explores the fundamental 
trade-offs between \mon, \con, and \ava 
when a user examines \res{s} in a visual interface under updates. 
We identified feasible property combinations---and their \approaches---based on the \macThe,
as well as new performance metrics, following 
it up by proving ordering relationships between various
\approaches for the metrics. 
We additionally designed new algorithms 
for efficiently maintaining different property combinations 
and processing updates. 
We implemented \tdb and its constituent \approaches 
in a popular BI tool, Superset. 
Our experiments demonstrated 
significant performance differences across our \approaches for various workloads,
illustrating the benefit of our framework and 
newly discovered property combinations. 
\ptr{

}
We believe our \tdb framework is the first step in a new research
direction around {\em bringing transactional notions to end-user analytics/BI}, 
with a human continuously ``in-the-loop''.
\ptr{With an increasing number of low/no-code BI tools becoming available,
especially on large, continuously changing datasets,
the need for ensuring correct and efficient user perception in these tools---via properties
such as \mon, \con, \ava---has never been greater.
There are many open questions that still remain.
From the UI design standpoint, 
we need effective ways to communicate the semantics 
of different properties and \approaches to different personas 
(i.e., dashboard designers vs. end-users) as well as
intuitive
ways for users to trade off \snMetric and \uaMetric. 
While our focus has been on introducing the \tdb framework,
we additionally want to explore the types of data-driven decisions
for which each \approach is best suited for, as well as  
the concrete use cases and verticals. 
Our work also opens up a larger research question: 
\emph{as groups of users with varying analytical abilities
more closely interact with data analysis tools, 
what critical user-facing properties are desired by users---and what are the right abstractions 
to model such interactions
while maintaining the related properties?}
While \txn{s} seem to be a natural abstraction, 
adapting \txn{s} for each user-facing data analysis scenario 
still requires work and it is unclear 
what the general abstraction should be across scenarios. 
\Tdb is a useful starting point in this regard.}

